\documentclass[12pt,draftcls,onecolumn]{IEEEtran}
\IEEEoverridecommandlockouts
\usepackage{cite}
\usepackage{amsmath}
\usepackage{amssymb}
\usepackage{amsfonts}
\usepackage{algorithmic}
\usepackage{algorithm}
\usepackage{color}
\usepackage{xcolor}   
\usepackage{amsthm}    
\usepackage{graphicx}  
\usepackage{multirow}  
\usepackage{etoolbox}
\usepackage{mathrsfs}
\usepackage{epsfig}
\usepackage[outdir=./]{epstopdf}
\usepackage{psfrag}
\usepackage{subfigure}
\usepackage{times}
\usepackage{stmaryrd}
\usepackage{subfig} 
\usepackage{float}
\usepackage{url}
\usepackage{enumerate}
\usepackage{longtable}
\usepackage{makecell}
\usepackage{multirow}
\usepackage{bm}
\usepackage{multicol}
\usepackage{setspace}
\newcommand{\subparagraph}{}
\usepackage{titlesec}
\titlespacing{\section}{0pt}{1pt}{1pt}
\titlespacing{\subsection}{0pt}{1pt}{1pt}
\titlespacing{\subsubsection}{0pt}{1pt}{1pt}

\DeclareMathOperator*{\argmax}{argmax}
\newtheorem{Thm}{Theorem}
\newtheorem{Lem}{Lemma}
\newtheorem{Cor}{Corollary}

\newtheorem{Prob}{Problem}
\newtheorem{Rem}{Remark}

\newcommand{\mcthr}{\textcolor{black}}

\newcommand{\mcblk}{\textcolor{black}}
\newcommand{\diag}{\mathrm{diag}}

\makeatletter
\allowdisplaybreaks[4]

\IEEEoverridecommandlockouts
\def\BibTeX{{\rm B\kern-.05em{\sc i\kern-.025em b}\kern-.08em
		T\kern-.1667em\lower.7ex\hbox{E}\kern-.125emX}}

\title{Analysis and Optimization of  A Double-IRS  Cooperatively Assisted System with A Quasi-Static Phase Shift Design}
\author{
	\IEEEauthorblockN{Gengfa Ding, Feng Yang, Lianghui Ding, and Ying Cui
		\thanks{The authors are with the Department of Electronic Engineering, Shanghai Jiao Tong University, Shanghai 200240, China. The paper will be presented in part in SPAWC 2022~\cite{Ding2207:Quasi}.}
	}
}
\begin{document}	
	\maketitle
	\vspace*{-15mm}
	\begin{abstract}
		\vspace*{-2mm}
		The analysis and optimization of single intelligent reflecting surface (IRS)-assisted systems have been extensively studied, whereas little is known regarding multiple-IRS-assisted systems. This paper investigates the analysis and optimization of a double-IRS cooperatively assisted downlink system \mcthr{(D-IRS-C)}, where a multi-antenna base station (BS) serves a single-antenna user with the help of two multi-element IRSs, connected by an inter-IRS channel. The channel between any two nodes is modeled with Rician fading. The BS adopts the instantaneous CSI-adaptive maximum-ratio transmission (MRT) beamformer, and the two IRSs adopt a cooperative quasi-static phase shift design. The goal is to maximize the average achievable rate, which can be reflected by the average channel power of the equivalent channel between the BS and user at \mcthr{low channel estimation and phase adjustment costs} and computational complexity. First, we obtain tractable expressions of the average channel power of the equivalent channel in the general (Rician factor), pure line of sight (LoS), and pure non-line of sight (NLoS) regimes, respectively. Then, we jointly optimize the phase shifts of the two IRSs to maximize the average channel power of the equivalent channel in these regimes. The optimization problems are challenging non-convex problems. We  obtain globally optimal closed-form solutions for some cases and propose computationally efficient iterative algorithms  to obtain stationary points for the other cases. Next, we compare the computational complexity for optimizing the phase shifts and the  optimal average channel power of \mcthr{D-IRS-C} with those of  \mcthr{a counterpart double-IRS non-cooperatively assisted system (D-IRS-NC)} and a counterpart single-IRS-assisted system \mcthr{(S-IRS)} at a large number of reflecting elements in the three regimes. Finally, we numerically demonstrate notable gains of the proposed  solutions over the existing solutions at different system parameters. To our knowledge, this is the first work that optimizes the quasi-static phase shift design of \mcthr{D-IRS-C} and characterizes its advantages over the optimal quasi-static phase shift design of the counterpart \mcthr{D-IRS-NC} and \mcthr{S-IRS}.
	\end{abstract}
	\vspace*{-6mm}
	\begin{IEEEkeywords}
		\vspace*{-4mm}
		Intelligent reflecting surface (IRS), double IRSs, cooperation, quasi-static phase shift design, coordinate descent, optimization.
	\end{IEEEkeywords}
	\section{Introduction} \label{sec:Introduction}
	Stringent requirements for future wireless networks, such as ultra-high data rate and energy efficiency, cannot be fully achieved with the existing wireless communication technologies. Intelligent reflecting surface (IRS), which consists of nearly passive, low-cost, reflecting elements with reconfigurable parameters, has been recently recognized as a promising solution for improving spectrum and energy efficiency~\cite{ZhangRui2020survey}. There have been extensive studies on IRS. In what follows, we restrict our attention to the optimization of base station (BS) beamforming and IRS phase shifts for an IRS-assisted single-cell network. In the existing works on optimal designs for IRS-assisted systems, BS beamforming designs are usually adaptive to  instantaneous channel state information (CSI), whereas the phase shift designs can be classified into instantaneous CSI-adaptive phase shift designs (which adapt to instantaneous CSI)~\cite{ZhangRui2019SDR,Liang2019WSR,Yu2019BCDMM,Huang2019EE,Yu2020SCA,Yuan2020MultiIRS} and quasi-static phase shift designs (which adapt to CSI statistics and do not change over time slots during a certain  period)~\cite{JinShi2019MISO,Cui2020Interference,jia2021robust,Cui2020Outage,Zhao2021TTS,Cui2019Outage,Kwan2021MISO}. Notice that a quasi-static (also termed statistical) phase shift design yields a low phase adjustment cost at some performance sacrifice compared with an instantaneous CSI-adaptive phase shift design. Considering the practical implementation issue, a quasi-static phase shift design may be more valuable~\cite{JinShi2019MISO,Cui2020Interference,jia2021robust,Cui2020Outage,Zhao2021TTS,Cui2019Outage,Kwan2021MISO}.
	
	Early research on IRS investigates the optimal design for an IRS-assisted system with a single IRS. For example, in~\cite{ZhangRui2019SDR,Liang2019WSR,Yu2019BCDMM,Huang2019EE,Cui2020Interference,jia2021robust,JinShi2019MISO,Cui2020Outage,Zhao2021TTS}, the authors optimize the BS beamformer and IRS phase shifts to minimize the transmit power~\cite{ZhangRui2019SDR}, outage probability~\cite{Cui2020Outage}, and average transmit power~\cite{Zhao2021TTS} and maximize the weighted sum rate~\cite{Liang2019WSR}, secrecy rate~\cite{Yu2019BCDMM}, energy efficiency~\cite{Huang2019EE}, and ergodic rate~\cite{JinShi2019MISO,Cui2020Interference,jia2021robust}. Specifically, \cite{ZhangRui2019SDR,Liang2019WSR,Yu2019BCDMM,Huang2019EE} adopt instantaneous CSI-adaptive phase shift designs, whereas~\cite{JinShi2019MISO,Cui2020Interference,jia2021robust,Cui2020Outage,Zhao2021TTS} consider quasi-static phase shift designs.
		
	Later study regarding IRS concentrates on the optimal design for an IRS-assisted system with more than one IRS.  For instance, in~\cite{Yu2020SCA,Yuan2020MultiIRS,Cui2019Outage,Kwan2021MISO}, the authors optimize the BS beamformer and IRS phase shifts to minimize the outage probability~\cite{Cui2019Outage} and maximize the sum secrecy rate~\cite{Yu2020SCA}, received signal power \cite{Yuan2020MultiIRS}, and ergodic achievable rate~\cite{Kwan2021MISO}. In particular, \cite{Yu2020SCA,Yuan2020MultiIRS} adopt instantaneous CSI-adaptive phase shift designs, whereas \cite{Cui2019Outage,Kwan2021MISO} consider quasi-static phase shift designs. Notice that the abovementioned works on multi-IRS-assisted systems \cite{Yu2020SCA,Yuan2020MultiIRS,Cui2019Outage,Kwan2021MISO} ignore channels between any two IRSs, referred to as inter-IRS channels, and hence cannot capture their interaction. To achieve the full potential of multi-IRS-assisted systems, \cite{You2021doubleIRSSISO,ZhangRui2020doubleIRSMIMOCE,Han2020doubleSISO,Zhangrui2021Multireflection,ZhangRui2021doubleIRSMIMO,Han2021doubleMIMO,Dong2021doubleIRSsecure,Mei2022MBMH} explicitly \mcthr{model inter-IRS channels in double-IRS-assisted systems} \cite{You2021doubleIRSSISO,ZhangRui2020doubleIRSMIMOCE,Han2020doubleSISO,Zhangrui2021Multireflection,ZhangRui2021doubleIRSMIMO,Han2021doubleMIMO,Dong2021doubleIRSsecure} and \mcthr{multi-IRS-assisted systems \cite{Mei2022MBMH}} and \mcthr{consider IRS cooperation}. Specifically, \mcthr{in \cite{You2021doubleIRSSISO,ZhangRui2020doubleIRSMIMOCE}, the authors  propose  cascaded channel estimation methods for  double-IRS cooperatively assisted systems}. In~\cite{Han2020doubleSISO,Zhangrui2021Multireflection}, the authors show that the power gain with respect to (w.r.t.) the total number of  reflecting elements of a double-IRS-assisted system is higher in order than that of the counterpart single-IRS-assisted system. In~\cite{ZhangRui2021doubleIRSMIMO,Han2021doubleMIMO,Dong2021doubleIRSsecure,Mei2022MBMH}, the authors optimize the BS beamformer and IRS phase shifts to maximize the signal-to-interference-plus-noise ratio (SINR)~\cite{ZhangRui2021doubleIRSMIMO}, capacity~\cite{Han2021doubleMIMO}, secrecy rate~\cite{Dong2021doubleIRSsecure}, and \mcthr{received signal power~\cite{Mei2022MBMH}}.
	
	The existing works on double-IRS cooperatively assisted systems~\cite{Han2020doubleSISO,Zhangrui2021Multireflection,You2021doubleIRSSISO,ZhangRui2020doubleIRSMIMOCE,ZhangRui2021doubleIRSMIMO,Han2021doubleMIMO,Dong2021doubleIRSsecure,Mei2022MBMH} \mcthr{have two main limitations}. Firstly, \cite{Han2020doubleSISO,Zhangrui2021Multireflection,You2021doubleIRSSISO,ZhangRui2020doubleIRSMIMOCE,ZhangRui2021doubleIRSMIMO,Han2021doubleMIMO,Dong2021doubleIRSsecure,Mei2022MBMH} simply assume that the direct \mcthr{channel} between the BS and user is entirely blocked for tractability. Secondly, \cite{Han2020doubleSISO,Zhangrui2021Multireflection,You2021doubleIRSSISO,ZhangRui2020doubleIRSMIMOCE,ZhangRui2021doubleIRSMIMO,Han2021doubleMIMO,Dong2021doubleIRSsecure,Mei2022MBMH} solely consider instantaneous CSI-adaptive phase shift designs, which \mcthr{incur higher channel estimation and phase adjustment costs and computational complexities.} Therefore, the analysis and optimization of the double-IRS cooperatively assisted systems with a \mcthr{more} general channel model \mcthr{(e.g., capturing the direct channel)} and \mcthr{a more cost-effective} phase shift design remain open.
	
	In this paper, we shall shed some light on the above issue. Specifically, we consider a \mcthr{double-IRS cooperatively assisted system (D-IRS-C)}, where a multi-antenna BS serves a single-antenna user with the help of two multi-element IRSs connected by an inter-IRS channel. The antennas at the BS and the reflecting elements at the IRSs are arranged in uniform rectangular arrays (URAs).  The channel between any \mcthr{two nodes} is modeled with Rician fading. In particular, the line of sight (LoS) components do not change during the considered time, and the non-line of sight (NLoS) components vary from time slot to time slot. The BS adopts the instantaneous CSI-adaptive maximum-ratio transmission (MRT) beamformer. The two IRSs adopt a cooperative quasi-static phase shift design to improve the average achievable rate at \mcthr{low channel estimation and phase adjustment costs} and computational complexity. As the average achievable rate can be approximately  reflected by the average channel power of the equivalent channel between the BS and user with a negligible approximation error, this paper focuses on the analysis and optimization of the average channel power\footnote{If not specified otherwise, the average channel power means the average channel power of the equivalent channel between the BS and user.}  rather than the average achievable rate. The main contributions of this paper are summarized as follows.
	\begin{itemize}
		\item \textbf{Analysis of Average Channel Power:} First, we characterize the influences of the phase shifts of the two IRSs on the average channel power  and divide the channel conditions into four cases accordingly. Then, we obtain the tractable expressions of the average channel power in the general \mcthr{(Rician factor)}, pure LoS, and pure NLoS regimes, respectively.
		\item \textbf{Optimization of Average Channel Power:} First, we jointly optimize the phase shifts of the two IRSs to maximize the average channel power  in the \mcthr{general} and pure LoS regimes, respectively. The corresponding optimization problems are challenging non-convex problems. We obtain globally optimal closed-form solutions for some cases and propose computationally efficient iterative algorithms  to obtain stationary points for the other cases based on the coordinate descent (CD) and block coordinate descent (BCD) methods. Then, in each regime, we characterize the optimal average channel power when the total number of reflecting elements is large.
		\item \textbf{Comparison with \mcthr{Counterpart IRS-Assisted Systems}:} First, we analyze and optimize the average channel powers of \mcthr{a counterpart double-IRS non-cooperatively assisted system (D-IRS-NC) and a counterpart single-IRS-assisted system (S-IRS)}. Then, we compare the computational complexities for calculating the quasi-static phase shift designs and optimal average channel powers of \mcthr{D-IRS-C, D-IRS-NC, and S-IRS}. Specifically, we show that \mcthr{D-IRS-C} can achieve a better performance and computational complexity tradeoff \mcthr{than D-IRS-NC and S-IRS when the total number of reflecting elements is sufficiently large.}
		\item \textbf{Numerical Results:} We numerically demonstrate notable gains of the proposed solutions over the existing quasi-static phase shift designs. Furthermore, we numerically show that for \mcthr{D-IRS-C}, the proposed quasi-static phase shift design is more desirable than the respective instantaneous CSI-adaptive phase shift design as long as the LoS components are sufficiently large.
	\end{itemize}
		
	\textbf{Notation}: Boldface lower-case letters (e.g., $\mathbf{x}$), boldface upper-case letters (e.g., $\mathbf{X}$), non-boldface letters (e.g., $x$), and calligraphic upper-case letters (e.g., $\mathcal{X}$) denote vectors, matrices, scalars, and  sets, respectively. $\mathbf{X}^H$, $\mathbf{X}^T$, $\mathrm{tr}(\mathbf{X})$, and $\|\mathbf{X}\|_F$ denote the conjugate transpose, transpose, trace, and Frobenius norm of a matrix, respectively. $\mathrm{vec}(\mathbf{X})$ denotes the vectorization of a matrix. $\mathrm{diag}(\mathbf{x})$ denotes a square diagonal matrix with the elements in $\mathbf{x}$ on its main diagonal. $\mathrm{Diag}(\mathbf{X})$ denotes the leading diagonal of matrix $\mathbf{X}$. ${\left\| \mathbf{x} \right\|_2}$  denotes the Euclidean norm of a vector. $\mathfrak{R}\{x\}$ and $\left| x \right|$ denote the real part and modulus of a complex number, respectively.  $\angle (\mathbf{x})$ denotes the argument of each element of $\mathbf{x}$.   $\mathbb{E}\left[\cdot\right]$ denotes the expectation w.r.t. all random variables in the brackets. $\otimes$ denotes the kronecker product. $\mathbb{C}^{a \times b}$ and $\mathbb{R}^{a \times b}$ denote the space of $a \times b$ complex-valued and real-valued matrices, respectively.  $\mathcal{CN}(\mu,\sigma^2)$ denotes a circularly symmetric complex Gaussian  random variable with mean   $\mu$ and variance $\sigma^2$. $\mathbf{1}_T$ and $\mathbf{0}_T$ denote the $T$ dimentional vectors with all elements being 1 and 0, respectively. $X \overset{d}{\sim} Y$ represents that random variables $X$ and $Y$ follow the same distribution.

	\section{System Model} \label{sec:System}
\begin{figure}[t]
		\begin{center}
			\includegraphics[width=5cm]{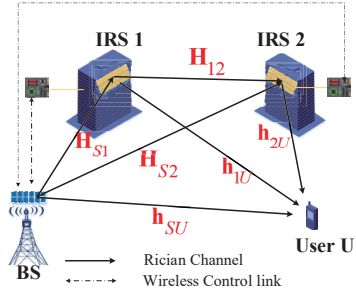}
		\end{center}
		\vspace*{-5mm}
		\caption{\small{Double-IRS cooperatively assisted system \mcthr{(D-IRS-C)}.}}
		\label{fig:systemModel}
		\vspace*{-10mm}
	\end{figure}

	As shown in Fig.~\ref{fig:systemModel}, we consider a double-IRS cooperatively assisted (downlink) system \mcthr{(D-IRS-C)}, where a multi-antenna BS, represented by $S$, communicates to a single-antenna user, represented by $U$, with the help of two multi-element IRSs, indexed by 1 and 2, which are connected by an inter-IRS channel.\footnote{\mcthr{The proposed framework can be extended to an IRS cooperatively assisted system with multiple IRSs and users. In this paper, we consider two IRSs and one user to obtain first-order design insights.}}  \mcthr{We assume that IRS 1 and IRS 2 are placed closer to the BS and user, respectively. \mcthr{We consider four channels, i.e., the direct channel $SU$, cascaded channels $(S1,1U)$, $(S2,2U)$, and $(S1,12,2U)$.\footnote{\mcthr{As in~\cite{ZhangRui2021doubleIRSMIMO,Han2021doubleMIMO}, we ignore the cascaded channel $(S2,21,1U)$ and the cascaded channels passing each IRS more than once, e.g., the cascaded channel $(S1,12,21,1U)$, as they have  much larger path loss. Note that~\cite{ZhangRui2021doubleIRSMIMO,Han2021doubleMIMO} also ignore the direct channel $SU$ for simplicity.}} The locations of the BS and IRSs are fixed and known, and the user's location is static during a certain period and known.\footnote{\mcthr{The user's location can be estimated using standard localization methods such as time difference of arrival (TDOA) techniques~\cite{He2017TDoA}}.}} \mcblk{The BS is equipped with a URA of $M_S \times N_S$ antennas. Each IRS $l \in  \mathcal{L}$ is equipped with a URA of $M_l \times N_l$ reflecting elements,} \mcthr{where ${\mathcal{L}} \triangleq \{1,2\}$ denotes the set of IRS indices}.\footnote{\mcthr{As in \cite{Han2020doubleSISO,ZhangRui2020doubleIRSMIMOCE,ZhangRui2021doubleIRSMIMO,Han2021doubleMIMO}, the optimization of the element allocation between the two IRSs and the IRSs' locations, which can be approximately tackled using exhaustive search or heuristic methods~\cite{Zeng2021IRSoptLoc}, are out of the scope of this paper.}}} Let $T_i \triangleq M_iN_i$, $i = S,1,2$, denote the numbers of the BS's antennas and IRSs' reflecting elements. Let $T\triangleq T_1+T_2$ denote the total number of reflecting elements in the system. Let $\mathcal{M}_i \triangleq \{{1,2,...,M_i}\}$, $\mathcal{N}_i \triangleq \{{1,2,...,N_i}\}$, and $\mathcal{T}_i \triangleq \{{1,2,...,T_i}\}$ denote the corresponding sets of indices. The phase shifts of each IRS's reflecting elements can be determined by a smart controller attached to it. The BS communicates to the two IRS controllers to configure the two IRSs' phase shifts via separate reliable wireless links so that both IRSs jointly assist the downlink transmission from the BS to the user.	
	
	We consider a narrow-band system and adopt the block-fading model for small-scale fading. Let $\mathbf{H}_{Sl} \in \mathbb{C}^{T_l \times T_S}$, $\mathbf{H}_{12} \in \mathbb{C}^{T_2 \times T_1}$, $\mathbf{h}_{SU}^H \in \mathbb{C}^{1 \times T_S}$, and $\mathbf{h}_{lU}^H  \in \mathbb{C}^{1 \times T_l}$ represent the channel matrix between the BS and IRS $l \in \mathcal{L}$,  the channel matrix between IRS 1 and IRS 2, the channel vector between the BS and  user,  and the channel vector between IRS $l \in \mathcal{L}$ and the user, respectively. Notice that in contrast with the existing works on double-IRS cooperatively assisted systems~\cite{Han2020doubleSISO,Zhangrui2021Multireflection,You2021doubleIRSSISO,ZhangRui2020doubleIRSMIMOCE,ZhangRui2021doubleIRSMIMO,Han2021doubleMIMO,Dong2021doubleIRSsecure}, we consider the direct channel between the BS and user. As the IRSs are usually far above the ground where scattering is relatively weak, we adopt the Rician fading model\footnote{The Rician fading model is more general than the Rayleigh fading model. \mcthr{This paper considers the Rician fading model to obtain first-order design insights for quasi-static phase shift design. Note that quasi-static phase shift design generally takes effect as long as channel components contain deterministic phases (e.g., in clustered delay line (CDL) models \cite{Jiang2013CDL}).}} for all small-scale fading channels~\cite{Cui2020Interference,jia2021robust,JinShi2019MISO,Cui2020Outage,Zhao2021TTS,Cui2019Outage,Kwan2021MISO}. Specifically,
	\begin{align}
		\mathbf{H}_{ab}   & = \sqrt{\alpha_{ab}} \bigg( \sqrt{\frac{K_{ab}}{K_{ab}+1}} \bar{\mathbf{H}}_{ab} + \sqrt{\frac{1}{K_{ab}+1}} \tilde{\mathbf{H}}_{ab} \bigg), ab \in \{S1,S2,12\}, \label{eq:Hab}
		\\
		\mathbf{h}_{ab}^H & = \sqrt{\alpha_{ab}} \bigg( \sqrt{\frac{K_{ab}}{K_{ab}+1}} \bar{\mathbf{h}}_{ab}^H + \sqrt{\frac{1}{K_{ab}+1}} \tilde{\mathbf{h}}_{ab}^H \bigg), ab \in \{1U,2U,SU\}, \label{eq:hab}
	\end{align}
	where \mcthr{$\alpha_{ab} > 0 $} represents the large-scale fading power;  $K_{ab} \geq 0 $ represents the Rician factor; $\tilde{\mathbf{H}}_{ab} \in \mathbb{C}^{T_b \times T_a}$ and $\tilde{\mathbf{h}}_{ab}^H \in \mathbb{C}^{1 \times T_a}$  represent the random normalized NLoS components in a slot with elements independently and identically distributed (i.i.d.) according to $\mathcal{CN}(0,1)$;  $\bar{\mathbf{H}}_{ab} \in \mathbb{C}^{T_b \times T_a}$ and $\bar{\mathbf{h}}_{ab}^H  \in \mathbb{C}^{1 \times T_a}$ represent the deterministic normalized LoS components with unit-modulus elements. Note that $\bar{\mathbf{H}}_{ab}$ and  $\bar{\mathbf{h}}_{ab}^H$ do not change during the considered period, as the locations of the BS, IRSs, and  user are assumed to be invariant~\cite{Cui2020Interference,jia2021robust,JinShi2019MISO,Cui2020Outage,Cui2019Outage}.
	
	Let $\lambda$ and $d\left(\leq \frac{\lambda}{2}\right)$ denote the wavelength of transmission signals and the distance between two adjacent reflecting elements or antennas in each row and each column of the URAs. Define $f(x^{(h)},x^{(v)},m,n)\triangleq 2 \pi \frac{d}{\lambda }\sin x^{(v)} ((m - 1)\cos x^{(h)} + (n - 1)\sin x^{(h)})$, $\mathbf{A} (x^{(h)},x^{(v)},M,N) \triangleq \left( e^{jf (x^{(h)},x^{(v)},m,n)}\right)_{m=1,...,M,n=1,...,N}$, and $\mathbf{a}(x^{(h)},x^{(v)},M,N)  \triangleq  \mathrm{vec} \left( \mathbf{A} (x^{(h)},x^{(v)},M,N) \right) \in \mathbb{C}^{MN }$.  Then, $\bar{\mathbf{H}}_{ab}$ and $\bar{\mathbf{h}}_{ab}$ are modeled as~\cite{Cui2020Interference,jia2021robust,JinShi2019MISO,Cui2020Outage}:
	\begin{align}
		\bar{\mathbf{H}}_{ab} & = \mathbf{a}_{A,ab} \mathbf{a}^H_{D,ab},  ab \in \{S1,S2,12\},\label{eq:Hbar}
		\\
		\bar{\mathbf{h}}_{ab} & = \mathbf{a}_{D,ab},  ab \in \{1U,2U,SU\},\label{eq:hbar}
	\end{align}
	where $\mathbf{a}_{A,ab} \triangleq \mathbf{a}(\delta_{ab}^{(h)}, \delta_{ab}^{(v)},M_b,N_b)$ and $\mathbf{a}_{D,ab}\triangleq \mathbf{a}(\varphi_{ab}^{(h)},\varphi_{ab}^{(v)},M_a,N_a)$. Here, $f (\theta^{(h)},\theta^{(v)},m,n)$ represents the difference of the corresponding phase changes over the LoS component; $\delta_{ab}^{(h)}\left(\delta_{ab}^{(v)}\right)$ represents the azimuth (elevation) angle of arrival (AoA) of a signal from node $a$ to the URA at node $b$; $\varphi_{ab}^{(h)} \left(\varphi_{ab}^{(v)}\right)$ represents the azimuth (elevation) angle of departure (AoD) of a signal from the URA at node $a$ to node $b$.
	
	To reduce the \mcthr{channel estimation and} phase adjustment costs, we consider a quasi-static phase shift design~\cite{Cui2020Interference,jia2021robust,JinShi2019MISO,Cui2020Outage,Zhao2021TTS,Cui2019Outage,Kwan2021MISO}. To be specific, the phase shifts of the two IRSs do not change with the NLoS components from slot to slot and remain constant during the considered period. Let $\bm \theta_l \triangleq \left(\theta_{l,m,n}\right)_{m\in \mathcal{M}_l,n\in\mathcal{N}_l} \in \mathbb{R}^{M_l \times N_l}$  denote the constant phase shifts of IRS $l$, where the phase shift of its $(m,n)$-th element satisfies $\theta_{l,m,n} \in \left[ 0,2 \pi \right)$. For notation convenience, we introduce $\bm \phi_l \triangleq \left(\phi_{l,t}\right)_{t\in\mathcal{T}_l} = \mathrm{vec} \left(\bm\theta_{l}\right) \in \mathbb{R}^{T_l}$ to represent the  phase shifts of IRS $l$, where its $t$-th element satisfies:
	\begin{equation}\label{eq:unitconstrint}
		\phi_{l,t} \in \left[ 0,2 \pi \right), t \in \mathcal{T}_l, l\in \mathcal{L}.
	\end{equation}
	
	The equivalent channel between the BS and user,  denoted by $\mathbf{h}_e^H(\bm\phi_1,\bm\phi_2) \in \mathbb{C}^{1 \times T_S}$, can be expressed as:
	\begin{equation}\label{eq:he}
		\mathbf{h}_e^H(\bm\phi_1,\bm\phi_2) =
		\mathbf{h}_{SU}^H + \sum_{l\in\mathcal{L}} \mathbf{h}_{lU}^H \diag{(\mathbf{v}_l^H)} \mathbf{H}_{Sl} +\mathbf{h}_{2U}^H \diag{(\mathbf{v}_2^H)} \mathbf{H}_{12} \diag{(\mathbf{v}_1^H)} \mathbf{H}_{S1},
	\end{equation}
	where $\mathbf{v}_l \triangleq \left(e^{-j\phi_{l,t}}\right)_{t \in \mathcal{T}_l} \in \mathbb{C}^{T_l}$. Note that $\mathbf{h}_{lU}^H \diag{(\mathbf{v}_l^H)} \mathbf{H}_{Sl}$ represents the cascaded channel $ (Sl,lU)$, and  $\mathbf{h}_{2U}^H \diag{(\mathbf{v}_2^H)} \mathbf{H}_{12} \diag{(\mathbf{v}_1^H)} \mathbf{H}_{S1}$ represents the cascaded channel $ (S1,12,2U)$. \mcthr{We adopt the following assumptions on CSI: $\alpha_{ab}$, $K_{ab}$, $ab \in \{S1,S2,SU,1U,2U,12\}$ are known; $\delta_{ab}^{(h)}\left(\delta_{ab}^{(v)}\right)$, $\varphi_{ab}^{(h)} \left(\varphi_{ab}^{(v)}\right)$, $ab \in \{S1,S2,SU,1U,2U,12\}$ are known, implying that $\bar{\mathbf{H}}_{ab}$, $ab \in \{S1,S2,12\}$ and $\bar{\mathbf{h}}_{ab}$, $ab \in \{1U,2U,SU\}$ (i.e., CSI statistics) are known; $\mathbf{h}_e(\bm\phi_1,\bm\phi_2)$ (i.e., instantaneous CSI) is known to the BS and user at each slot for given $\bm \phi_1$ and $\bm \phi_2$.}
	
	\begin{Rem}[\mcthr{CSI Estimation}]
		\mcthr{Note that $\alpha_{ab}$ and $K_{ab}$ can be obtained by standard offline channel measurement \cite{SUn2013CSIMesu}. Given the BS, IRSs, and user's locations, $\delta_{ab}^{(h)}\left(\delta_{ab}^{(v)}\right)$ and $\varphi_{ab}^{(h)} \left(\varphi_{ab}^{(v)}\right)$ can be easily obtained using angle estimation techniques~\cite{Hu2020IRSAoA} at the beginning of the considered communication period. Thus, the cost for estimating CSI statistics for quasi-static phase shift design is negligible. With time division duplexing (TDD), at each slot, $\mathbf{h}_e(\bm\phi_1,\bm\phi_2)$ can be obtained by setting the phase shifts to the optimized values, sending one pilot symbol from the user ($T_S$ pilot symbols from the BS sequentially), and estimating the received signal at the BS (user), using standard channel estimation methods~\cite{Biguesh2006MIMOCE}. Thus, the  cost for estimating instantaneous CSI for quasi-static phase shift design  is much lower than that for instantaneous CSI-adaptive phase shift design~\cite{ZhangRui2020doubleIRSMIMOCE,ZhangRui2021doubleIRSMIMO,Han2021doubleMIMO}.\footnote{\mcthr{Instantaneous CSI  $\mathbf{h}_{SU}$, $ \text{diag}(\mathbf{h}_{1U})\mathbf{H}_{S1}$, $ \text{diag}(\mathbf{h}_{2U})\mathbf{H}_{S2}$, and $ \text{diag}( h_{2U,t} \mathbf{H}_{12,:,t}) \mathbf{H}_{S1}$, $t=1,..,T_1$, where $h_{2U,t}$ denotes the $t$-th element of $\mathbf{h}_{2U}$, are required per slot in~\cite{ZhangRui2020doubleIRSMIMOCE,ZhangRui2021doubleIRSMIMO,Han2021doubleMIMO}. To estimate such instantaneous CSI, totally $\left\lceil \frac{\left(T_{2}+1\right) T_{1}}{T_S}\right\rceil+T_{1}+T_{2}+2$ (if $T_S<T_1$) or $2T_2+T_1+3$ (if $T_S>T_1$) pilot symbols have to be sent \cite{ZhangRui2020doubleIRSMIMOCE}.}}}
	\end{Rem}

	The BS serves the user via linear beamforming. Let ${\mathbf{w}} \in \mathbb{C}^{T_S}$ with $\Vert {\mathbf{w}}\Vert_2^2 = 1$ represent the normalized linear beamforming vector at the BS. Then, the received signal at the user  can be expressed as:
	\begin{align}
		Y = \mathbf{h}_e^H(\bm\phi_1,\bm\phi_2) \mathbf{w} \sqrt{P_S} X_S + Z,
	\end{align}
	where $P_S$ represents the transmit power of the BS, $X_S \in \mathbb{C}$ with $\mathbb{E}\left[ |X_S|^2\right]=1$ represents an information symbol for the user, and $Z \sim \mathcal{CN}(0, \sigma^2)$ denotes the additive white gaussian noise (AWGN).
	
 	We consider coding within each slot. To maximize the achievable rate (singal-to-noise ratio, SNR) at each slot, we adopt the instantaneous CSI-adaptive MRT beamformer at the BS~\cite{Cui2020Interference,jia2021robust,JinShi2019MISO,Cui2020Outage,Zhao2021TTS,Kwan2021MISO}:\footnote{It is obvious that $\mathbf{w}$ in \eqref{eq:w_MRT} is optimal for the maximization of the  instantaneous SNR w.r.t. $\mathbf{w}$ under $\|\mathbf{w}\|_2^2=1$ at each slot and hence  is optimal for the average rate maximization. }
	\begin{align}\label{eq:w_MRT}
		\mathbf{w}=\frac{\mathbf{h}_e(\bm\phi_1,\bm\phi_2)}{\|\mathbf{h}_e(\bm\phi_1,\bm\phi_2)\|_2}.
	\end{align}
	Thus, for given $\bm\phi_1$ and $\bm\phi_2$, the average achievable rate of \mcthr{D-IRS-C} over slots, $C(\bm\phi_1,\bm\phi_2)$ (bit/s/Hz), is given by:
	\begin{align}\label{eq:C}
		C(\bm\phi_1,\bm\phi_2) \triangleq \mathbb{E} \left[ \log_2 \left( 1 + \frac{P_S}{\sigma^2} \left\| \mathbf{h}_e(\bm\phi_1,\bm\phi_2) \right\|_2^2\right) \right].
	\end{align}
	
	Therefore, we would like to analyze and maximize $C(\bm\phi_1,\bm\phi_2)$. For tractability, we approximate $C(\bm\phi_1,\bm\phi_2)$ with its upper bound, $\log_2\left( 1+ \frac{P_S}{\sigma^2} \gamma( \bm \phi_1, \bm \phi_2) \right)$, obtained based on Jensen's inequality as in~\cite{JinShi2019MISO,Cui2020Interference,jia2021robust}. Here, $\gamma\left( \bm \phi_1,\bm \phi_2\right)  \triangleq \mathbb{E} \left[\| \mathbf{h}_e(\bm\phi_1,\bm\phi_2) \|_2^2 \right]$ represents the average channel power \mcthr{of D-IRS-C.} \mcthr{Later in Section \ref{sec:Numerical},  we will numerically show that the upper bound is a good approximation of $C(\bm\phi_1,\bm\phi_2)$. In Section~\ref{sec:Analysis} and Section~\ref{sec:Optimization},}  we analyze and maximize $\gamma\left( \bm \phi_1, \bm \phi_2\right)$ instead of $\log_2\left( 1+ \frac{P_S}{\sigma^2} \gamma( \bm \phi_1, \bm \phi_2) \right)$, which is an increasing function of $\gamma( \bm \phi_1, \bm \phi_2)$. Note that in what follows,  we let $ab$, $\overline{ab}$, and $\widetilde{ab}$ represent the Rician channel, LoS channel, and NLoS channel between node $a$ and node $b$, respectively. The Rician channel $ab$ can be viewed as the composition of the LoS channel $\overline{ab}$ and NLoS channel $\widetilde{ab}$.

	\section{Analysis} \label{sec:Analysis}
	
	In this section, we analyze the average channel power \mcthr{of D-IRS-C} in the general and special regimes, respectively.
	
	\subsection{Analysis in General Regime}
	
	In this part, we consider the analysis of the average channel power \mcthr{of D-IRS-C} in the general  regime. First, we characterize the influences of $\bm \phi_1$ and $\bm \phi_2$ on $\gamma(\bm\phi_1,\bm\phi_2)$.
	\vspace{-2mm}
	\begin{Lem}[\mcthr{Influences of Phase Shifts of D-IRS-C}]\label{lem:Influence}
		(i) If $K_{S1}(K_{1U}+K_{12})=0$ (i.e., $K_{S1} = 0$ or $K_{1U} = K_{12} = 0$), then  $\gamma(\bm\phi_1,\bm\phi_2)$ does not change with $\bm \phi_1$.
		(ii) If $K_{2U}(K_{12}+K_{S2})=0$ (i.e., $K_{2U} = 0$ or $K_{12} = K_{S2} = 0$), then  $\gamma(\bm\phi_1,\bm\phi_2)$ does not change with $\bm \phi_2$.
	\end{Lem}
	\vspace{-2mm}
	\begin{IEEEproof}		
		Please refer to Appendix A. %\ref{proof:Influence}
	\end{IEEEproof}
	
\begin{figure}[t]
		\begin{center}
			\includegraphics[width=10cm]{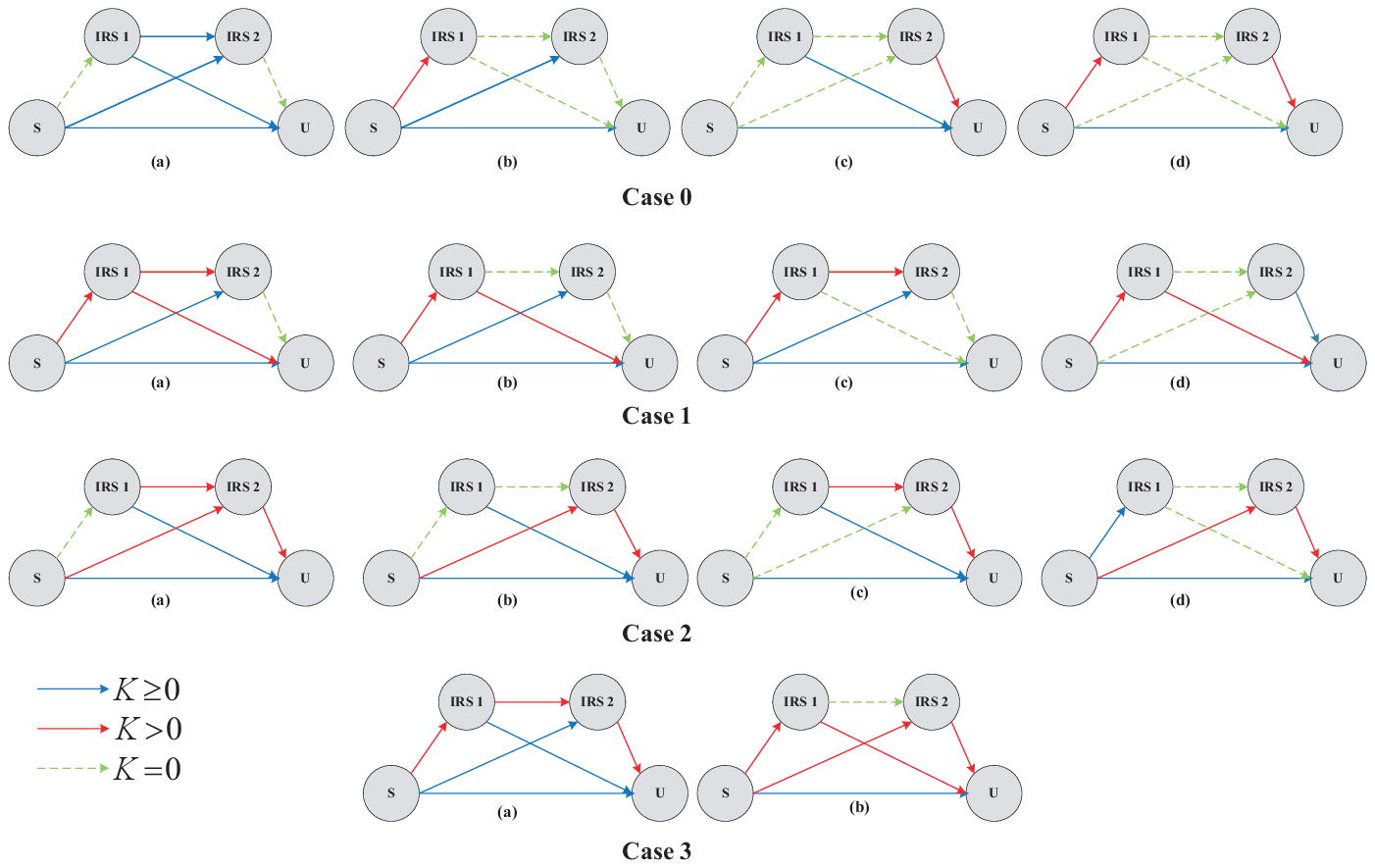}
		\end{center}
		\vspace*{-5mm}
		\caption{\small{\mcthr{The channel setups of D-IRS-C in the four cases.}}}
		\label{fig:cases}
		\vspace*{-10mm}
	\end{figure}

	\mcthr{Lemma~\ref{lem:Influence} indicates that IRS $l$ takes effect if and only if there exist two LoS channels through which signals can arrive at IRS $l$ from a neighbor node and depart from IRS $l$ to a neighbor node.} Based on Lemma~\ref{lem:Influence}, we can divide the channel conditions into four cases according to the values of Rician factors, as shown in Fig.~\ref{fig:cases}.\footnote{\mcthr{The special cases listed in this paper are those which not only correspond to special channel setups but also have closed-form solutions or iterative algorithms with closed-form updates in each iteration.}}
	\begin{itemize}
		\item \textbf{Case 0} ($K_{S1} = 0$ or $K_{1U} = K_{12} = 0$ and  $K_{12}=K_{S2} = 0$ or $K_{2U} = 0$):  $\gamma(\bm\phi_1,\bm\phi_2)$ does not change with $\bm \phi_1$ or $\bm \phi_2$ and hence is rewritten as  $\gamma^{(0)}$.
		\item \textbf{Case 1} ($K_{S1} > 0$,  $K_{1U} + K_{12} > 0$,  and $K_{12}=K_{S2} = 0$ or $K_{2U} = 0$):  $\gamma(\bm\phi_1,\bm\phi_2)$ changes only with $\bm \phi_1$ and hence is rewritten as  $\gamma^{(1)}(\bm\phi_1)$.
		\item \textbf{Case 2} ($K_{S1} = 0$ or $K_{1U} = K_{12} = 0$, $K_{12}+K_{S2} > 0$, and $K_{2U} > 0$):  $\gamma(\bm\phi_1,\bm\phi_2)$ changes only with $\bm \phi_2$ and hence is rewritten as  $\gamma^{(2)}(\bm\phi_2)$.
		\item \textbf{Case 3} ($K_{S1} > 0$, $K_{1U} + K_{12} > 0$, $K_{12}+K_{S2} > 0$, and $K_{2U} > 0$):  $\gamma(\bm\phi_1,\bm\phi_2)$ changes  with both $\bm \phi_1$ and  $\bm \phi_2$ and  is also written as  $\gamma^{(3)}(\bm\phi_1,\bm\phi_2)$. 
	\end{itemize}
	
	Then, we characterize the average channel power \mcthr{of D-IRS-C} in the four cases of channel conditions. For ease of exposition, we introduce several new notations.  Firstly, for $ab \in \{S1,S2,12,1U,2U,SU\}$, we define $L_{\overline{ab}} \triangleq \frac{K_{ab}\alpha_{ab}}{K_{ab}+1}$ and $L_{\widetilde{ab}} \triangleq \frac{\alpha_{ab}}{K_{ab}+1}$, which can be interpreted as the large-scale  fading powers of the LoS channel $\overline{ab}$  and  NLoS channel $\widetilde{ab}$, respectively.  Obviously, $L_{\overline{ab}}$ and $L_{\widetilde{ab}}$ both increase with $\alpha_{ab}$, $L_{\overline{ab}}$ increases with $K_{ab}$, and $L_{\widetilde{ab}}$ decreases with $K_{ab}$. Secondly, we define $L_{a_1b_1,a_2b_2} \triangleq \prod \limits_{i=1}^{2}L_{a_ib_i}, (a_1b_1,a_2b_2) \in  \bigcup\limits_{l\in\mathcal{L}} \left(\{\overline{Sl},\widetilde{Sl}\}\times\{\overline{lU},\widetilde{lU}\}\right)$ and  $L_{a_1b_1,a_2b_2,a_3b_3} \triangleq \prod\limits_{i=1}^{3} L_{a_ib_i}, (a_1b_1,a_2b_2,a_3b_3) \in \{\overline{S1},\widetilde{S1}\}  \times \{\overline{12},\widetilde{12}\} \times \{\overline{2U},\widetilde{2U}\}$, which can be interpreted as the large-scale fading powers of the  \mcthr{cascaded channels $(a_1b_1,a_2b_2)$ and $(a_1b_1,a_2b_2,a_3b_3)$}, respectively. Thirdly, for $a \in \{S,1\}, l \in \mathcal{L}, b \in  \{2,U\}$, we define:
	\begin{align}
		\boldsymbol{\Delta}_{\overline{al},\overline{lb}} & \triangleq \angle\left( \diag(\mathbf{a}_{D,lb}^H)\mathbf{a}_{A,al}\right) \in \mathbb{R}^{ T_l}, \label{eq:Delta}
		\\
		r_{\overline{al},\overline{ab}}                   & \triangleq \mathbf{a}_{D,al}^H \mathbf{a}_{D,ab} \in \mathbb{C},\label{eq:ralab}
		\\
		r_{\overline{ab},\overline{lb}}                   & \triangleq \mathbf{a}_{A,ab}^H \mathbf{a}_{A,lb} \in \mathbb{C}.\label{eq:rablb}
	\end{align}
	Note that $\boldsymbol{\Delta}_{\overline{al},\overline{lb}}$ represents the sum of the phase changes over the LoS channel $\overline{al}$ and LoS channel $\overline{lb}$; $r_{\overline{al},\overline{ab}}$ depends on $\varphi_{al}^{(h)},\varphi_{al}^{(v)},\varphi_{ab}^{(h)}$, and $\varphi_{ab}^{(h)}$, which are determined only by the placement of the URA at node $a$ and the locations of  IRS $l$ and node $b$; $r_{\overline{12},\overline{S2}}$ depends on $\delta_{12}^{(h)},\delta_{12}^{(v)},\delta_{S2}^{(h)}$, and $\delta_{S2}^{(v)}$, which are  determined only by the placement of the URA at IRS 2 and the locations of the BS and IRS~1. Finally, we define:
	\begin{align*}
		\mathbf{A}_{11} & \triangleq \mathrm{diag}\left(\bar{\mathbf{h}}_{1U}^H\right) \bar{\mathbf{H}}_{S1}\bar{\mathbf{H}}_{S1}^H\mathrm{diag}\left(\bar{\mathbf{h}}_{1U}\right) \in \mathbb{C} ^{T_1 \times T_1},
		\mathbf{A}_{12}  \triangleq \diag(\mathbf{a}_{D,12}^H)\bar{\mathbf{H}}_{S1}\bar{\mathbf{H}}_{S1}^H\diag(\mathbf{a}_{D,12}) \in \mathbb{C} ^{T_1 \times T_1},
		\\
		\mathbf{A}_{21} & \triangleq \mathrm{diag}\left(\bar{\mathbf{h}}_{2U}^H\right) \bar{\mathbf{H}}_{S2}\bar{\mathbf{H}}_{S2}^H\mathrm{diag}\left(\bar{\mathbf{h}}_{2U}\right) \in \mathbb{C} ^{T_2\times T_2},
		\mathbf{A}_{22}  \triangleq \diag(\bar{\mathbf{h}}_{2U}^H)\bar{\mathbf{H}}_{12}\bar{\mathbf{H}}_{12}^H\diag(\bar{\mathbf{h}}_{2U}) \in \mathbb{C} ^{T_2 \times T_2},
		\\
		\mathbf{A}_3    & \triangleq \mathrm{diag}\left(\bar{\mathbf{h}}_{2U}^H\right)\bar{\mathbf{H}}_{12}\mathrm{diag}\left(\mathbf{a}_{A,S1}\right) \in \mathbb{C} ^{T_2 \times T_1},
		\\
		\mathbf{b}_{11} & \triangleq \mathrm{diag} \left( \bar{\mathbf{h}}_{1U}^H \right) \bar{\mathbf{H}}_{S1} \bar{\mathbf{h}}_{SU}\in \mathbb{C} ^{T_1},
		\mathbf{b}_{12}  \triangleq \mathrm{diag}\left(\mathbf{a}_{D,12}^H\right) \bar{\mathbf{H}}_{S1} \bar{\mathbf{H}}_{S2}^H\mathbf{a}_{A,12} \in \mathbb{C} ^{T_1},
		\\
		\mathbf{b}_{21} & \triangleq \mathrm{diag} \left( \bar{\mathbf{h}}_{2U}^H \right) \bar{\mathbf{H}}_{S2} \bar{\mathbf{h}}_{SU}\in \mathbb{C} ^{T_2},
		\mathbf{b}_{22}  \triangleq \mathrm{diag} \left(\bar{\mathbf{h}}_{2U}^H\right) \bar{\mathbf{H}}_{12} \bar{\mathbf{h}}_{1U}  \in \mathbb{C} ^{T_2},
		\\
		\mathbf{B}_{1}  & \triangleq \mathrm{diag} \left( \bar{\mathbf{h}}_{2U}^H \right) \bar{\mathbf{H}}_{12} \in \mathbb{C} ^{T_2 \times T_1},
		\\
		\mathbf{B}_{2}  & \triangleq \bar{\mathbf{H}}_{S1} \bar{\mathbf{H}}_{S1}^H \mathrm{diag} \left( \bar{\mathbf{h}}_{1U} \right) \in \mathbb{C} ^{T_1 \times T_1},
		\mathbf{B}_{3}   \triangleq \bar{\mathbf{H}}_{S1} \bar{\mathbf{H}}_{S2}^H \mathrm{diag} \left( \bar{\mathbf{h}}_{2U} \right) \in \mathbb{C} ^{T_1 \times T_2},
		\\
		\mathbf{B}_{4}  & \triangleq \mathrm{diag} \left(  \bar{\mathbf{h}}_{2U}^H \right) \bar{\mathbf{H}}_{12} \mathrm{diag} \left( \bar{\mathbf{H}}_{S1} \bar{\mathbf{h}}_{SU} \right) \in \mathbb{C} ^{T_2 \times T_1},
		\mathbf{B}_{5}   \triangleq \mathrm{diag} \left(\bar{\mathbf{h}}_{2U}^H\right) \bar{\mathbf{H}}_{S2}  \bar{\mathbf{H}}_{S1}^H  \mathrm{diag} \left(\bar{\mathbf{h}}_{1U}\right) \in \mathbb{C} ^{T_2 \times T_1}.
	\end{align*}
		
	Based on the above notations, \mcthr{we characterize the average channel power of D-IRS-C in the general regime as follows.}
	\begin{Thm}[Average Channel Power \mcthr{of D-IRS-C} in General Regime]\label{thm:analyGeneralK}
		\begin{align}
			\gamma^{(0)} =                      & \alpha_{SU}T_S + \sum_{l=1}^{2}(L_{\widetilde{Sl},\overline{lU}} + L_{\overline{Sl},\widetilde{lU}} +L_{\widetilde{Sl},\widetilde{lU}})T_ST_l \nonumber
			\\
			& +\big(L_{\widetilde{S1},\overline{12},\widetilde{2U}} +L_{\overline{S1},\widetilde{12},\overline{2U}} +L_{\overline{S1},\widetilde{12},\widetilde{2U}}+L_{\widetilde{S1},\widetilde{12},\overline{2U}}+L_{\widetilde{S1},\widetilde{12},\widetilde{2U}}\big)T_ST_1T_2, \label{eq:gamma0}
			\\
			\gamma^{(1)}(\bm\phi_1) =           & \mathbf{v}_1^H \left(L_{\overline{S1},\overline{1U}} \mathbf{A}_{11} + L_{\overline{S1},\overline{12},\widetilde{2U}}T_2 \mathbf{A}_{12}\right)\mathbf{v}_1 \nonumber
			\\
			& +2\mathfrak{R}\left\{\mathbf{v}_1^H\left(\sqrt { L_{\overline{SU}}  L_{\overline{S1},\overline{1U}} } \mathbf{b}_{11} +  \sqrt{L_{\overline{S2},\widetilde{2U}} L_{\overline{S1},\overline{12},\widetilde{2U}}} \mathbf{b}_{12}\right)\right\} + \gamma^{(0)}, \label{eq:gamma1}
			\\
			\gamma^{(2)}(\bm\phi_2) =           & \mathbf{v}_2^H \left( L_{\overline{S2},\overline{2U}} \mathbf{A}_{21} +L_{\widetilde{S1},\overline{12},\overline{2U}}T_S\mathbf{A}_{22} \right)\mathbf{v}_2 \nonumber
			\\
			& +2\mathfrak{R}\left\{ \mathbf{v}_2^H \left( \sqrt{ L_{\overline{SU}}  L_{\overline{S2},\overline{2U}}}\mathbf{b}_{21}+ \sqrt{L_{\widetilde{S1},\overline{1U}} L_{\widetilde{S1},\overline{12},\overline{2U}}}T_S \mathbf{b}_{22} \right)\right\} + \gamma^{(0)}, \label{eq:gamma2}
			\\
			\gamma^{(3)}(\bm\phi_1,\bm\phi_2) = & \gamma^{(1)}(\bm\phi_1) + \gamma^{(2)}(\bm\phi_2) - \gamma^{(0)} + L_{\overline{S1},\overline{12},\overline{2U}}T_S \mathbf{v}_2^H \mathbf{A}_3\mathbf{v}_1^*\mathbf{v}_1^T\mathbf{A}_3^H\mathbf{v}_2 \nonumber
			\\
			& +2\mathfrak{R}\Big\{\mathbf{v}_1^H \mathrm{diag}(\mathbf{v}_2^H\mathbf{B}_1)\left(\sqrt{L_{\overline{S1},\overline{1U}}L_{\overline{S1},\overline{12},\overline{2U}}} \mathbf{B}_2\mathbf{v}_1 + \sqrt{L_{\overline{S2},\overline{2U}}L_{\overline{S1},\overline{12},\overline{2U}}}\mathbf{B}_3\mathbf{v}_2\right) \nonumber
			\\
			& +\mathbf{v}_2^H \left(\sqrt{L_{\overline{SU}}  L_{\overline{S1},\overline{12},\overline{2U}} } \mathbf{B}_4\mathbf{v}_1^* + \sqrt{ L_{\overline{S1},\overline{1U}} L_{\overline{S2},\overline{2U}} }  \mathbf{B}_5\mathbf{v}_1\right)  \Big\}. \label{eq:gamma3}
		\end{align}
	\end{Thm}
	\begin{IEEEproof}
		Please refer to Appendix B.%\ref{proof:analyGeneralK}
	\end{IEEEproof}
	
	Theorem~\ref{thm:analyGeneralK} indicates that the average channel power \mcthr{of D-IRS-C in Case 0 of the general regime}  does not depend on $\bm \phi_1$ or $\bm \phi_2$, and hence a quasi-static phase shift design is void in Case~0.

	\subsection{Analysis in Pure LoS Regime}
	In this part, we analyze the average channel power when the Rician factors go to infinity, corresponding to the regime of pure LoS channels.\footnote{\mcthr{Pure LoS channels usually exist in the indoor scenario  or far above the ground  where scattering is relatively weak \cite{EISI2020}.}}  Apparently, Cases 0, 1, and 2 are void in the pure LoS regime, and the pure LoS regime can be regarded as a special case of Case 3 of the general regime. Based on Theorem \ref{thm:analyGeneralK}, we have the following result.
	\begin{Cor}[Average Channel Power \mcthr{of D-IRS-C} in Pure LoS Regime]\label{col:analyLargeK}
		As $K_{S1}$, $K_{S2}$, $K_{12}$, $K_{SU}$, $K_{1U}$, $K_{2U} \rightarrow \infty$, $ \gamma^{(3)} (\bm \phi_1,\bm \phi_2)\rightarrow \bar{\gamma}^{(3)}(\bm \phi_1,\bm \phi_2)$, where
		\begin{align}
			\bar{\gamma}^{(3)}(\bm \phi_1,\bm \phi_2) & \triangleq    \alpha_{SU}T_S + \alpha_{S1}\alpha_{1U} \mathbf{v}_1^H  \mathbf{A}_{11}\mathbf{v}_1
			+ \alpha_{S2}\alpha_{2U} \mathbf{v}_2^H  \mathbf{A}_{21}\mathbf{v}_2  + \alpha_{S1}\alpha_{12}\alpha_{2U}T_S\mathbf{v}_2^H \mathbf{A}_3\mathbf{v}_1^*\mathbf{v}_1^T\mathbf{A}_3^H\mathbf{v}_2 \nonumber
			\\
			& +2\mathfrak{R}\Big\{
			\sqrt{ \alpha_{SU}\alpha_{S1}\alpha_{1U}} \mathbf{v}_1^H\mathbf{b}_{11}+\sqrt{ \alpha_{SU}\alpha_{S2}\alpha_{2U}} \mathbf{v}_2^H\mathbf{b}_{21} \nonumber
			\\
			& \quad \quad\quad + \mathbf{v}_1^H \mathrm{diag}(\mathbf{v}_2^H\mathbf{B}_1)\Big( \sqrt{\alpha_{S1}\alpha_{1U}\alpha_{S1}\alpha_{12}\alpha_{2U}} \mathbf{B}_2\mathbf{v}_1  +\sqrt{\alpha_{S2}\alpha_{2U}\alpha_{S1}\alpha_{12}\alpha_{2U}} \mathbf{B}_3\mathbf{v}_2 \Big) \nonumber
			\\
			& \quad \quad\quad +\mathbf{v}_2^H \Big(\sqrt{\alpha_{SU}\alpha_{S1} \alpha_{12}\alpha_{2U}} \mathbf{B}_4\mathbf{v}_1^* + \sqrt{\alpha_{S1}\alpha_{1U}\alpha_{S2}\alpha_{2U}} \mathbf{B}_5\mathbf{v}_1\Big)\Big\}. \label{eq:gamma3Infty}
		\end{align}
	\end{Cor}
	
	As the Rician factors go to infinity, the large-scale fading power of each channel or cascaded channel  \mcthr{consisting} of at least one NLoS channel goes zero. Thus, Theorem~\ref{thm:analyGeneralK} readily implies Corollary~\ref{col:analyLargeK}.
	
	\subsection{Analysis in Pure NLoS Regime}
	In this part, we analyze the average channel power \mcthr{of D-IRS-C} when the Rician factors are zero,  corresponding to the regime of pure NLoS channels. Obviously, Cases 1, 2, and 3 are void in the pure NLoS regime, and the pure NLoS regime  can be regarded as a special case of Case 0 in the general regime. From Theorem \ref{thm:analyGeneralK}, we have the following result.
	\begin{Cor}[Average Channel Power \mcthr{of D-IRS-C} in Pure NLoS Regime]\label{col:analySmallK}
		If  $K_{S1}=K_{S2}=K_{12}=K_{SU}= K_{1U}=K_{2U}=0$, then
		$\gamma^{(0)} = \tilde{\gamma}^{(0)}$, where
		\begin{align}
			\tilde{\gamma}^{(0)}\triangleq \alpha_{SU}T_S +  \alpha_{S1}\alpha_{1U}T_ST_1 + \alpha_{S2}\alpha_{2U}T_ST_2+\alpha_{S1}\alpha_{12}\alpha_{2U}T_ST_1T_2.
		\end{align}
	\end{Cor}
	
	When the Rician factors become zero, the large-scale fading power of each channel or cascaded channel \mcthr{consisting} of at least one LoS channel becomes zero. As a result, Theorem~\ref{thm:analyGeneralK} readily implies Corollary~\ref{col:analySmallK}. \mcthr{Analogously,} a quasi-static phase shift design  is ineffective \mcthr{for D-IRS-C} in the pure NLoS regime.

	\section{Optimization} \label{sec:Optimization}
	In this section, we  maximize the average channel power \mcthr{of D-IRS-C} w.r.t. the phase shifts in the general and special regimes, respectively.
	
	\subsection{Optimization in General Regime}
	In this part, we maximize the average channel power \mcthr{of D-IRS-C} w.r.t. the phase shifts in the general  regime. As shown in Theorem~\ref{thm:analyGeneralK}, $\gamma^{(0)}$ is irrelevant to $\bm\phi_1$ and $\bm\phi_2$ in the general  regime. Thus, we optimize the phase shifts only for Cases 1, 2, and 3 in the general regime in the sequel.

	\subsubsection{\mcthr{Optimization for Case $l=1,2$ of General Regime}}
	For $l = 1,2$, as $\gamma^{(l)}(\bm \phi_l)$ depends only on $\bm\phi_l$, we optimize $\bm\phi_l$ in  Case $l$.
	\begin{Prob}[Optimization for \mcthr{D-IRS-C in Case $l=1,2$ of} General Regime]\label{prob:caselGeneral}
		For Case $l=1,2$,
		\begin{align}
			\gamma^{(l)\star} \triangleq \max_{\bm \phi_l} & \quad  \gamma^{(l)}(\bm \phi_l) \nonumber
			\\ s.t.&\quad \phi_{l,t} \in [0,2\pi),t\in \mathcal{T}_l,
		\end{align}
		where $\gamma^{(1)}(\bm \phi_1)$ and $\gamma^{(2)}(\bm \phi_2)$ are given by \eqref{eq:gamma1} and \eqref{eq:gamma2}, respectively. Let $\bm \phi_l^{\star}$ denote an optimal solution for Case $l$, where $l\in\mathcal{L}$.
	\end{Prob}
	
	Problem \ref{prob:caselGeneral} is a challenging non-convex problem with a very complex objection function and $T_l$ optimization variables. However, by the triangle inequality and Cauchy-Schwartz inequality, we can obtain  globally optimal closed-form  solutions of Problem \ref{prob:caselGeneral} for Case 1 with $K_{1U} = 0$ or $K_{12} = 0$ and for Case 2 with $K_{S2} = 0$ or $K_{12} = 0$, respectively.
	For ease of exposition, define $\Lambda(x) = x - 2\pi\lfloor\frac{x}{2\pi}\rfloor \in [0,2\pi)$, $x \in \mathbb{R}$. Note that $\Lambda(\cdot)$ can be used to provide phase shifts $\bm \phi_l$, $l\in\mathcal{L}$ satisfying \eqref{eq:unitconstrint} \cite{Cui2020Interference}.
	\begin{Thm}[Optimal Solution of Problem \ref{prob:caselGeneral}]\label{thm:caselGeneralSpecialCase}
		(i) The unique optimal solution of Problem \ref{prob:caselGeneral} for Case 1 with $K_{1U} = 0$ or $K_{12} = 0$ is given by:
		\begin{align}\label{eq:phi1starCasel}
			\bm \phi_1^{\star} =
			\left\{
			\begin{aligned}
				& \Lambda(-\bm \Delta_{\overline{S1},\overline{12}} - \angle(r_{\overline{S1},\overline{S2}}r_{\overline{S2},\overline{12}}) \mathbf{1}_{T_1}), & K_{1U} = 0,
				\\
				& \Lambda(-\bm \Delta_{\overline{S1},\overline{1U}}-\angle(r_{\overline{S1},\overline{SU}})\mathbf{1}_{T_1}),                                   & K_{12} = 0.
			\end{aligned}
			\right.
		\end{align}
		(ii) The unique optimal solution of Problem \ref{prob:caselGeneral} for Case 2 with $K_{S2} = 0$ or $K_{12} = 0$ is given by:
		\begin{align}\label{eq:phi2starCasel}
			\bm \phi^{\star}_2 = \left\{
			\begin{aligned}
				& \Lambda(-\bm \Delta_{\overline{12},\overline{2U}} - \angle(r_{\overline{12},\overline{1U}})\mathbf{1}_{T_2}), & K_{S2} = 0,
				\\
				& \Lambda(-\bm \Delta_{\overline{S2},\overline{2U}} - \angle(r_{\overline{S2},\overline{SU}})\mathbf{1}_{T_2}), & K_{12} = 0.
			\end{aligned}
			\right.
		\end{align}
	\end{Thm}
	\begin{IEEEproof}
		Please refer to Appendix C.% \ref{proof:caselGeneralSpecialCase}.
	\end{IEEEproof}
	
	From Theorem~\ref{thm:caselGeneralSpecialCase}, we can draw the following conclusions \mcthr{for D-IRS-C}. Firstly, in Case 1 with $K_{1U} =0$, the optimal phase changes over the cascaded LoS channel $(\overline{S1},\overline{12})$ are identical to the phase changes over the LoS channel $\overline{S2}$. Secondly, in Case 2 with $K_{S2}=0$, the optimal phase changes over the cascaded LoS channel $(\overline{12},\overline{2U})$ are identical to the phase changes over the LoS channel $\overline{1U}$. Finally, in Case 1 and Case 2 both with $K_{12}=0$, the optimal phase changes over the cascaded LoS  channel $(\overline{Sl},\overline{lU})$ are  identical to the phase changes over the LoS channel $\overline{SU}$.

	However, for Case 1 with \mcthr{$K_{12}$, $K_{1U}> 0$} and Case 2 with \mcthr{$K_{S2}$, $K_{12}> 0$}, we cannot derive a globally optimal solution with the same method. Instead, we propose a computationally efficient iterative algorithm to obtain  stationary points of  Problem~\ref{prob:caselGeneral} for Case 1 with  \mcthr{$K_{12}$, $K_{1U}> 0$} and  Case 2 with \mcthr{$K_{S2}$, $K_{12}> 0$}, respectively, based on the coordinate descent method. The idea is that at each step of one iteration, $\phi_{l,t}, t \in \mathcal{T}_l$ are sequentially updated by analytically solving the coordinate optimizations, each with a single variable. Specifically, in each coordinate optimization, we maximize $\gamma^{(l)}(\bm \phi_l)$ w.r.t. $\phi_{l,t}$ for some $t \in \mathcal{T}_l$ with the other phase shifts being fixed. For ease of exposition, we rewrite $\bm \phi_l$ as $( \phi_{l,t},\bm \phi_{l,-t})$, where $\bm \phi_{l,-t}$ represents the elements of $\bm \phi_l$ except for $\phi_{l,t}$. Then, given $\bm \phi_{l,-t}$ obtained in the previous step, the coordinate optimizations  w.r.t. $\phi_{l,t}$ for Case 1 with \mcthr{$K_{12}$, $K_{1U}> 0$} and Case 2 with \mcthr{$K_{S2}$, $K_{12}> 0$} are formulated as follows.
	\begin{Prob}[Coordinate Optimization  w.r.t. $\phi_{l,t}$ \mcthr{for D-IRS-C in Case $l=1,2$ of} General Regime]\label{prob:caselGeneralCD} For Case $l=1,2$ and for all $t \in \mathcal{T}_l$,
		\begin{align*}
			\phi_{l,t}^{\dagger} \triangleq \argmax_{ \phi_{l,t} \in [0,2\pi) } & \quad \gamma^{(l)}(\phi_{l,t}, \bm \phi_{l,-t}).
		\end{align*}
	\end{Prob}
	
	Problem \ref{prob:caselGeneralCD} is a single variable optimization problem and hence is much simpler than \mbox{Problem~\ref{prob:caselGeneral}}. Define:
	\begin{align}
		& \mathbf{A}_1  \triangleq L_{\overline{S1},\overline{1U}} \mathbf{A}_{11} + L_{\overline{S1},\overline{12},\widetilde{2U}}T_2 \mathbf{A}_{12},
		\quad	\quad	\quad 	\quad 	\  \mathbf{A}_2  \triangleq  L_{\overline{S2},\overline{2U}} \mathbf{A}_{21} +L_{\widetilde{S1},\overline{12},\overline{2U}}T_S\mathbf{A}_{22},\label{eq:Al}
		\\
		& \mathbf{b}_1  \triangleq \sqrt { L_{\overline{SU}}  L_{\overline{S1},\overline{1U}} } \mathbf{b}_{11} +  \sqrt{L_{\overline{S2},\widetilde{2U}} L_{\overline{S1},\overline{12},\widetilde{2U}}} \mathbf{b}_{12},
		\mathbf{b}_2  \triangleq \sqrt{ L_{\overline{SU}}  L_{\overline{S2},\overline{2U}}}\mathbf{b}_{21}+ \sqrt{L_{\widetilde{S1},\overline{1U}} L_{\widetilde{S1},\overline{12},\overline{2U}}}T_S \mathbf{b}_{22}.\label{eq:bl}
	\end{align}
	After some basic algebraic manipulations, we can derive a closed-form optimal solution of Problem~\ref{prob:caselGeneralCD}.
	\vspace{-2mm}
	\begin{Lem}[Optimal Solution of Problem~\ref{prob:caselGeneralCD}] \label{lem:caselGeneralCDslu}
		For Case $l=1,2$ and for all $t \in \mathcal{T}_l$, the unique optimal solution of Problem~\ref{prob:caselGeneralCD} is given by:
		\begin{align}\label{eq:philtdaggerCasel}
			\phi_{l,t}^{\dagger} = \Lambda \left(-\angle\left(\sum_{k \in \mathcal{T}_l, k \ne t} A_{l,t,k} e^{-j\phi_{l,k}} + b_{l,t}\right)\right),
		\end{align}
		where $A_{l,t,k}$ and $b_{l,t}$ represent the $(t,k)$-th element of $\mathbf{A}_l$ given by \eqref{eq:Al} and the $t$-th element of $\mathbf{b}_l$ given by \eqref{eq:bl}, respectively.
	\end{Lem}
\vspace{-2mm}
	\begin{IEEEproof}
		Please refer to Appendix D. %~\ref{proof:caselGeneralCDslu}.
	\end{IEEEproof}
	
	The details of the coordinate descent-based algorithm for Case 1 with \mcthr{$K_{12}$, $K_{1U}> 0$} and Case 2 with \mcthr{$K_{S2}$, $K_{12}> 0$} are summarized in Algorithm~\ref{alg:generalCasel}. Since Problem~\ref{prob:caselGeneralCD} can be solved analytically, the computation time of Algorithm~\ref{alg:generalCasel} is relatively short. Specifically, the computational complexity of Step 4 of Algorithm \ref{alg:generalCasel} is $\mathcal{O}(T_l)$. Hence the overall computational complexity of each iteration of Algorithm~\ref{alg:generalCasel} is $\mathcal{O}(T_l^2)$. Furthermore, we know that every limit point generated by Algorithm~\ref{alg:generalCasel} is a stationary point of Problem~\ref{prob:caselGeneralCD}, as Problem~\ref{prob:caselGeneralCD} has a unique optimal solution~\cite[Proposition 2.7.1]{Bertsekas1998NP}.
	
	\begin{algorithm}[t]
		\caption{ Coordinate Descent Algorithm for Obtaining A Stationary Point of Problem~\ref{prob:caselGeneral}}
		\label{alg:generalCasel}
		\begin{algorithmic}[1]
			\begin{small}
				\STATE{{\bf Initialization:} Select $\bm\phi_l$ satisfying the constraints in \eqref{eq:unitconstrint} as the initial point.}
				\REPEAT
				\FOR {$t=1,...,T_l$}
				\STATE {Calculate $\phi_{l,t}^{\dagger}$ according to \eqref{eq:philtdaggerCasel}, and set $\phi_{l,t} = \phi_{l,t}^{\dagger}$.}
				\ENDFOR
				\UNTIL{some convergence criterion is met.}
			\end{small}
		\end{algorithmic}
	\end{algorithm}
	\setlength{\textfloatsep}{12pt}

	\subsubsection{\mcthr{Optimization for Case 3 of General Regime}}
	As $\gamma^{(3)}(\bm \phi_1, \bm \phi_2)$ depends on  $\phi_1$ and $\phi_2$, we jointly optimize $\phi_1$  and 	$\phi_2$ in Case 3.
	\begin{Prob}[Optimization for \mcthr{D-IRS-C in Case 3 of} General Regime]\label{prob:case3General}
		For Case 3,
		\begin{align*}
			\gamma^{(3)\star} \triangleq \max_{\bm \phi_1,\bm \phi_2} & \quad \gamma^{(3)}(\bm \phi_1, \bm \phi_2)
			\\ s.t.														&\quad \eqref{eq:unitconstrint},
		\end{align*}
		where $\gamma^{(3)}(\bm \phi_1, \bm \phi_2)$ is given by \eqref{eq:gamma3}.  Let $(\bm \phi_1^{\star},\bm \phi_2^{\star})$ denote an optimal solution.
	\end{Prob}
	
	Apparently, Problem \ref{prob:case3General} is a more challenging non-convex problem than Problem \ref{prob:caselGeneral}, as it has a more complex objective function and more optimization variables. However, by the triangle inequality and Cauchy-Schwartz inequality, we can still obtain a  globally optimal closed-form solution of Problem \ref{prob:case3General} for Case 3 with  $K_{S2}= K_{1U} = 0$ or $K_{12} = K_{SU} = 0$.
	
	\begin{Thm}[Optimal Solution of Problem \ref{prob:case3General}]\label{thm:case3GeneralSpecialCase}
		(i) For Case 3 with $K_{S2} = K_{1U} = 0$, any $(\phi^{\star}_1, \phi^{\star}_2)$ with $\bm \phi_1^\star = \Lambda\left(-\bm \Delta_{\overline{S1},\overline{12}} + \psi\mathbf{1}_{T_1} - \frac{1}{2}\angle\left(r_{\overline{S1},\overline{SU}}\right)\mathbf{1}_{T_1}\right)$ and $\bm \phi_2^\star = \Lambda\left(-\bm \Delta_{\overline{12},\overline{2U}} - \psi\mathbf{1}_{T_2} - \frac{1}{2}\angle\left(r_{\overline{S1},\overline{SU}}\right)\mathbf{1}_{T_2}\right)$, for all $\psi \in \mathbb{R}$, is an optimal solution of Problem~\ref{prob:case3General}.
		(ii) For Case 3 with $K_{12} = K_{SU} = 0$, any $(\phi^{\star}_1, \phi^{\star}_2)$ with $\phi^{\star}_1 = \Lambda(-\bm \Delta_{\overline{S1},\overline{1U}}+\psi\mathbf{1}_{T_1})$, $\phi^{\star}_2=\Lambda(-\bm \Delta_{\overline{S2},\overline{2U}}+\psi\mathbf{1}_{T_2} + \angle(r_{\overline{S1},\overline{S2}})\mathbf{1}_{T_2})$, for all $\psi \in \mathbb{R}$, is an optimal solution of Problem~\ref{prob:case3General}.
	\end{Thm}
	\begin{IEEEproof}
		Please refer to Appendix E.
	\end{IEEEproof}
	
	Theorem \ref{thm:case3GeneralSpecialCase} indicates two facts \mcthr{for D-IRS-C} in the general  regime. Firstly, in Case 3 with $K_{S2} = K_{1U} = 0$, the optimal phase changes over the cascaded LoS channel $(\overline{S1},\overline{12},\overline{2U})$ are the same as the phase changes over the LoS channel $\overline{SU}$. Secondly, in Case 3 with $K_{12} = K_{SU} = 0$, the optimal phase changes over the cascaded LoS channels $(\overline{S1},\overline{1U})$ and $(\overline{S2},\overline{2U})$ are identical.
	
	However, for Case 3 with $K_{S2} > 0$, $K_{1U} > 0$, and $K_{12} + K_{SU} > 0$, we cannot derive a globally  optimal solution with the same method. Instead, we propose a computationally efficient iterative algorithm based on the coordinate descent method. The idea is that at each  iteration, $\phi_{l,t}, t \in \mathcal{T}_l,l \in \mathcal{L}$ are sequentially updated by analytically solving the coordinate optimizations. Specifically, in each coordinate optimization, we maximize $\gamma(\bm\phi_1,\bm\phi_2)$ w.r.t. $\phi_{l,t}$ for some $t \in \mathcal{T}_l,l \in \mathcal{L}$ with the other phase shifts being fixed. For ease of exposition, we rewrite $(\bm \phi_1,\bm \phi_2)$ as $( \phi_{l,t},\bm \phi_{l,-t}, \bm \phi_{-l})$, where $\bm \phi_{l,-t}$ represents the elements of $\bm \phi_l$ except for $\phi_{l,t}$, and $-l$ represents the element in $\mathcal{L} \backslash \{l\}$. Then, given $\bm \phi_{l,-t}$ and $\bm \phi_{-l}$ obtained in the previous step, the coordinate optimization w.r.t. $\phi_{l,t}$  is  formulated as follows.
	
	\begin{Prob}[Coordinate Optimization  w.r.t. $\phi_{l,t}$ for \mcthr{D-IRS-C in Case 3 of} General Regime]\label{prob:case3GeneralCD} For Case 3 and for all $t\in \mathcal{T}_l, l\in\mathcal{L}$,
		\begin{align*}
			\phi_{l,t}^{\dagger} \triangleq \argmax_{ \phi_{l,t} \in [0,2\pi) } & \gamma(\phi_{l,t},\bm\phi_{l,-t},\bm\phi_{-l}).
		\end{align*}
	\end{Prob}
	
	Problem~\ref{prob:case3GeneralCD} is a single variable non-convex optimization problem and hence is much simpler than Problem~\ref{prob:case3General}. Define:
	\begin{subequations}\label{eq:Cl}
		\begin{align}
			\mathbf{C}_1 & \triangleq \mathbf{A}_1 + L_{\overline{S1},\overline{12},\overline{2U}}T_S\mathbf{A}_3^T\mathbf{v}_2^*\mathbf{v}_2^T\mathbf{A}_3^*  + 2\sqrt{L_{\overline{S1},\overline{1U}}L_{\overline{S1},\overline{12},\overline{2U}}}\mathfrak{R} \{\mathrm{diag}(\mathbf{v}_2^H\mathbf{B}_1)\mathbf{B}_2\} \in \mathbb{C}^{T_1 \times T_1},
			\\
			\mathbf{C}_2 & \triangleq \mathbf{A}_2+ L_{\overline{S1},\overline{12},\overline{2U}}T_S\mathbf{A}_3\mathbf{v}_1^*\mathbf{v}_1^T\mathbf{A}_3^H + 2\sqrt{L_{\overline{S2},\overline{2U}}L_{\overline{S1},\overline{12},\overline{2U}}}\mathfrak{R}\{ \mathbf{B}_1\diag(\mathbf{v}_1^H)\mathbf{B}_3\} \in \mathbb{C}^{T_2 \times T_2},
		\end{align}
	\end{subequations}
	\vspace{-5mm}
	\begin{subequations}\label{eq:dl}
		\begin{align}
			& \!\!\!\!\!\!\!\!\!\!\!\!\!\!\!\!\!\!\!\!\!\!\!\!\!\!\! \mathbf{d}_1  \triangleq \mathbf{b}_1 + \sqrt{L_{\overline{S2},\overline{2U}}L_{\overline{S1},\overline{12},\overline{2U}}}\mathrm{diag}(\mathbf{v}_2^H\mathbf{B}_1)\mathbf{B}_3\mathbf{v}_2 \nonumber
			\\
			& \!\!\!\!\!\!\!\!\!\!\!\!\!+ \sqrt{L_{\overline{SU}}  L_{\overline{S1},\overline{12},\overline{2U}} } \mathbf{B}_4^T\mathbf{v}_2^* + \sqrt{ L_{\overline{S1},\overline{1U}} L_{\overline{S2},\overline{2U}} }  \mathbf{B}_5^H\mathbf{v}_2 \in \mathbb{C}^{T_1},
			\\
			& \!\!\!\!\!\!\!\!\!\!\!\!\!\!\!\!\!\!\!\!\!\!\!\!\!\!\! \mathbf{d}_2  \triangleq \mathbf{b}_2 + \sqrt{L_{\overline{S1},\overline{1U}}L_{\overline{S1},\overline{12},\overline{2U}}}\mathbf{B}_1\diag(\mathbf{v}_1^H)\mathbf{B}_2\mathbf{v}_1
			\nonumber
			\\
			& \!\!\!\!\!\!\!\!\!\!\!\!\!+ \sqrt{L_{\overline{SU}}  L_{\overline{S1},\overline{12},\overline{2U}} } \mathbf{B}_4\mathbf{v}_1^* + \sqrt{ L_{\overline{S1},\overline{1U}} L_{\overline{S2},\overline{2U}} }  \mathbf{B}_5\mathbf{v}_1 \in \mathbb{C}^{T_2}.
		\end{align}
	\end{subequations}
	After some basic algebraic manipulations, we can derive a closed-form optimal solution of Problem \ref{prob:case3GeneralCD}.
	\begin{Lem}[Optimal Solution of Problem \ref{prob:case3GeneralCD}]\label{lem:case3GeneralCDslu} For Case 3 and all $t\in \mathcal{T}_l$, \mbox{$l\in\mathcal{L}$}, the unique optimal solution of Problem~\ref{prob:case3GeneralCD} is given by:
		\begin{align}\label{eq:philtdagger}
			\phi_{l,t}^{\dagger} = \Lambda\bigg(-\angle\bigg(\sum_{k \in \mathcal{T}_l,k \ne t} C_{l,t,k} e^{-j\phi_{l,k}} + d_{l,t}\bigg)\bigg),
		\end{align}
		where $C_{l,t,k}$ represents the $(t,k)$-th element of $\mathbf{C}_l$ given by \eqref{eq:Cl}, and $d_{l,t}$ represents the $t$-th element of $\mathbf{d}_l$ given by \eqref{eq:dl}.
	\end{Lem}
	\begin{IEEEproof}
		Please refer to Appendix F.
	\end{IEEEproof}	
	
	The details of the coordinate descent-based algorithm are summarized in Algorithm \ref{alg:generalCase3}. The computational complexities of Step 4 and Step 6 of Algorithm \ref{alg:generalCase3} are $\mathcal{O}(T_1T_2T_l)$ and $\mathcal{O}(T_l)$, $l\in\mathcal{L}$. Hence, the computational complexity of each iteration of Algorithm \ref{alg:generalCase3} is $\mathcal{O}(T_1^2T_2 + T_2^2T_1)$. If $T_1 = cT$ and $T_2 = (1-c)T$ for some $c\in(0,1)$, then the computational complexity of each iteration of Algorithm \ref{alg:generalCase3} becomes $\mathcal{O}(T^3)$. As Problem \ref{prob:case3GeneralCD} can be solved analytically, the computation efficiency of Algorithm \ref{alg:generalCase3} is relatively high. Furthermore, we know  that every limit point generated by Algorithm \ref{alg:generalCase3} is a stationary point of Problem~\ref{prob:case3General}, as Problem \ref{prob:case3GeneralCD} has a unique optimal solution~\cite[Proposition 2.7.1]{Bertsekas1998NP}.
	
	\begin{algorithm}[t]
		\caption{Coordinate Descent Algorithm for Obtaining A Stationary Point of Problem \ref{prob:case3General}}
		\label{alg:generalCase3}
		\begin{multicols}{2}
			\begin{small}
				\begin{algorithmic}[1]
					\STATE{{\bf Initialization:} Select $\bm\phi_1$ and $\bm\phi_2$ satisfying the constraints in \eqref{eq:unitconstrint} as the initial point. }
					\REPEAT
					\FOR {$l=1,2$}
					\STATE {Calculate $\mathbf{C}_l$ and $\mathbf{d}_l$ according to \eqref{eq:Cl} and \eqref{eq:dl}, respectively.}
					\FOR {$t=1,..,T_l$}
					\STATE {Calculate $\phi_{l,t}^{\dagger}$ according to \eqref{eq:philtdagger}, and set $\phi_{l,t} = \phi_{l,t}^{\dagger}$.}
					\ENDFOR
					\ENDFOR
					\UNTIL{some convergence criterion is met.}
				\end{algorithmic}
			\end{small}
		\end{multicols}
	\end{algorithm}
	\setlength{\textfloatsep}{12pt}
	
	\subsubsection{Optimal Average Channel Power in General Regime at Large Number of Reflecting Elements}	
	We characterize the  average channel power \mcthr{of D-IRS-C} in Case 0 and the optimal average channel power \mcthr{of D-IRS-C}  in Cases 1, 2, and 3 of the general regime at Large $T_1$, $T_2$, and $T$.	
	\begin{Thm}[Optimal Average Channel Power \mcthr{of D-IRS-C} in General Regime]\label{thm:doubleAsyT}
		(i) $\gamma^{(0)}       \stackrel{T_1,T_2\rightarrow\infty}{\sim} (L_{\widetilde{S1},\overline{12},\widetilde{2U}} +L_{\overline{S1},\widetilde{12},\overline{2U}}+L_{\overline{S1},\widetilde{12},\widetilde{2U}}+L_{\widetilde{S1},\widetilde{12},\overline{2U}}+L_{\widetilde{S1},\widetilde{12},\widetilde{2U}})T_ST_1T_2$, $\gamma^{(1)\star}  \stackrel{T_1,T_2\rightarrow\infty}{\sim} L_{\overline{S1},\overline{12},\widetilde{2U}} T_ST_1^2T_2$, $\gamma^{(2)\star}  \stackrel{T_1,T_2\rightarrow\infty}{\sim} L_{\widetilde{S1},\overline{12},\overline{2U}} T_ST_1T_2^2$, and $\gamma^{(3)\star}  \stackrel{T_1,T_2\rightarrow\infty}{\sim} L_{\overline{S1},\overline{12},\overline{2U}}T_ST_1^2T_2^2$.
		(ii) If $T_1 = cT$ and $T_2 = (1-c)T$ for some $c\in(0,1)$, then
		$\gamma^{(0)}       \stackrel{T\rightarrow\infty}{\sim} (L_{\widetilde{S1},\overline{12},\widetilde{2U}} +L_{\overline{S1},\widetilde{12},\overline{2U}}+L_{\overline{S1},\widetilde{12},\widetilde{2U}}+L_{\widetilde{S1},\widetilde{12},\overline{2U}}+L_{\widetilde{S1},\widetilde{12},\widetilde{2U}})c(1-c)T_ST^2$, $\gamma^{(1)\star}  \stackrel{T\rightarrow\infty}{\sim} L_{\overline{S1},\overline{12},\widetilde{2U}} c^2(1-c)T_ST^3$, $\gamma^{(2)\star}  \stackrel{T\rightarrow\infty}{\sim} L_{\widetilde{S1},\overline{12},\overline{2U}} c(1-c)^2T_ST^3$, and $\gamma^{(3)\star}  \stackrel{T\rightarrow\infty}{\sim} L_{\overline{S1},\overline{12},\overline{2U}}c^2(1-c)^2T_ST^4$.\footnote{$f(x,y) \stackrel{x,y\rightarrow\infty}{\sim} g(x,y)$ means $\lim_{x,y\rightarrow\infty} \frac{f(x,y)}{g(x,y)}=1$, and $f(x) \stackrel{x\rightarrow\infty}{\sim} g(x)$ means $\lim_{x\rightarrow\infty} \frac{f(x)}{g(x)}=1$.}
	\end{Thm}
	\begin{IEEEproof}
		Please refer to Appendix G.%\ref{proof:doubleAsyT}.
	\end{IEEEproof}
	
 \mcthr{By Theorem~\ref{thm:doubleAsyT} and~\cite[Theorem 1]{Han2021doubleMIMO}}, for \mcthr{D-IRS-C in the general regime}, the optimal quasi-static phase shift design for Case 3 achieves the same average power gain in order w.r.t. the total number of  reflecting elements $T$ (i.e., $\Theta(T^4)$) as the optimal instantaneous CSI-adaptive phase shift design but with \mcthr{lower channel estimation and phase  adjustment costs}. \mcthr{Besides, for D-IRS-C in the general regime, the optimal average channel powers in Case 1 and Case 2 (i.e., $\Theta(T^3)$) are equivalent in order and are higher in order than the average channel power in Case 0 (i.e., $\Theta(T^2)$) and lower in order than the average channel power in Case 3 (i.e., $\Theta(T^4)$).}
	
	\subsection{Optimization in Pure LoS Regime}
	In this part, \mcthr{we maximize $\bar{\gamma}^{(3)}(\bm \phi_1,\bm \phi_2)$ w.r.t. $\bm \phi_1$ and $\bm \phi_2$ in the pure LoS regime.}
	
	\begin{Prob}[Optimization  \mcthr{for D-IRS-C} in Pure LoS Regime]\label{prob:case3LargeK}
		\begin{align*}
			\bar{\gamma}^{(3)\star} \triangleq & \quad \max_{\bm \phi_1,\bm \phi_2} \bar{\gamma}^{(3)}(\bm \phi_1,\bm \phi_2)
			\\
			s.t.                               & \quad  \eqref{eq:unitconstrint},
		\end{align*}
		where $\bar{\gamma}^{(3)}(\bm \phi_1,\bm \phi_2)$ is given by \eqref{eq:gamma3Infty}. Let $\left(\bar{\bm \phi}_1^{\star},\bar{\bm \phi}_2^{\star}\right)$ denote an optimal solution.
	\end{Prob}
	
	As $\bar{\gamma}^{(3)}(\bm \phi_1, \bm \phi_2)$ has a simpler form than $\gamma^{(3)}(\bm \phi_1, \bm \phi_2)$, Problem \ref{prob:case3LargeK} is more tractable than Problem~\ref{prob:case3General}. We propose a computationally efficient iterative algorithm to obtain a stationary point of Problem \ref{prob:case3LargeK} using the block coordinate descent method. Specifically, we divide the optimization variables into $T_2+1$ disjoint blocks, i.e., $\bm \phi_1$,  $\phi_{2,t}, t\in \mathcal{T}_2$, with the first block consisting of $T_1$ coordinates and each of the remaining blocks consisting of only one coordinate, and sequentially update each block by analytically solving the corresponding block coordinate and coordinate optimization problems. Here, we treat $\bm \phi_1$ as a single block, as the optimization problem w.r.t. $\bm \phi_1$ can be analytically solved, with a lower computational complexity than the separate optimization problems w.r.t. $\bm \phi_{1,t},t\in\mathcal{T}_1$, which will be seen shortly. The block coordinate optimization  w.r.t. $\bm \phi_1$ and the coordinate optimizations  w.r.t. $\bm \phi_{2,t},t\in\mathcal{T}_2$ are  formulated as follows.

	\begin{Prob}[Block Coordinate Optimization  w.r.t. $\bm \phi_1$ \mcthr{for D-IRS-C} in Pure LoS Regime]\label{prob:suboptphi1} In the pure LoS regime,
		\begin{align*}
			\bar{\bm \phi}_1^{\dagger} \triangleq & \quad \argmax_{\bm \phi_1} \bar{\gamma}^{(3)}(\bm \phi_1,\bm \phi_2)
			\\
			s.t.                                  & \quad  \phi_{1,t} \in [0,2\pi),t\in\mathcal{T}_1.
		\end{align*}
	\end{Prob}
	
	\begin{Prob}[Coordinate Optimization w.r.t. $\bm \phi_{2,t}$ \mcthr{for D-IRS-C} in Pure LoS Regime]\label{prob:suboptphi2t} In the pure LoS regime,  for all $t \in \mathcal{T}_2$,
		\begin{align*}
			\bar{\phi}_{2,t}^{\dagger} \triangleq & \quad \argmax_{\phi_{2,t} \in [0,2\pi)} \bar{\gamma}^{(3)}( \phi_{2,t},\phi_{2,-t}, \bm \phi_1).
		\end{align*}
	\end{Prob}
	
	Define:
	\begin{align}
		\bar{\mathbf{C}}_2 & \triangleq \alpha_{S2}\alpha_{2U}\mathbf{A}_{21}+ \alpha_{S1}\alpha_{12}\alpha_{2U}T_S \mathbf{A}_3\mathbf{v}_1^*\mathbf{v}_1^T\mathbf{A}_3^H \nonumber
		\\&\quad+ 2\sqrt{\alpha_{S2}\alpha_{2U}\alpha_{S1}\alpha_{12}\alpha_{2U}} \mathfrak{R} \{\mathbf{B}_1\diag(\mathbf{v}_1^H)\mathbf{B}_3\} \in \mathbb{C}^{T_2 \times T_2}, \label{eq:C2bar}
		\\
		\bar{\mathbf{d}}_2 & \triangleq \sqrt{\alpha_{SU}\alpha_{S2}\alpha_{2U}} \mathbf{b}_{21}+ \sqrt{\alpha_{S1}\alpha_{1U}\alpha_{S1}\alpha_{12}\alpha_{2U}}\mathbf{B}_1\diag(\mathbf{v}_1^H)\mathbf{B}_2\mathbf{v}_1 \nonumber
		\\
		& \quad+ \sqrt{\alpha_{SU} \alpha_{S1} \alpha_{1U}} \mathbf{B}_4\mathbf{v}_1^* + \sqrt{\alpha_{S1}\alpha_{1U}\alpha_{S2}\alpha_{2U}} \mathbf{B}_5\mathbf{v}_1 \in \mathbb{C}^{T_2}.\label{eq:d2bar}
	\end{align}
	After some basic algebraic manipulations, we can derive  closed-form optimal solutions of Problem \ref{prob:suboptphi1} and  Problem~\ref{prob:suboptphi2t} using the triangle inequality and Cauchy-Schwartz inequality.
	\begin{Lem}[Optimal Solutions of Problem~\ref{prob:suboptphi1} and Problem~\ref{prob:suboptphi2t}] \label{lem:case3LargeK}
		(i)	The unique optimal solution of Problem~\ref{prob:suboptphi1} is given by:
		\begin{align}\label{eq:suboptphi1}
			\bar{\bm \phi}_1^{\dagger} & = \Lambda \Big(-\angle\left( \diag\left( \sqrt{\alpha_{S1}\alpha_{1U}} \bar{\mathbf{h}}_{1U}^H + \sqrt{\alpha_{S1}\alpha_{1U}\alpha_{S1}\alpha_{12}\alpha_{2U}} \mathbf{v}_2^H \mathbf{B}_1\right) \mathbf{a}_{A,S1}\right) \nonumber
			\\
			& \quad\quad\ \  - \angle\big(\mathbf{a}^H_{D,S1} ( \sqrt{\alpha_{SU}} \bar{\mathbf{h}}_{SU} + \sqrt{\alpha_{S2}\alpha_{2U}}\bar{\mathbf{H}}_{S2}^H  \mathrm{diag}(\mathbf{v}_2)\bar{\mathbf{h}}_{2U} )\big)\mathbf{1}_{T_1} \Big).
		\end{align}
		(ii) For all $t \in \mathcal{T}_2$, the unique optimal solution of Problem~\ref{prob:suboptphi2t} is given by:
		\begin{align}\label{eq:suboptphi2t}
			\bar{\phi}_{2,t}^{\dagger} = \Lambda\left(-\angle\left(\sum_{k \in \mathcal{T}_2, k \ne t} \bar{C}_{2,t,k} e^{-j\phi_{2,k}} + \bar{d}_{2,t}\right)\right),
		\end{align}
		where $\bar{C}_{2,t,k}$  represents the $(t,k)$-th element of $\bar{\mathbf{C}}_2$ given by \eqref{eq:C2bar},  and $\bar{d}_{2,t}$ represents the $t$-th element of $\bar{\mathbf{d}}_2$ given by \eqref{eq:d2bar}.
	\end{Lem}
	\begin{IEEEproof}
		Please refer to Appendix H.%~\ref{proof:case3LargeK}.
	\end{IEEEproof}
	
	\begin{algorithm}[t]
		\caption{Block Coordinate Descent Algorithm for Obtaining A Stationary Point of Problem~\ref{prob:case3LargeK}}
		\label{alg:generalCase3LargeK}
		\vspace{-3mm}
		\begin{multicols}{2}
			\begin{small}		
				\begin{algorithmic}[1]
					\STATE{{\bf Initialization:} Select $\bm\phi_1$ and $\bm\phi_2$ satisfying the constraints in \eqref{eq:unitconstrint} as the initial point. }
					\REPEAT
					\STATE{Calculate $\bar{\bm \phi}_1^{\dagger}$ according to \eqref{eq:suboptphi1}, and set $\bm \phi_1= \bar{\bm \phi}_1^{\dagger}$.}
					\STATE{Calculate $\bar{\mathbf{C}}_2$ and $\bar{\mathbf{d}}_2$ according to \eqref{eq:C2bar} and \eqref{eq:d2bar}, respectively.}
					\FOR { $t=1,..,T_2$}
					\STATE {Calculate $\bar{\phi}_{2,t}^{\dagger}$ according to \eqref{eq:suboptphi2t}, and set $\phi_{2,t} = \bar{\phi}_{2,t}^{\dagger}$.}
					\ENDFOR
					\UNTIL{some convergence criterion is met.}
				\end{algorithmic}
			\end{small}
		\end{multicols}
		\vspace{-2mm}
	\end{algorithm}
	\setlength{\textfloatsep}{12pt}
	
	The details of the \mcthr{block coordinate descent} algorithm are summarized in Algorithm~\ref{alg:generalCase3LargeK}. The computational complexities of Step 3, Step 4, and Step 6 are $\mathcal{O}(T_1T_2)$, $\mathcal{O}(T_1T_2^2)$, and $\mathcal{O}(T_2)$, respectively. Hence the computational complexity of each iteration of Algorithm~\ref{alg:generalCase3LargeK} is $\mathcal{O}(T_1T_2^2)$. \mcthr{If $T_1 = cT$ and $T_2 = (1-c)T$ for some $c\in(0,1)$, then the computational complexity of each iteration of Algorithm~\ref{alg:generalCase3LargeK} becomes $\mathcal{O}(T^3)$.} Since Problem \ref{prob:suboptphi1} and Problem \ref{prob:suboptphi2t} have  unique optimal solutions, respectively, every limit point generated by Algorithm \ref{alg:generalCase3LargeK} is a stationary point of Problem~\ref{prob:case3LargeK} \cite[Proposition 2.7.1]{Bertsekas1998NP}.
	
%	\begin{Rem}[Comparison between Algorithm \ref{alg:generalCase3LargeK} and Algorithm~\ref{alg:generalCase3}]
%		The overall computational complexity for updating $\bm \phi_1$ in each iteration of Algorithm~\ref{alg:generalCase3LargeK} is $\mathcal{O}(T_1T_2)$, whereas the overall computational complexity for updating $\bm \phi_{1,t}, t\in\mathcal{T}_1$ in each iteration of Algorithm \ref{alg:generalCase3} is $\mathcal{O}(T_1^2T_2)$. Besides, the overall computational complexity for updating $\bm \phi_{2,t}$ in each iteration of Algorithm~\ref{alg:generalCase3LargeK} is identical to that of Algorithm \ref{alg:generalCase3}, for all $t\in\mathcal{T}_2$. Thus, the computational complexity of Algorithm~\ref{alg:generalCase3LargeK} is lower than that of Algorithm \ref{alg:generalCase3}.
%	\end{Rem}

	The following lemma characterizes how the optimal average channel power of the pure LoS regime scales $T_1$, $T_2$, and $T$, when they are large.
	\begin{Lem}[Optimal Average Channel Power \mcthr{of D-IRS-C} in Pure LoS Regime]\label{thm:doubleAsyTLargeK}
		(i) $
		\bar{\gamma}^{(3)\star}  \stackrel{T_1,T_2\rightarrow\infty}{\sim} \alpha_{S1}\alpha_{12}\alpha_{2U}T_ST_1^2T_2^2.
		$
		(ii) If $T_1 = cT$ and $T_2 = (1-c)T$ for some $c\in(0,1)$, then
		$
		\bar{\gamma}^{(3)\star}  \stackrel{T\rightarrow\infty}{\sim} \alpha_{S1}\alpha_{12}\alpha_{2U} c^2(1-c)^2T_ST^4.
		$
	\end{Lem}
	\begin{IEEEproof}
		Following the proof of Theorem~\ref{thm:doubleAsyT}, we can show Lemma~\ref{thm:doubleAsyTLargeK}.
	\end{IEEEproof}
	
%	Lemma \ref{thm:doubleAsyTLargeK}  indicates that as $T_1,T_2\rightarrow\infty$, $\bar{\gamma}^{(3)\star}$ increases quadratically with $T_1$ and $T_2$. Furthermore, as $T\rightarrow\infty$, $\bar{\gamma}^{(3)\star}$ increases quartically with~$T$.

	\subsection{Optimization in Pure NLoS Regime}
	As shown in Corollary~\ref{col:analySmallK}, the average channel power in the pure NLoS regime, i.e., $\tilde{\gamma}^{(0)}$, does not change with $\bm \phi_1$ or $\bm \phi_2$. Thus, there is no need to further optimize the phase shifts in this regime. However, the deployment of reflecting elements still influences  $\tilde{\gamma}^{(0)}$. The following lemma characterizes how $\tilde{\gamma}^{(0)}$ scales  $T_1$, $T_2$, and $T$, when they are large.
	\begin{Lem}[Optimal Average Channel Power \mcthr{of D-IRS-C} in Pure NLoS Regime]\label{thm:doubleAsyTSmallK}
		(i) $\tilde{\gamma}^{(0)} \stackrel{T_1,T_2\rightarrow \infty}{\sim} \alpha_{S1}\alpha_{12}\alpha_{2U}T_ST_1T_2$.
		(ii) If $T_1 = cT$ and $T_2 = (1-c)T$ for some $c\in(0,1)$, then $\tilde{\gamma}^{(0)} \stackrel{T \rightarrow \infty}{\sim} \alpha_{S1}\alpha_{12}\alpha_{2U}c(1-c)T_ST^2$.
	\end{Lem}
	\begin{IEEEproof}By showing that $\lim_{T_1,T_2\rightarrow\infty} \frac{\tilde{\gamma}^{(0)}}{\alpha_{S1}\alpha_{12}\alpha_{2U}T_ST_1T_2} = 1$ and $\lim_{T_1,T_2\rightarrow\infty} \frac{\tilde{\gamma}^{(0)}}{\alpha_{S1}\alpha_{12}\alpha_{2U}c(1-c)T_ST^2}=1$, we can show $\tilde{\gamma}^{(0)} \stackrel{T_1,T_2\rightarrow \infty}{\sim} \alpha_{S1}\alpha_{12}\alpha_{2U}T_ST_1T_2$ and  $\tilde{\gamma}^{(0)} \stackrel{T \rightarrow \infty}{\sim} \alpha_{S1}\alpha_{12}\alpha_{2U}c(1-c)T_ST^2$, respectively.\end{IEEEproof}
	
%	Lemma~\ref{thm:doubleAsyTSmallK} indicates that as $T_1,T_2\rightarrow\infty$, $\tilde{\gamma}^{(0)}$ increases  linearly with $T_1$ and $T_2$. Furthermore, as $T\rightarrow\infty$,  $\tilde{\gamma}^{(0)}$ increases quadratically with $T$.
	
	\section{Comparison} \label{sec:Comparision}
\begin{figure}[t]
		\centering
		\subfigure[\small{\mcthr{Double-IRS non-cooperatively assisted system (D-IRS-NC).}}]{\resizebox{5.6 cm}{!}{\includegraphics[width=5.4cm]{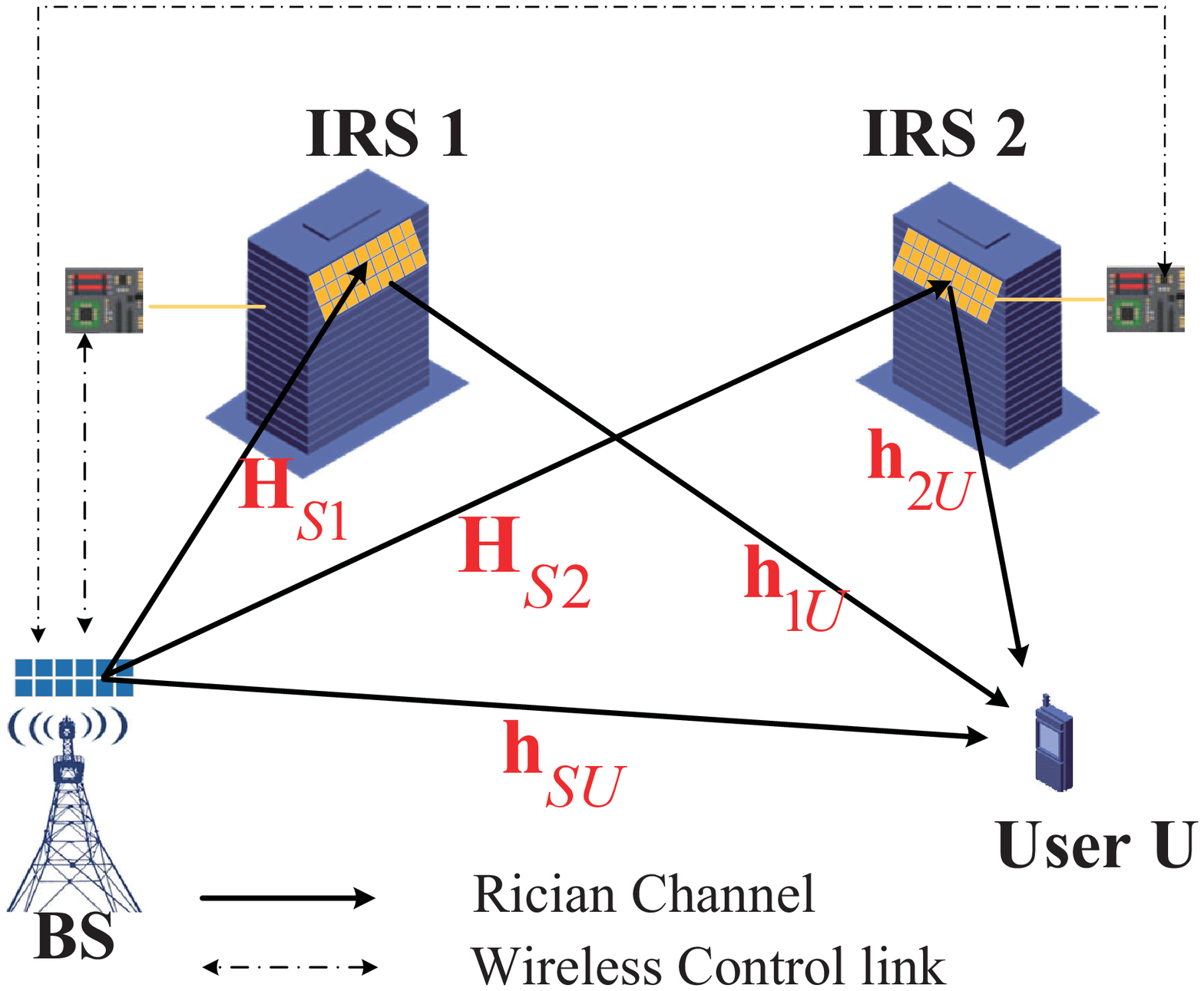}}}\qquad
		\subfigure[\small{\mcthr{Single-IRS-assisted system (S-IRS).}}]{\resizebox{5.6 cm}{!}{\includegraphics[width=5.4cm]{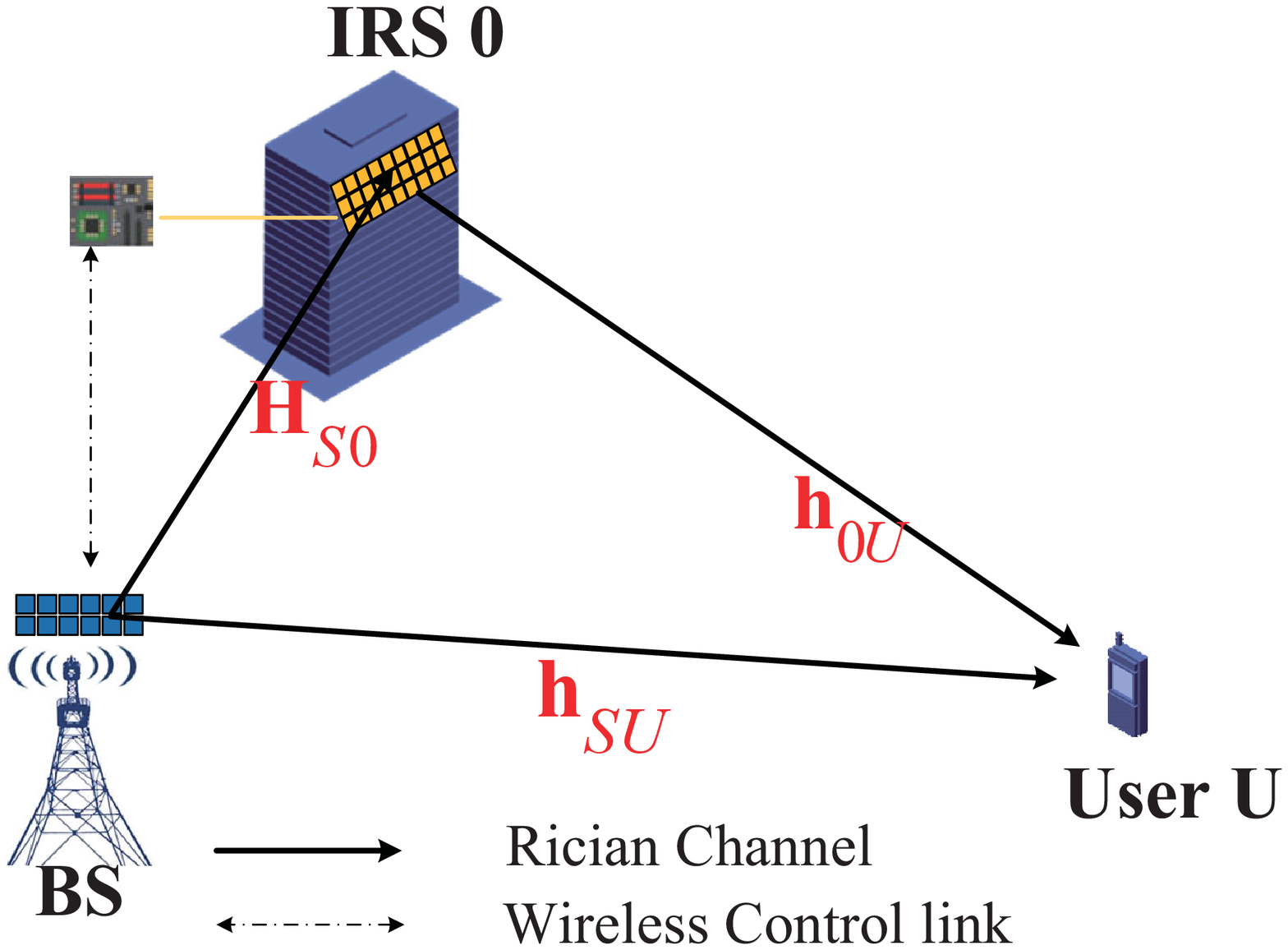}}}	
	\vspace{-2mm}
	\caption{\small{\mcthr{Counterpart IRS-assisted systems.}}}
	\label{fig:doublenonandsingle}
	\end{figure}

%	namely double-IRS non-cooperatively assisted system (D-IRS-NC) and single-IRS-assisted (S-IRS) system
	\mcthr{In this section, we consider two counterpart IRS-assisted systems, namely double-IRS non-cooperatively assisted system (D-IRS-NC) and single-IRS-assisted system (S-IRS), as shown in Fig.~\ref{fig:doublenonandsingle}. The assumptions adopted for the two systems are similar to those for D-IRS-C in Section~\ref{sec:System}. First, we analyze and optimize the average channel powers of their quasi-static phase shift designs. Then, we compare their optimal quasi-static phase shift designs with the optimal quasi-static phase shift design of D-IRS-C in computational complexity and average channel power.}
	\subsection{\mcthr{Double-IRS Non-cooperatively Assisted System (D-IRS-NC)}}
	\mcthr{In this part, we analyze and optimize the average channel power of the quasi-static phase shift design of D-IRS-NC, which is almost the same as D-IRS-C but has no inter-IRS channel. More specifically, there exist three channels, i.e., the direct channel $SU$ and cascaded channels $(S1,1U)$, $(S2,2U)$, represented by $\mathbf{h}_{SU}^H$, $\mathbf{h}_{1U}^H \diag{(\mathbf{v}_1^H)} \mathbf{H}_{S1}$, and $\mathbf{h}_{2U}^H \diag{(\mathbf{v}_2^H)} \mathbf{H}_{S2}$, respectively, as shown in Fig.~\ref{fig:doublenonandsingle} (a). Thus, for D-IRS-NC, the equivalent channel between the BS and user, denoted by $\mathbf{h}_{DNC,e}^{H}(\bm \phi_1,\bm \phi_2) \in \mathbb{C}^{1 \times T_S}$, is given by:
	\begin{align}\label{eq:hednc}
	\mathbf{h}_{DNC,e}^{H}(\bm \phi_1,\bm \phi_2) =\mathbf{h}_{SU}^H + \sum_{l\in\mathcal{L}} \mathbf{h}_{lU}^H \diag{(\mathbf{v}_l^H)} \mathbf{H}_{Sl},
	\end{align} 
	where $\mathbf{H}_{Sl}$ and  $\mathbf{h}_{SU}^H$, $\mathbf{h}_{lU}^H$ are given by~\eqref{eq:Hab} and~\eqref{eq:hab}, respectively. Note that $\mathbf{h}_{DNC,e}^{H}(\bm \phi_1,\bm \phi_2)$ in~\eqref{eq:hednc} is identical to $\mathbf{h}_{e}^{H}(\bm \phi_1,\bm \phi_2)$ in~\eqref{eq:he} with $\alpha_{12}=0$. Let $\gamma_{DNC}(\bm \phi_1,\bm \phi_2) \triangleq \mathbb{E}\left[ \|\mathbf{h}_{DNC,e}(\bm \phi_1,\bm \phi_2)\|_2^2\right]$ represent the average channel power of the equivalent channel of D-IRS-NC.}

\mcthr{First, we characterize the influences of $\bm \phi_1$ and $\bm \phi_2$ on $\gamma_{DNC}(\bm \phi_1,\bm \phi_2)$.
	\begin{Lem}[Influences of Phase Shifts of D-IRS-NC]\label{lem:InfluenceNon}
		For $l \in \mathcal{L}$, if $K_{Sl}K_{lU}=0$ (i.e., $K_{Sl} = 0$ or $K_{lU} = 0$), then  $\gamma_{DNC}(\bm\phi_1,\bm\phi_2)$ does not change with $\bm \phi_l$.
	\end{Lem}
\begin{IEEEproof}
	Following the proof of Lemma~\ref{lem:Influence}, we can show Lemma~\ref{lem:InfluenceNon}.
\end{IEEEproof}Based on Lemma~\ref{lem:InfluenceNon}, we can also divide the channel conditions of D-IRS-NC into four cases.
\begin{itemize}
\item \textbf{Case 0} ($K_{S1} = 0$ or $K_{1U} = 0$ and  $K_{S2} = 0$ or $K_{2U} = 0$):  $\gamma_{DNC}(\bm\phi_1,\bm\phi_2)$ does not change with $\bm \phi_1$ or $\bm \phi_2$ and hence is rewritten as  $\gamma_{DNC}^{(0)}$.
\item \textbf{Case 1} ($K_{S1}$,  $K_{1U} > 0$ and $K_{S2} = 0$ or $K_{2U} = 0$):  $\gamma_{DNC}(\bm\phi_1,\bm\phi_2)$ changes only with $\bm \phi_1$ and hence is rewritten as  $\gamma_{DNC}^{(1)}(\bm\phi_1)$.
\item \textbf{Case 2} ($K_{S1} = 0$ or $K_{1U} = 0$ and $K_{S2}$, $K_{2U} > 0$):  $\gamma_{DNC}(\bm\phi_1,\bm\phi_2)$ changes only with $\bm \phi_2$ and hence is rewritten as  $\gamma_{DNC}^{(2)}(\bm\phi_2)$.
\item \textbf{Case 3} ($K_{S1} $, $K_{1U}$, $K_{S2}$, $K_{2U} > 0$):  $\gamma_{DNC}(\bm\phi_1,\bm\phi_2)$ changes  with both $\bm \phi_1$ and  $\bm \phi_2$ and  is also written as  $\gamma_{DNC}^{(3)}(\bm\phi_1,\bm\phi_2)$.
\end{itemize}}

\mcthr{Next, we characterize the average channel power of D-IRS-NC in the general, pure LoS, and pure NLoS regimes.
\begin{Thm}[Average Channel Power of D-IRS-NC in General Regime]\label{thm:analyGeneralKnon}
	\begin{align}
		\gamma_{DNC}^{(0)} =                      & \alpha_{SU}T_S + \sum_{l=1}^{2}(L_{\widetilde{Sl},\overline{lU}} + L_{\overline{Sl},\widetilde{lU}} +L_{\widetilde{Sl},\widetilde{lU}})T_ST_l
		\label{eq:gamma0non}
		\\
		\gamma_{DNC}^{(1)}(\bm\phi_1) =           &  L_{\overline{S1},\overline{1U}} \mathbf{v}_1^H\mathbf{A}_{11}\mathbf{v}_1 +2\mathfrak{R}\left\{\sqrt { L_{\overline{SU}}  L_{\overline{S1},\overline{1U}} }\mathbf{v}_1^H \mathbf{b}_{11}\right\}   + \gamma_{DNC}^{(0)}, 
		\label{eq:gamma1non}
		\\
		\gamma_{DNC}^{(2)}(\bm\phi_2) =           & L_{\overline{S2},\overline{2U}} \mathbf{v}_2^H\mathbf{A}_{21} \mathbf{v}_2
		+2\mathfrak{R}\left\{\sqrt{ L_{\overline{SU}}  L_{\overline{S2},\overline{2U}}}\mathbf{v}_2^H\mathbf{b}_{21}\right\} + \gamma_{DNC}^{(0)}, \label{eq:gamma2non}
		\\
		\gamma_{DNC}^{(3)}(\bm\phi_1,\bm\phi_2) = & \gamma_{DNC}^{(1)}(\bm\phi_1) + \gamma_{DNC}^{(2)}(\bm\phi_2) + \sqrt{L_{\overline{S1},\overline{1U}}L_{\overline{S2},\overline{2U}}}\mathbf{v}_2^H\mathbf{B}_5\mathbf{v}_1 - \gamma_{DNC}^{(0)}. \label{eq:gamma3non}
	\end{align}
\end{Thm}
\begin{IEEEproof}
	Following the proof of Theorem~\ref{thm:analyGeneralK}, we can show Theorem~\ref{thm:analyGeneralKnon}.
\end{IEEEproof}
\mcthr{Similarly, $\gamma_{DNC}^{(0)}$, $\gamma_{DNC}^{(1)}(\bm \phi_1)$, $\gamma_{DNC}^{(2)}(\bm \phi_2)$, and $\gamma_{DNC}^{(3)}(\bm \phi_1,\bm \phi_2)$  in Theorem~\ref{thm:analyGeneralKnon} are identical to $\gamma^{(0)}$, $\gamma^{(1)}(\bm \phi_1)$, $\gamma^{(2)}(\bm \phi_2)$, and $\gamma^{(3)}(\bm \phi_1,\bm \phi_2)$ in Theorem~\ref{thm:analyGeneralK} with $\alpha_{12}=0$ (i.e., $L_{\overline{12}} = L_{\widetilde{12}}=0$), respectively. According to the definitions of the four cases, Cases 0, 1, and 2 of D-IRS-NC are void in the pure LoS regime, and Cases 1, 2, and 3 of D-IRS-NC are  void in the pure NLoS regime. Besides, Theorem~\ref{thm:analyGeneralKnon} readily implies the following.}
\begin{Cor}[Average Channel Power of D-IRS-NC in Pure LoS Regime]\label{col:analyLargeKnon}
	As $K_{S1}$, $K_{S2}$, $K_{SU}$, $K_{1U}$,  $K_{2U} \rightarrow \infty$, $ \gamma_{DNC}^{(3)} (\bm \phi_1,\bm \phi_2)\rightarrow \bar{\gamma}_{DNC}^{(3)}(\bm \phi_1,\bm \phi_2)$, where
	\begin{align}\label{eq:gamma3barnon}
		\bar{\gamma}_{DNC}^{(3)}&(\bm \phi_1,\bm \phi_2) \triangleq \alpha_{SU}T_S+ \alpha_{S1}\alpha_{1U}\mathbf{v}_1^H\mathbf{A}_{11}\mathbf{v}_1+\alpha_{S2}\alpha_{2U}\mathbf{v}_2^H\mathbf{A}_{21}\mathbf{v}_2
		\nonumber
		\\
		&+2\mathfrak{R}\left\{\sqrt { \alpha_{SU}  \alpha_{S1}\alpha_{1U} }\mathbf{v}_1^H \mathbf{b}_{11} + \sqrt{ \alpha_{SU}  \alpha_{S2}\alpha_{2U}}\mathbf{v}_2^H\mathbf{b}_{21} + \sqrt{\alpha_{S1}\alpha_{1U}\alpha_{S2}\alpha_{2U}}\mathbf{v}_2^H\mathbf{B}_5\mathbf{v}_1\right\}.
	\end{align}
\end{Cor}
\begin{Cor}[Average Channel Power of D-IRS-NC in Pure NLoS Regime]\label{col:analySmallKnon}
	If  $K_{S1}$, $K_{S2}$, $K_{SU}$, $K_{1U}$, $K_{2U}=0$, then
	$\gamma_{DNC}^{(0)} = \tilde{\gamma}_{DNC}^{(0)}$, where
	\begin{align}
		\tilde{\gamma}_{DNC}^{(0)}\triangleq \alpha_{SU}T_S +  \alpha_{S1}\alpha_{1U}T_ST_1 + \alpha_{S2}\alpha_{2U}T_ST_2.
	\end{align}
\end{Cor}	
Analogously, $\bar{\gamma}_{DNC}^{(3)}(\bm \phi_1,\bm \phi_2)$ in Corollary~\ref{col:analyLargeKnon} and $\tilde{\gamma}_{DNC}^{(0)}$ in Corollary~\ref{col:analySmallKnon} are identical to $\bar{\gamma}^{(3)}(\bm \phi_1,\bm \phi_2)$ in Corollary~\ref{col:analyLargeK} with $\alpha_{12}=0$ and $\tilde{\gamma}^{(0)}$ in Corollary~\ref{col:analySmallK} with $\alpha_{12}=0$, respectively. Noting that $\gamma_{DNC}^{(0)}$ and $\tilde{\gamma}_{DNC}^{(0)}$ do not change with $\bm \phi_1$ and $\bm \phi_2$, for D-IRS-NC, we optimize the phase shifts only in Cases 1, 2, and 3  of the general regime and in Case 3 of the pure LoS regime. Specifically, for D-IRS-NC, the maximization of $\gamma_{DNC}^{(l)}(\bm \phi_l)$ w.r.t. $\bm \phi_l$ in Case $l=1,2$ of the general regime is formulated as:
\begin{align}\label{prob:caselGeneralnon}
	\begin{aligned}
		\gamma_{DNC}^{(l)\star} \triangleq \max_{\bm \phi_l} & \quad  \gamma_{DNC}^{(l)}(\bm \phi_l)
	\\ s.t.&\quad \phi_{l,t} \in [0,2\pi),t\in \mathcal{T}_l,
	\end{aligned}
\end{align}
where $\gamma_{DNC}^{(1)}(\bm \phi_1)$ and $\gamma_{DNC}^{(2)}(\bm \phi_2)$ are given by \eqref{eq:gamma1non} and \eqref{eq:gamma2non}, respectively. Let $\bm \phi_{DNC,l}^{\star}$ denote an optimal solution for Case $l=1,2$.
The optimal solution of  the problem in~\eqref{prob:caselGeneralnon} is given below.
\begin{Lem}[Optimal Solution of Problem in \eqref{prob:caselGeneralnon}]\label{thm:caselGeneralnon}
The unique optimal solution of Problem in~\eqref{prob:caselGeneralnon} for Case $l=1,2$ is given by:
	\begin{align}
		\bm \phi_{DNC,l}^{\star} = \Lambda(-\bm \Delta_{\overline{Sl},\overline{lU}} - \angle(r_{\overline{Sl},\overline{SU}})\mathbf{1}_{T_l}).
	\end{align}
\end{Lem}
\begin{IEEEproof}
	Following the proof of Theorem~\ref{thm:caselGeneralSpecialCase}, we can show Lemma~\ref{thm:caselGeneralnon}.
\end{IEEEproof}
Lemma~\ref{thm:caselGeneralnon} can be viewed as a special case of Theorem~\ref{thm:caselGeneralSpecialCase}. Lemma~\ref{thm:caselGeneralnon} indicates that for Case $l=1,2$ of D-IRS-NC in the general regime, the optimal phase  changes over the cascaded LoS channel $(\overline{Sl},\overline{lU})$ are identical to the phase changes over the LoS channel $\overline{SU}$. Furthermore, the computational complexity for calculating $\bm \phi_{DNC,l}^{\star}$ based on Lemma~\ref{thm:caselGeneralnon} is $\mathcal{O}(T)$.}

\mcthr{For D-IRS-NC, the maximization of $\gamma_{DNC}^{(3)}(\bm \phi_1, \bm \phi_2)$ w.r.t. $\bm \phi_1$ and $\bm \phi_2$ in Case 3 of the general regime is formulated as:
	\begin{equation}\label{prob:case3Generalnon}
		\begin{aligned}
		\gamma_{DNC}^{(3)\star} \triangleq \max_{\bm \phi_1,\bm \phi_2} & \quad \gamma_{DNC}^{(3)}(\bm \phi_1, \bm \phi_2)
		\\ s.t.														&\quad \eqref{eq:unitconstrint},
		\end{aligned}
	\end{equation}
where $\gamma_{DNC}^{(3)}(\bm \phi_1, \bm \phi_2)$ is given by \eqref{eq:gamma3non}.  Let $(\bm \phi_{DNC,1}^{\star},\bm \phi_{DNC,2}^{\star})$ denote an optimal solution. Besides, for D-IRS-NC, the maximization of $\gamma_{DNC}^{(3)}(\bm \phi_1, \bm \phi_2)$ w.r.t. $\bm \phi_1$ and $\bm \phi_2$ in Case 3 of the pure LoS regime is formulated as:
	\begin{equation}\label{prob:PureLoSlnon}
	\begin{aligned}
		\bar{\gamma}_{DNC}^{(3)\star} \triangleq \max_{\bm \phi_1,\bm \phi_2} & \quad \bar{\gamma}_{DNC}^{(3)}(\bm \phi_1, \bm \phi_2)
\\ s.t.														&\quad \eqref{eq:unitconstrint},
	\end{aligned}
\end{equation}
where $\bar{\gamma}_{DNC}^{(3)}(\bm \phi_1, \bm \phi_2)$ is given by \eqref{eq:gamma3barnon}.  Let $(\bm \bar{\phi}_{DNC,1}^{\star},\bm \bar{\phi}_{DNC,2}^{\star})$ denote an optimal solution.}

\mcthr{
The optimal solutions of the problem in~\eqref{prob:case3Generalnon} and the problem in~\eqref{prob:PureLoSlnon} are given below.
\begin{Lem}[Optimal Solutions of Problem in~\eqref{prob:case3Generalnon} and Problem in \eqref{prob:PureLoSlnon}] \label{thm:PureLoSlnon}
	The unique optimal solutions of the problem in~\eqref{prob:case3Generalnon} and the problem in~\eqref{prob:PureLoSlnon} are given by:
	\begin{align}
		\bm \phi_{DNC,1}^{\star} = \bm \bar{\bm\phi}_{DNC,1}^{\star} &= \Lambda(-\bm \Delta_{\overline{S1},\overline{1U}} - \angle(r_{\overline{S1},\overline{SU}})\mathbf{1}_{T_1}),
		\\
		\bm \phi_{DNC,2}^{\star} = \bm \bar{\bm\phi}_{DNC,2}^{\star} &= \Lambda(-\bm \Delta_{\overline{S2},\overline{2U}} - (\angle(r_{\overline{S2},\overline{SU}}) - \angle(r_{\overline{S1},\overline{S2}}))\mathbf{1}_{T_2}).
	\end{align}
\end{Lem}
\begin{IEEEproof}
	Following the proof of Theorem~\ref{thm:caselGeneralSpecialCase}, we can show Lemma~\ref{thm:PureLoSlnon}.
\end{IEEEproof}
Lemma~\ref{thm:PureLoSlnon} indicates that for D-IRS-NC in  Case 3 of the general and pure LoS regimes, the optimal phase changes over the cascaded LoS channels $(\overline{S1},\overline{1U})$  and $(\overline{S2},\overline{2U})$ and the LoS channel $\overline{SU}$ are identical. Furthermore, the computational complexities for calculating $\bm \phi_{DNC,1}^{\star}$ ($\bm \bar{\bm \phi}_{DNC,1}^{\star}$) and $\bm \phi_{DNC,2}^{\star}$ ($\bm \bar{\bm \phi}_{DNC,2}^{\star}$) based on Lemma~\ref{thm:PureLoSlnon} are $\mathcal{O}(T)$.}

\mcthr{Finally, we characterize the optimal average channel power of D-IRS-NC in the general, pure LoS, and pure NLoS regimes at large $T_1$, $T_2$, and $T$.
\begin{Thm}[Optimal Average Channel Power of D-IRS-NC]\label{thm:doubleAsyTnon}
	(i)
	$
	\gamma_{DNC}^{(0)}       \stackrel{T_1,T_2\rightarrow\infty}{\sim} \sum_{l\in\mathcal{L}}(L_{\widetilde{Sl},\overline{lU}} + L_{\overline{Sl},\widetilde{lU}} +L_{\widetilde{Sl},\widetilde{lU}})T_ST_l$,
	$
	\gamma_{DNC}^{(1)\star}  \stackrel{T_1,T_2\rightarrow\infty}{\sim}  L_{\overline{S1},\overline{1U}} T_ST_1^2$,
	$\gamma_{DNC}^{(2)\star}  \stackrel{T_1,T_2\rightarrow\infty}{\sim} L_{\widetilde{S2},\overline{2U}}T_ST_2^2$, and
	$\gamma_{DNC}^{(3)\star}  \stackrel{T_1,T_2\rightarrow\infty}{\sim}
	T_S(\sum_{l\in\mathcal{L}}\sqrt{L_{\overline{Sl},\overline{lU}}}T_l)^2$;
	$\tilde{\gamma}_{DNC}^{(0)}\stackrel{T_1,T_2\rightarrow\infty}{\sim}  \sum_{l\in\mathcal{L}} \alpha_{Sl}\alpha_{lU}T_ST_l$, and 
	$\bar{\gamma}_{DNC}^{(3)}(\bm \phi_1,\bm \phi_2) \stackrel{T_1,T_2\rightarrow\infty}{\sim} T_S(\sum_{l\in\mathcal{L}}$ $\sqrt{\alpha_{Sl}\alpha_{lU}}T_l)^2.$
	(ii) If $T_1 = cT$ and $T_2 = (1-c)T$ for some $c\in(0,1)$, then
	$
	\gamma_{DNC}^{(0)}       \stackrel{T\rightarrow\infty}{\sim} c(L_{\widetilde{S1},\overline{1U}} + L_{\overline{S1},\widetilde{1U}} +L_{\widetilde{S1},\widetilde{1U}})T_ST + (1-c)(L_{\widetilde{S2},\overline{2U}} + L_{\overline{S2},\widetilde{2U}} +L_{\widetilde{S2},\widetilde{2U}})T_ST$,
	$
	\gamma_{DNC}^{(1)\star}  \stackrel{T\rightarrow\infty}{\sim} cL_{\overline{S1},\overline{1U}} T_ST^2$,
	$
	\gamma_{DNC}^{(2)\star}  \stackrel{T\rightarrow\infty}{\sim} (1-c)L_{\overline{S2},\overline{2U}}T_ST^2$, and
	$
	\gamma_{DNC}^{(3)\star}  \stackrel{T\rightarrow\infty}{\sim} (c\sqrt{L_{\overline{S1},\overline{1U}}}+(1-c)\sqrt{L_{\overline{S2},\overline{2U}}})^2T_ST^2$;
	$\tilde{\gamma}_{DNC}^{(0)}\stackrel{T\rightarrow\infty}{\sim}  (c\alpha_{S1}\alpha_{1U} + (1-c)\alpha_{S2}\alpha_{2U})T_ST$,
	and
	$\bar{\gamma}_{DNC}^{(3)}(\bm \phi_1,\bm \phi_2) \stackrel{T\rightarrow\infty}{\sim} (c\sqrt{\alpha_{S1}\alpha_{1U}} + (1-c)\sqrt{{\alpha_{S2}\alpha_{2U}}})^2T_ST^2.$
\end{Thm}
\begin{IEEEproof}
	Following the proof of Theorem~\ref{thm:doubleAsyT}, we can show Theorem~\ref{thm:doubleAsyTnon}.
\end{IEEEproof}
\mcthr{By Theorem~\ref{thm:doubleAsyTnon} and~\cite[Proposition 2]{Yuan2020MultiIRS}, for D-IRS-NC, the optimal quasi-static phase shift design achieves the same average power gain in order  w.r.t. $T$ (i.e., $\Theta(T^2)$)} as the optimal instantaneous CSI-adaptive phase shift design in~\cite{Yuan2020MultiIRS}.} \mcthr{Besides, for D-IRS-NC in the general regime, the optimal average channel powers in Cases 1, 2, and 3 (i.e., $\Theta(T^2)$) are equivalent in order and are higher in order than the average channel power in Case 0 (i.e., $\Theta(T)$).}

\begin{Rem}[\mcthr{Quasi-static Phase Shift Design for D-IRS-NC}]
	\mcthr{The results in Section~\ref{sec:Comparision}.A  extend the analysis and optimization results on quasi-static phase shift design for a  multi-IRS non-cooperatively assisted system in~\cite{Cui2019Outage},  where the BS is equipped with a single antenna. Moreover, note that the optimal average channel power of D-IRS-NC with a quasi-static phase shift design and a large number of reflecting elements has not been characterized in the existing literature.}
\end{Rem}
\subsection{\mcthr{Single-IRS-Assisted System (S-IRS)}}
In this part, \mcthr{ we analyze and optimize the average channel power of the quasi-static phase shift design of S-IRS} where the BS serves the user with the help of one IRS indexed by 0 and equipped with a URA of $M_0 \times N_0$ elements, \mcthr{as shown in Fig.~\ref{fig:doublenonandsingle} (b)}. Define $\mathcal{M}_0 \triangleq \{1,..,M_0\}$, $\mathcal{N}_0 \triangleq\{1,..,N_0\}$, and $\mathcal{T}_0 \triangleq\{1,..,M_0N_0\}$. Let $\bm\phi_{0} \triangleq \left( \phi_{0,t} \right)_{t \in \mathcal{T}_0} \in \mathbb{R}^{T_0 \times 1}$ denote the constant phase shifts of IRS 0, where its $t$-th element $\phi_{0,t}$ satisfies:
	\begin{align}\label{eq:singleConstriant}
		\phi_{0,t} \in [0,2\pi), t\in \mathcal{T}_0.
	\end{align}
	Accordingly, denote $\mathbf{v}_0 \triangleq \left(e^{-j\phi_{0,t}}\right)_{t \in \mathcal{T}_0} \in \mathbb{C}^{T}$. For a fair comparison, we let $M_0 N_0 = T$ and \mcthr{place IRS 0 at IRS 1's location in D-IRS-C}. The other setups \mcthr{of S-IRS} remain the same as  \mcthr{those of D-IRS-C}. Let $\mathbf{H}_{S0}\in \mathbb{C}^{T \times T_S}$ and $\mathbf{h}_{0U}^H\in \mathbb{C}^{1 \times T_S}$ represent the Rician channel between the BS and IRS 0 and the Rician channel between IRS 0 and the user, respectively. Specifically,  we have:
	\begin{align*}
		\mathbf{H}_{S0}   & = \sqrt{\alpha_{S0}} \bigg( \sqrt{\frac{K_{S0}}{K_{S0}+1}} \bar{\mathbf{H}}_{S0} + \sqrt{\frac{1}{K_{S0}+1}} \tilde{\mathbf{H}}_{S0} \bigg),
		\\
		\mathbf{h}_{0U}^H & = \sqrt{\alpha_{0U}} \bigg( \sqrt{\frac{K_{0U}}{K_{0U}+1}} \bar{\mathbf{h}}_{0U}^H + \sqrt{\frac{1}{K_{0U}+1}} \tilde{\mathbf{h}}_{0U}^H \bigg),
	\end{align*}
	where $\alpha_{S0}$,  $\alpha_{0U}>0$ represent the large-scale fading powers; $K_{S0}$, $K_{0U}\geq 0$ represent the Rician factors; $\tilde{\mathbf{H}}_{S0} \in \mathbb{C}^{T \times T_S}$ and $\tilde{\mathbf{h}}_{0U}^H\in \mathbb{C}^{1 \times T_S}$ represent the random normalized NLoS components in a slot with elements  i.i.d. according to $\mathcal{CN}(0,1)$;  $\bar{\mathbf{H}}_{S0} =\mathbf{a}_{A,S0}\mathbf{a}_{D,S0}^H\in \mathbb{C}^{T \times T_S}$ and $\tilde{\mathbf{h}}_{0U}^H=\mathbf{a}^H_{D,0U}\in \mathbb{C}^{1\times T_S}$ represent the deterministic normalized LoS components with unit-modulus elements. Then, \mcthr{for S-IRS}, the equivalent channel between the BS and user, \mcthr{denoted by $\mathbf{h}_{SGL,e}^H(\bm \phi_0) \in \mathbb{C}^{1 \times T_S}$, is given by:
	\begin{align*}
		\mathbf{h}_{SGL,e}^H(\bm \phi_0) = \mathbf{h}_{SU}^H + \mathbf{h}_{0U}^H\mathrm{diag}(\mathbf{v}_0^H)\mathbf{H}_{S0}.
	\end{align*}
	Let $\gamma_{SGL}(\bm \phi_0) \triangleq \mathbb{E}\left[ \|\mathbf{h}_{SGL,e}(\bm \phi_0)\|_2^2\right]$} represent the average channel power of the equivalent channel of S-IRS.
	
	For ease of illustration, denote $\mathbf{A}_0  \triangleq  \diag(\bar{\mathbf{h}}_{0U}^H)\bar{\mathbf{H}}_{S0}\bar{\mathbf{H}}_{S0}^H\diag(\bar{\mathbf{h}}_{0U})\in \mathbb{C}^{T \times T}$ and  $\mathbf{b}_0  \triangleq \diag(\bar{\mathbf{h}}_{0U}^H)$ $\bar{\mathbf{H}}_{S0}\bar{\mathbf{h}}_{SU}\in \mathbb{C}^{T}
	$. Now, we characterize the average channel power of \mcthr{S-IRS} in the general, pure LoS, and pure NLoS regimes.
	
	\begin{Lem}[Average Channel Power of \mcthr{S-IRS}]\label{thm:analyAvgPowSgl}
		(i) For any $K_{S0}, K_{0U}\geq 0$, $\gamma_{SGL}(\bm \phi_0)  = L_{\overline{S0},\overline{0U}} \mathbf{v}_0^H \mathbf{A}_0 \mathbf{v}_0 + 2\mathfrak{R}\left\{ \sqrt {L_{\overline{SU}} L_{\overline{S0},\overline{0U}}}\mathbf{v}_0^H\mathbf{b}_0 \right\}+\alpha_{SU}T_S +(L_{\widetilde{S0},\overline{0U}} + L_{\overline{S0},\widetilde{0U}} +L_{\widetilde{S0},\widetilde{0U}})T_ST.$
		(ii) As $K_{S0}, K_{0U}\rightarrow \infty$,  $\gamma_{SGL}(\bm \phi_0)\rightarrow \bar{\gamma}_{SGL}(\bm \phi_0)\triangleq \alpha_{S0}\alpha_{0U} \mathbf{v}_0^H \mathbf{A}_0 \mathbf{v}_0 +\alpha_{SU}T_S+ 2\mathfrak{R}\left\{ \sqrt {\alpha_{SU}\alpha_{S0}\alpha_{0U}}\mathbf{v}_0^H\mathbf{b}_0 \right\}$.
		(iii) If $K_{S0}=K_{0U}=0$,  then $\gamma_{SGL}(\bm \phi_0) = \tilde{\gamma}_{SGL}$, where $\tilde{\gamma}_{SGL}\triangleq \alpha_{SU}T_S + \alpha_{S0}\alpha_{0U}T_ST$.
	\end{Lem}
	\begin{IEEEproof}
		Following the proof of Theorem \ref{thm:analyGeneralK}, we can show Lemma~\ref{thm:analyAvgPowSgl}.
	\end{IEEEproof}
	
	\mcthr{By Corollary~\ref{col:analySmallKnon} and Lemma~\ref{thm:analyAvgPowSgl}, if $\alpha_{Sl}\alpha_{lU}=\alpha_{S0}\alpha_{0U},l=1,2$, then $\tilde{\gamma}_{SGL}=\tilde{\gamma}_{DNC}^{(0)}$.} Noting that $\tilde{\gamma}_{SGL}$  does not change with $\bm \phi_0$, we optimize the phase shifts only in the general  and pure LoS regimes. Specifically, in the general regime, the maximization of $\gamma_{SGL}(\bm \phi_0)$ w.r.t. $\bm \phi_0$ is formulated as:
	\begin{equation} \label{prob:sigleGeneral}
		\begin{aligned}
			\gamma_{SGL}^{\star}  \triangleq \max_{\bm \phi_0} & \quad  	\gamma_{SGL}(\bm \phi_0)
			\\ s.t.												&	\quad 	\eqref{eq:singleConstriant}.
		\end{aligned}
	\end{equation}
	Let $\bm \phi_0^{\star}$ denote an optimal solution of the problem in \eqref{prob:sigleGeneral}. In the pure LoS regime, the maximization of $\bar{\gamma}_{SGL}(\bm \phi_0)$ w.r.t. $\bm \phi_0$ is formulated as:
	\begin{equation} \label{prob:singleLoS}
		\begin{aligned}
			\bar{\gamma}_{SGL}^{\star}  \triangleq \max_{\bm \phi_0} & \quad  	\bar{\gamma}_{SGL}(\bm \phi_0)
			\\ s.t.												&	\quad 	\eqref{eq:singleConstriant}.
		\end{aligned}
	\end{equation}
	Let $\bar{\bm \phi}_0^{\star}$ denote an optimal solution of the problem in \eqref{prob:singleLoS}.
	
	Define $\bm \Delta_{\overline{S0},\overline{0U}} \triangleq \angle\left( \diag(\mathbf{a}_{D,0U}^H)\mathbf{a}_{A,S0}\right) \in \mathbb{R}^{T}$ and $
	r_{\overline{S0},\overline{SU}} \triangleq \mathbf{a}_{D,S0}^H\mathbf{a}_{D,SU} \in \mathbb{C}$. The optimal solutions of the problem in \eqref{prob:sigleGeneral} and problem in \eqref{prob:singleLoS} are given below.
	
	\begin{Lem}[Optimal Solutions of Problem in \eqref{prob:sigleGeneral} and Problem in \eqref{prob:singleLoS}] \label{lem:OptSluSingle} The unique optimal solutions of the problem in \eqref{prob:sigleGeneral} and problem in \eqref{prob:singleLoS} are given by:
		\begin{align}\label{eq:phi0star}
			\bm \phi_{0}^{\star} 	=\bar{\bm \phi}_{0}^{\star}= \Lambda \left(- \bm \Delta_{\overline{S0},\overline{0U}}-\angle\left( r_{\overline{S0},\overline{SU}}\right)\mathbf{1}_T\right).
		\end{align}
	\end{Lem}
	\begin{IEEEproof}
		Following the proof of Theorem~\ref{thm:caselGeneralSpecialCase}, we can show Lemma~\ref{lem:OptSluSingle}.
	\end{IEEEproof}
	
	Lemma \ref{lem:OptSluSingle} indicates that \mcthr{for S-IRS} in the general and pure LoS regimes, the optimal phase changes of over the cascaded LoS channel $(\overline{S0},\overline{0U})$ are identical to the phase changes of  over the  LoS channel $\overline{SU}$. Furthermore, the computational complexities for calculating $\bm \phi_0^{\star}$  and $\bar{\bm \phi}_0^{\star}$ based on Lemma~\ref{lem:OptSluSingle} are $\mathcal{O}(T)$.
	
	Finally, \mcthr{for S-IRS}, we characterize the optimal average channel power  in the general and pure LoS regimes and the average channel power in the pure NLoS regime at  large $T$.
	\begin{Lem}[Optimal Average Channel Power of \mcthr{S-IRS}]\label{thm:singleAsyT}
		$\gamma_{SGL}^{\star} \stackrel{T\rightarrow\infty}{\sim} L_{\overline{S0},\overline{0U}}T_ST^2$,
		$\bar{\gamma}_{SGL}^{\star} \stackrel{T\rightarrow\infty}{\sim} \alpha_{S0}\alpha_{0U}T_ST^2$, and
		$\tilde{\gamma}_{SGL} \stackrel{T\rightarrow\infty}{\sim} \alpha_{S0}\alpha_{0U}T_ST$.
	\end{Lem}
	\begin{IEEEproof}
		Following the proof of Theorem~\ref{thm:doubleAsyT}, we can show Lemma~\ref{thm:singleAsyT}.
	\end{IEEEproof}
	 \mcthr{By Lemma~\ref{thm:singleAsyT} and~\cite[Proposition 2]{ZhangRui2019SDR}}, for S-IRS, the optimal quasi-static phase shift design achieves the same average power gain in order w.r.t. $T$ (i.e., $\Theta(T^2)$) as the optimal instantaneous CSI-adaptive phase shift design in~\cite{ZhangRui2019SDR}. \mcthr{Besides, for S-IRS, the optimal average channel powers in the general and pure LoS regimes (i.e., $\Theta(T^2)$) are equivalent in order and are higher in order than the average channel power in the pure NLoS regime (i.e., $\Theta(T)$)}.

	\begin{Rem}[Quasi-static Phase Shift Design for \mcthr{S-IRS}]
		\mcthr{The results in Section~\ref{sec:Comparision}.B} extend the analysis and optimization results on quasi-static phase shift design for a less general S-IRS in~\cite{JinShi2019MISO}  where the direct channel is modeled as Rayleigh fading. Moreover, note that the optimal average channel power of S-IRS with a quasi-static phase shift design and a large number of reflecting elements has not been characterized in the existing literature.
	\end{Rem}

	\subsection{Comparison}
	\mcthr{In this part, we compare the optimal quasi-static phase shift design of D-IRS-C  with  those of  D-IRS-NC and S-IRS in computational complexity and average channel power. Specifically, based on the optimization and analytical results in Section~\ref{sec:Optimization}, Section~\ref{sec:Comparision}.A, and Section~\ref{sec:Comparision}.B, \mcthr{we summarize the computational complexities and average channel powers of the optimal quasi-static phase shift designs of D-IRS-C, D-IRS-NC, and S-IRS in the general, pure LoS, and pure NLoS regimes  in Table~\ref{tab:comparison}. \mcthr{Here, we are interested in the growth rates of the computational complexity and average channel power \mcthr{w.r.t. the total number of reflecting elements.}}}}\footnote{The computational complexity results are derived from those in Section~\ref{sec:Optimization}, \mcthr{Section~\ref{sec:Comparision}.A, and Section~\ref{sec:Comparision}.B} by letting $T_1=cT$ and $T_2 = (1-c)T$ for some $c\in(0,1)$ and $T\rightarrow\infty$. In some cases, there is no need to optimize the phase shifts, and hence we use $\mathcal{O}(1)$ for the computational complexity and consider the average channel power.}
	\begin{itemize}
		\item \textbf{ Computational Complexity:} In each case where the phase shift optimization is necessary, the closed-form optimal quasi-static phase shift design of \mcthr{D-IRS-C} (if it exists) has the same computational complexity \mcthr{in order} as the closed-form optimal \mcthr{quasi-static phase shift designs of D-IRS-NC and S-IRS}, whereas the numerical quasi-static phase shift design of  \mcthr{D-IRS-C} has higher computational complexity \mcthr{in order} than the closed-form optimal \mcthr{quasi-static phase shift designs of D-IRS-NC and S-IRS}.
		\item \textbf{ Optimal Average Channel Power:} \mcthr{ (i) In the general regime, the optimal average channel power of D-IRS-NC in Cases 1, 2, and 3 and that of S-IRS are identical in order.  Besides, the optimal average channel powers of D-IRS-NC and S-IRS are identical in order in the pure LoS and pure NLoS regimes, respectively. This is because the underlying difference between D-IRS-NC and S-IRS lies in path loss which does not influence the order of growth of the optimal average channel power w.r.t. $T$. (ii) In the general regime, the optimal average channel power of D-IRS-C in each case is higher in order than the optimal average channel power of D-IRS-NC in the same case and the optimal average channel power of S-IRS. \mcthr{In the pure LoS and pure NLoS regimes, respectively}, the optimal average channel power of D-IRS-C is higher in order than the optimal average channel powers of D-IRS-NC and S-IRS. This is because the additional inter-IRS channel in D-IRS-C can convey signals from the BS to the user.}
		\item \textbf{ Tradeoff:} \mcthr{D-IRS-C} achieves a better performance and computational complexity tradeoff than \mcthr{D-IRS-NC and S-IRS} in \mcthr{the cases with closed-form optimal quasi-static phase shift designs and the cases that do not require optimizing phase shifts} and \mcthr{achieves} a different performance and computational complexity tradeoff in the other cases. For quasi-static phase shift designs, as phase shifts remain constant during a certain period, the computational complexity for optimizing phase shifts is negligible. Therefore, \mcthr{D-IRS-C} is more desirable than \mcthr{D-IRS-NC and S-IRS} in practice.
	\end{itemize}

	\begin{table}[t]
		\caption{\mcthr{Comparisons.}}
		\label{tab:comparison}
		\centering
		\vspace{-3mm}
		\small
	\begin{tabular}{|l|ll|l|l|}
		\hline
		Regime                     & \multicolumn{2}{l|}{System}                             & Computational Complexity                                                                                                  & \begin{tabular}[c]{@{}l@{}}Optimal Average\\ Channel Power\end{tabular} \\ \hline
		\multirow{9}{*}{General}   & \multicolumn{1}{l|}{\multirow{4}{*}{D-IRS-C}}  & Case 0 & $\mathcal{O}(1)$                                                                                                          & $\Theta(T^2)$                                                           \\ \cline{3-5} 
		& \multicolumn{1}{l|}{}                          & Case 1 & \multirow{2}{*}{\begin{tabular}[c]{@{}l@{}}$\mathcal{O}(T^2)$ (numerical), \\ $\mathcal{O}(T)$ (analytical)\end{tabular}} & \multirow{2}{*}{$\Theta(T^3)$}                                          \\ \cline{3-3}
		& \multicolumn{1}{l|}{}                          & Case 2 &                                                                                                                           &                                                                         \\ \cline{3-5} 
		& \multicolumn{1}{l|}{}                          & Case 3 & \begin{tabular}[c]{@{}l@{}}$\mathcal{O}(T^3)$ (numerical),\\ $\mathcal{O}(T)$ (analytical)\end{tabular}                   & $\Theta(T^4)$                                                           \\ \cline{2-5} 
		& \multicolumn{1}{l|}{\multirow{4}{*}{\mcthr{D-IRS-NC}}} & Case 0 & $\mathcal{O}(1)$                                                                                                          & $\Theta(T)$                                                             \\ \cline{3-5} 
		& \multicolumn{1}{l|}{}                          & Case 1 & \multirow{3}{*}{$\mathcal{O}(T)$ (analytical)}                                                                            & \multirow{3}{*}{$\Theta(T^2)$}                                          \\ \cline{3-3}
		& \multicolumn{1}{l|}{}                          & Case 2 &                                                                                                                           &                                                                         \\ \cline{3-3}
		& \multicolumn{1}{l|}{}                          & Case 3 &                                                                                                                           &                                                                         \\ \cline{2-5} 
		& \multicolumn{2}{l|}{S-IRS}                              & $\mathcal{O}(T)$ (analytical)                                                                                             & $\Theta(T^2)$                                                           \\ \hline
		\multirow{3}{*}{Pure LoS}  & \multicolumn{2}{l|}{D-IRS-C (Case 3)}                            & $\mathcal{O}(T^3)$ (numerical)                                                                                            & $\Theta(T^4)$                                                           \\ \cline{2-5} 
		& \multicolumn{2}{l|}{\mcthr{D-IRS-NC (Case 3)}}                           & $\mathcal{O}(T)$ (analytical)                                                                                             & $\Theta(T^2)$                                                           \\ \cline{2-5} 
		& \multicolumn{2}{l|}{S-IRS}                              & $\mathcal{O}(T)$ (analytical)                                                                                             & $\Theta(T^2)$                                                           \\ \hline
		\multirow{3}{*}{Pure NLoS} & \multicolumn{2}{l|}{D-IRS-C (Case 3)}                            & $\mathcal{O}(1)$                                                                                                          & $\Theta(T^2)$                                                           \\ \cline{2-5} 
		& \multicolumn{2}{l|}{\mcthr{D-IRS-NC (Case 3)}}                           & $\mathcal{O}(1)$                                                                                                          & $\Theta(T)$                                                             \\ \cline{2-5} 
		& \multicolumn{2}{l|}{S-IRS}                              & $\mathcal{O}(1)$                                                                                                          & $\Theta(T)$                                                             \\ \hline
	\end{tabular}
	\end{table}

	\section{Numerical Results} \label{sec:Numerical}
\begin{figure}[t]
		\begin{center}
			\includegraphics[width=8cm]{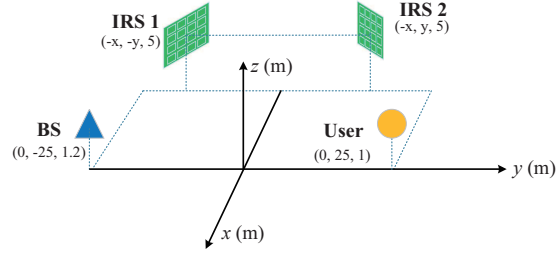}
		\end{center}
		\vspace{-5mm}
		\caption{\small{\mcthr{The system topology in Section~\ref{sec:Numerical}}.}}
		\label{fig:pathloss}
		\vspace{-5mm}
	\end{figure}

	\mcthr{In this section, we numerically evaluate the average rates of the phase shift designs for D-IRS-C, D-IRS-NC, S-IRS, and the counterpart system without any IRS ({\em W/O-IRS}) \cite{Pan2021Direct}. As shown in Fig.~\ref{fig:pathloss}, for all systems, the BS and user are located at $(0,-25,1.2)$ and $(0,25,1)$ (in m), respectively; for D-IRS-C and D-IRS-NC, IRS 1 and IRS 2 are located at $(-x,-y,5)$ and $(-x,y,5)$, respectively; for S-IRS, IRS 0 can be located at the locations of IRS 1, IRS 2, or the midpoint between IRS 1 and IRS 2, and the corresponding systems are termed  {\em S-IRS-Pos-1}, {\em S-IRS-Pos-2}, {\em S-IRS-Pos-Mid}, respectively.} 
	
	In the simulation, we set $d = \frac{\lambda}{2}$, \mcthr{$x=5$, $y = 20$, $M_S=N_S=2$, $M_1=N_1=M_2=N_2=10$}, $P_S = 5 \ \text{dBm}$, $\sigma^2 = -104 \ \text{dBm}$, $\varphi_{S1}^{(h)}=\varphi_{S1}^{(v)} = \pi \slash 6$, $\varphi_{S2}^{(h)}=\varphi_{S2}^{(v)} = \pi \slash 4$, $\varphi_{12}^{(h)}=\varphi_{12}^{(v)} = \pi \slash 5$, $\varphi_{SU}^{(h)}=\varphi_{SU}^{(v)} = \pi \slash 3$, $\varphi_{1U}^{(h)}=\varphi_{1U}^{(v)} = \pi \slash 8$, $\varphi_{2U}^{(h)}=\varphi_{2U}^{(v)} = \pi \slash 9$, $\delta_{S1}^{(h)}=\delta_{S1}^{(v)} = \pi \slash 6$, $\delta_{S2}^{(h)}=\delta_{S2}^{(v)} = \pi \slash 5$, $\delta_{12}^{(h)}=\delta_{12}^{(v)} = \pi \slash 4$,  $K_{S1}=K_{S2}=K_{1U}=K_{2U}=K_{SU}=K_{12}=K=10\ \text{dB}$, if not specified otherwise. We set $\alpha_{ab} = 1/(1000d_{ab}^{\bar{\alpha}_{ab}})$ (i.e., $-30+10\bar{\alpha}_{ab}\log_{10}(d_{ab})$ dB), $ab = S1,S2,12,1U,2U,SU$, where $\bar{\alpha}_{ab}$ represents the corresponding path loss exponent. \mcthr{We choose $\bar{\alpha}_{12}=2.2$, $\bar{\alpha}_{S1}$, $\bar{\alpha}_{2U}$, $\bar{\alpha}_{S2}$, $\bar{\alpha}_{1U}= 2.3$, and  $\bar{\alpha}_{SU}=3.7$~\cite{ZhangRui2019SDR,Cui2020Interference,ZhangRui2021doubleIRSMIMO}.\footnote{\mcthr{IRSs are usually placed far above the ground, and their locations are carefully selected. Thus, the inter-IRS channel experiences the fewest obstacles and weakest scattering, and the channels $S1$, $S2$, $1U$, and $2U$ experience fewer obstacles and weaker scattering. Besides, the BS and user are usually located on the ground. Thus, the direct channel $SU$ experiences the most obstacles and strongest scattering~\cite{ZhangRui2019SDR,Cui2020Interference,ZhangRui2021doubleIRSMIMO}.}}} \mcthr{We evaluate seven schemes, all adopting the instantaneous CSI-adaptive MRT beamformer at the BS. For any system {\em SYS} with IRS, the optimal quasi-static phase shift design in the general, pure LoS, and pure NLoS regimes are referred to {\em SYS-Gen}, {\em SYS-LoS}, and {\em SYS-NLoS}, respectively. Note that -Gen is sometimes omitted for notation simplicity. Besides, for D-IRS-C,  {\em D-IRS-C-Random} chooses the phase shifts uniformly at random~\cite{Cui2020Interference}, and {\em D-IRS-C-ICSI} adopts the optimal instantaneous CSI-adaptive phase shift design as in~\cite{ZhangRui2021doubleIRSMIMO}.}
	\begin{figure*}[t]
		\centering
		\begin{minipage}[t]{5.4cm}
			\includegraphics[width=5.4cm]{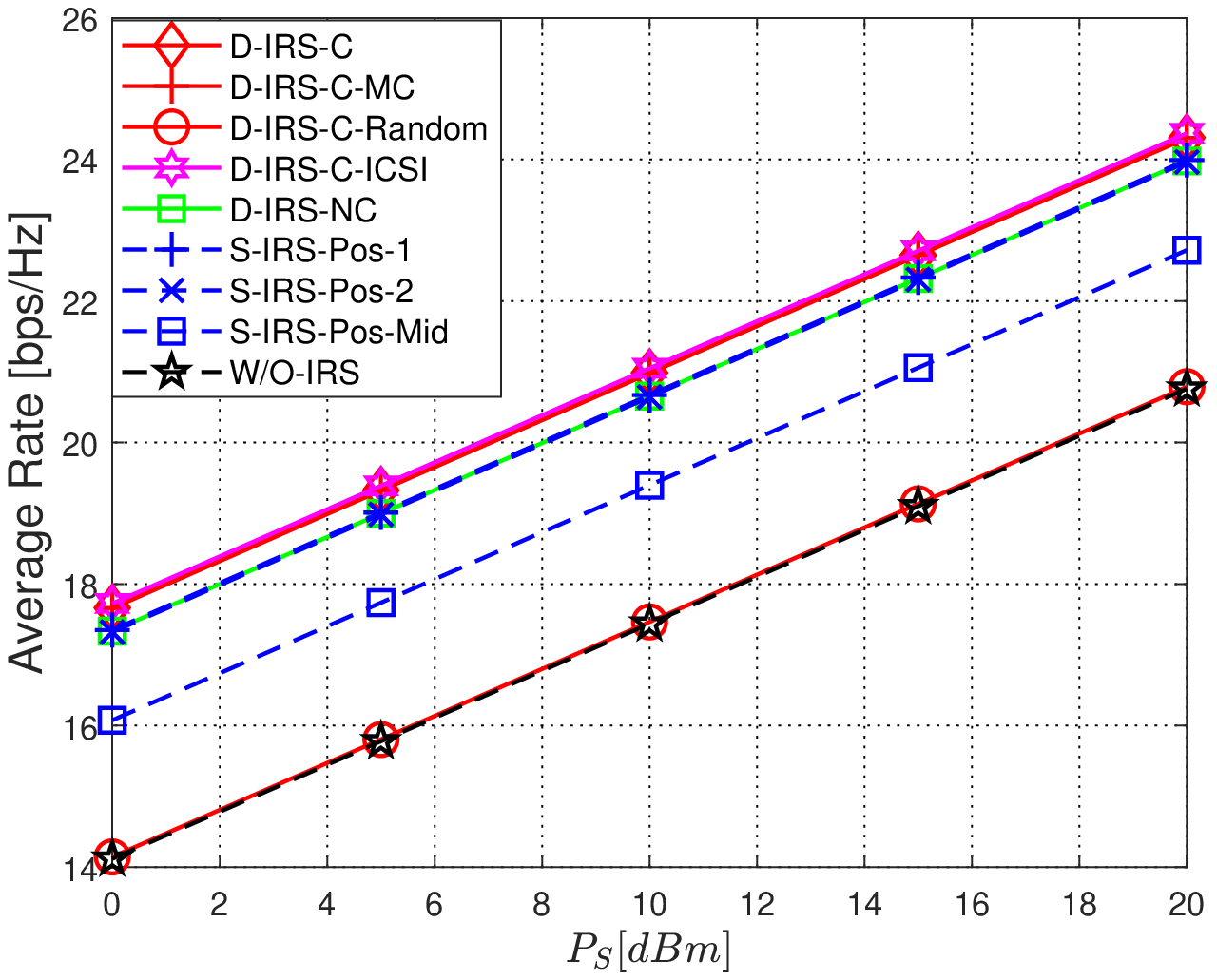}
			\caption{\small{\mcthr{Average rate versus $P_S$}.}}
			\label{fig:averagePS}
		\end{minipage}
		\begin{minipage}[t]{5.4cm}
			\includegraphics[width=5.4cm]{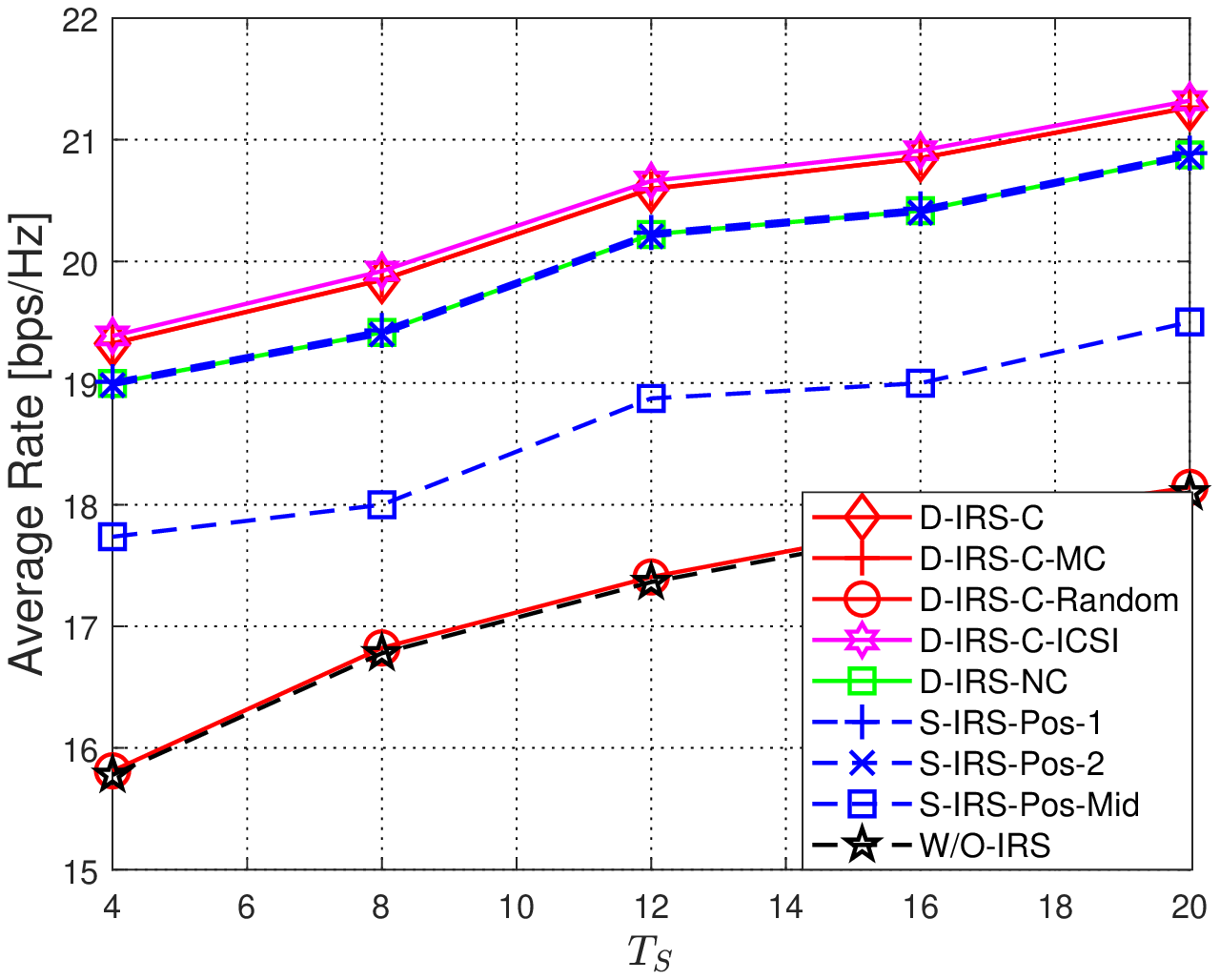}
			\caption{\small{\mcthr{Average rate versus $T_S$.}}}
			\label{fig:averageTS}
		\end{minipage}
		\begin{minipage}[t]{5.4cm}
			\includegraphics[width=5.4cm]{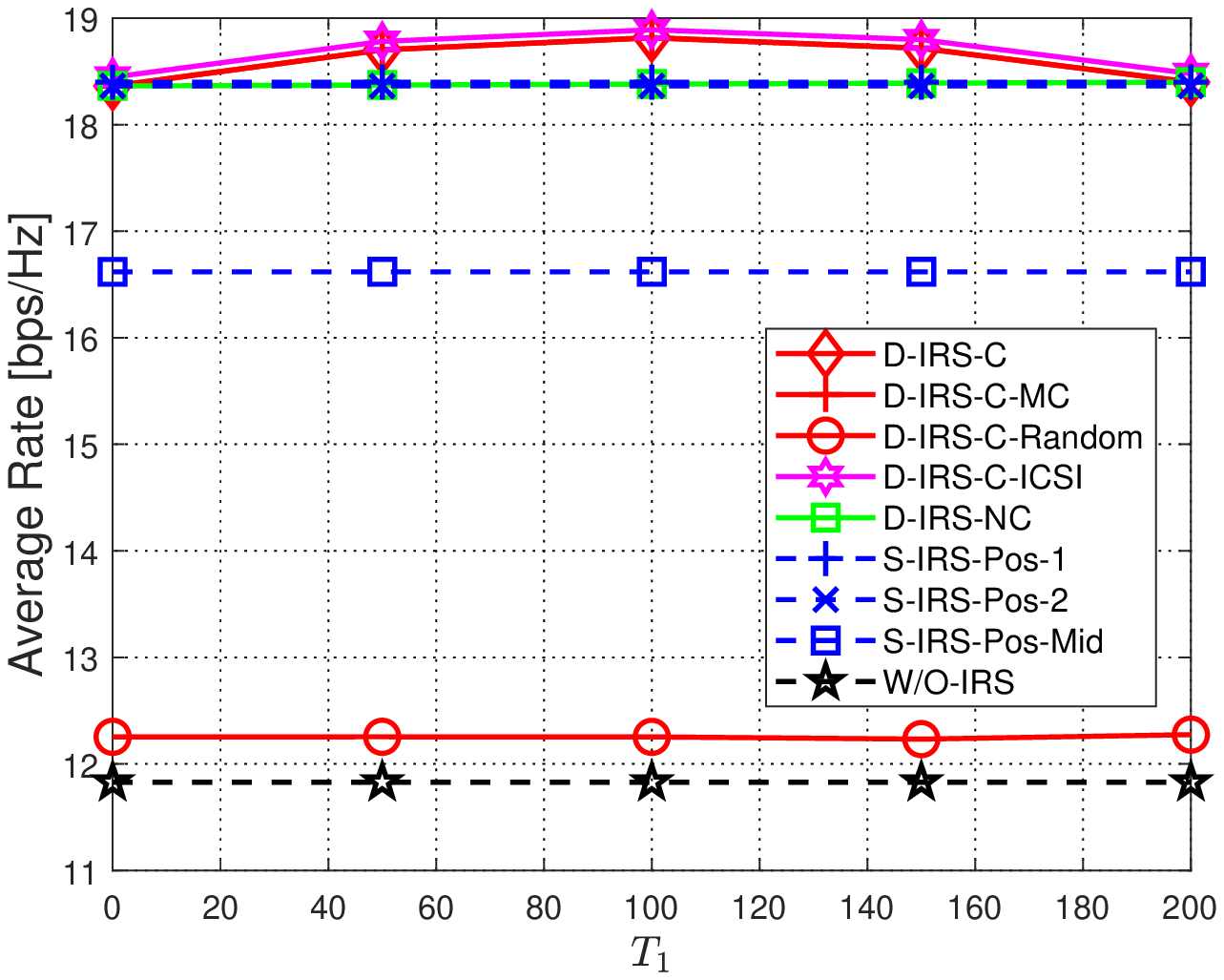}
			\caption{\small{\mcthr{Average rate versus $T_1$ at $T_2 = 200-T_1$.}}}
			\label{fig:averageT1}
		\end{minipage}
%				\vspace{-7mm}
	\end{figure*}
\begin{figure}[t]
	\centering
\begin{minipage}[t]{16cm}
		\includegraphics[width=16cm]{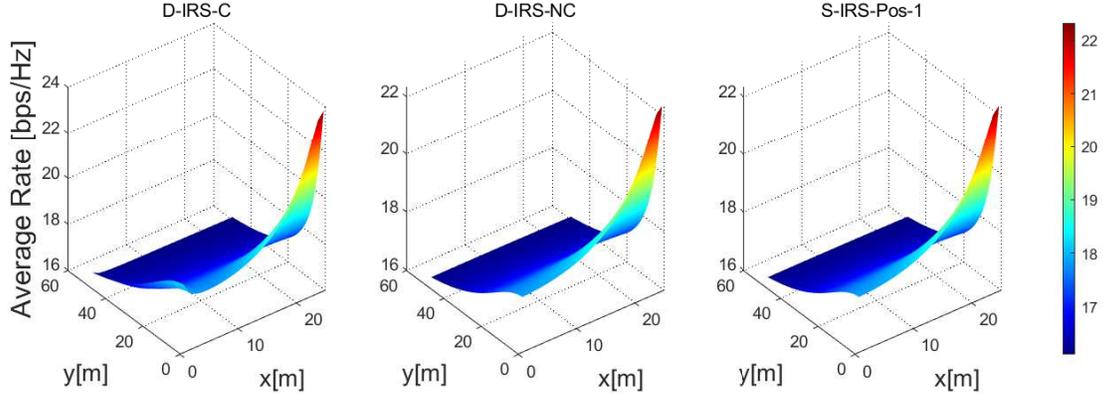}
		\caption{\small{\mcthr{Average rate versus $x$ and $y$}.}}
		\label{fig:averagedxdy}
		\vspace{-5mm}
	\end{minipage}
\end{figure}
	\begin{figure}[h]
		\begin{center}
			\subfigure[\small{\mcthr{General regime.}}]
			{\resizebox{5.4 cm}{!}{\includegraphics{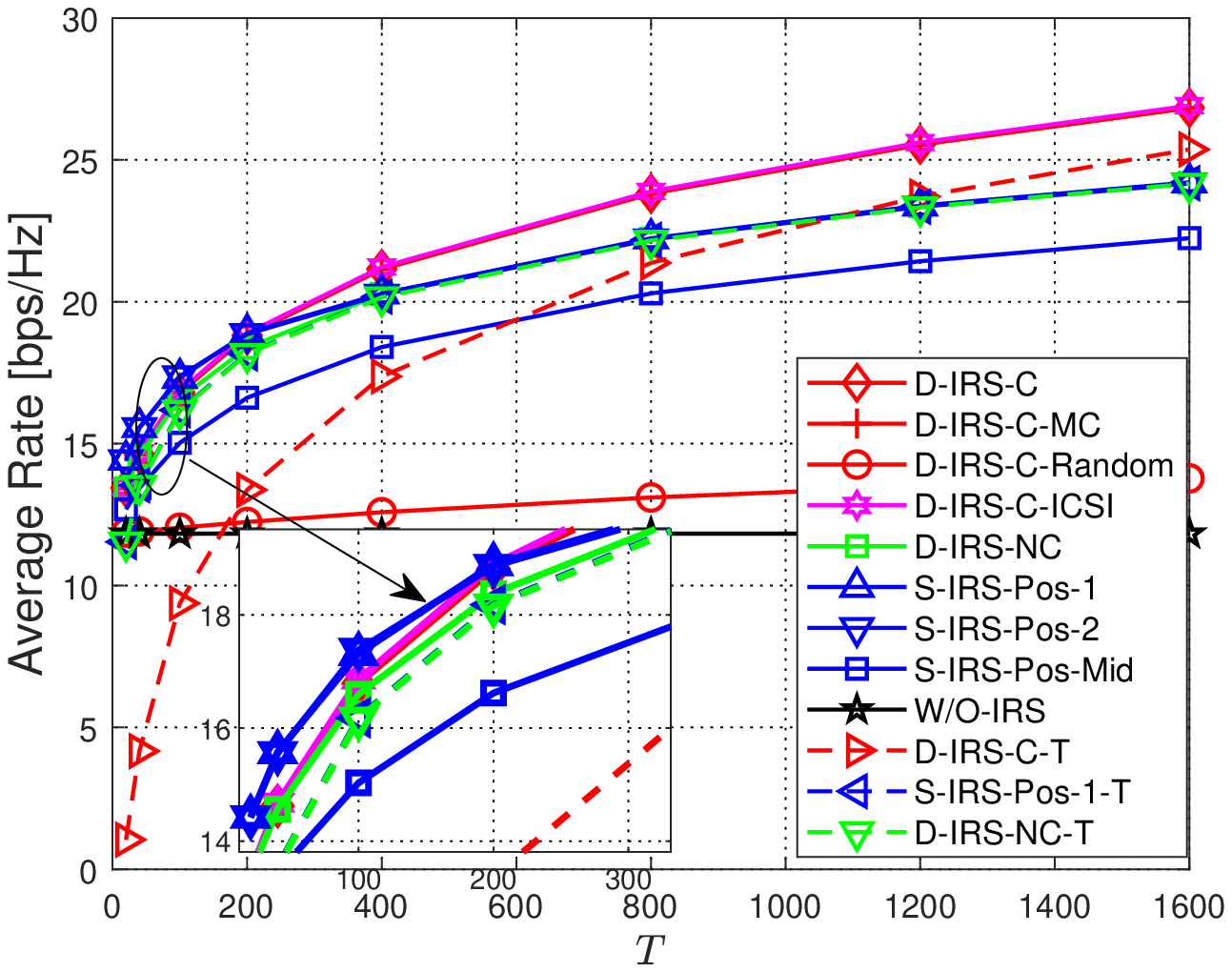}}}
			\subfigure[\small{\mcthr{Pure LoS regime.}}]
			{\resizebox{5.4 cm}{!}{\includegraphics{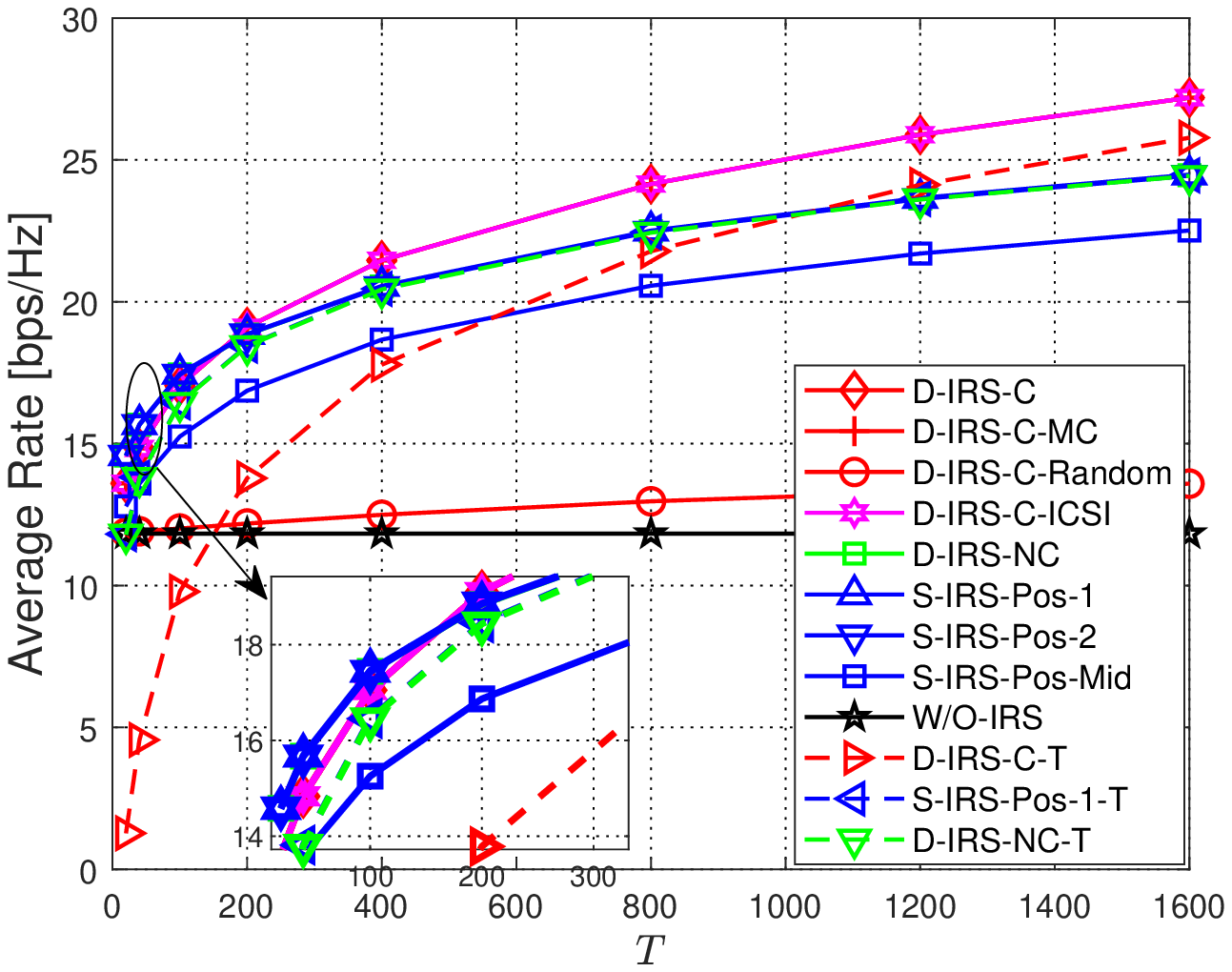}}} 
			\subfigure[\small{\mcthr{Pure NLoS regime.}}]
			{\resizebox{5.4 cm}{!}{\includegraphics{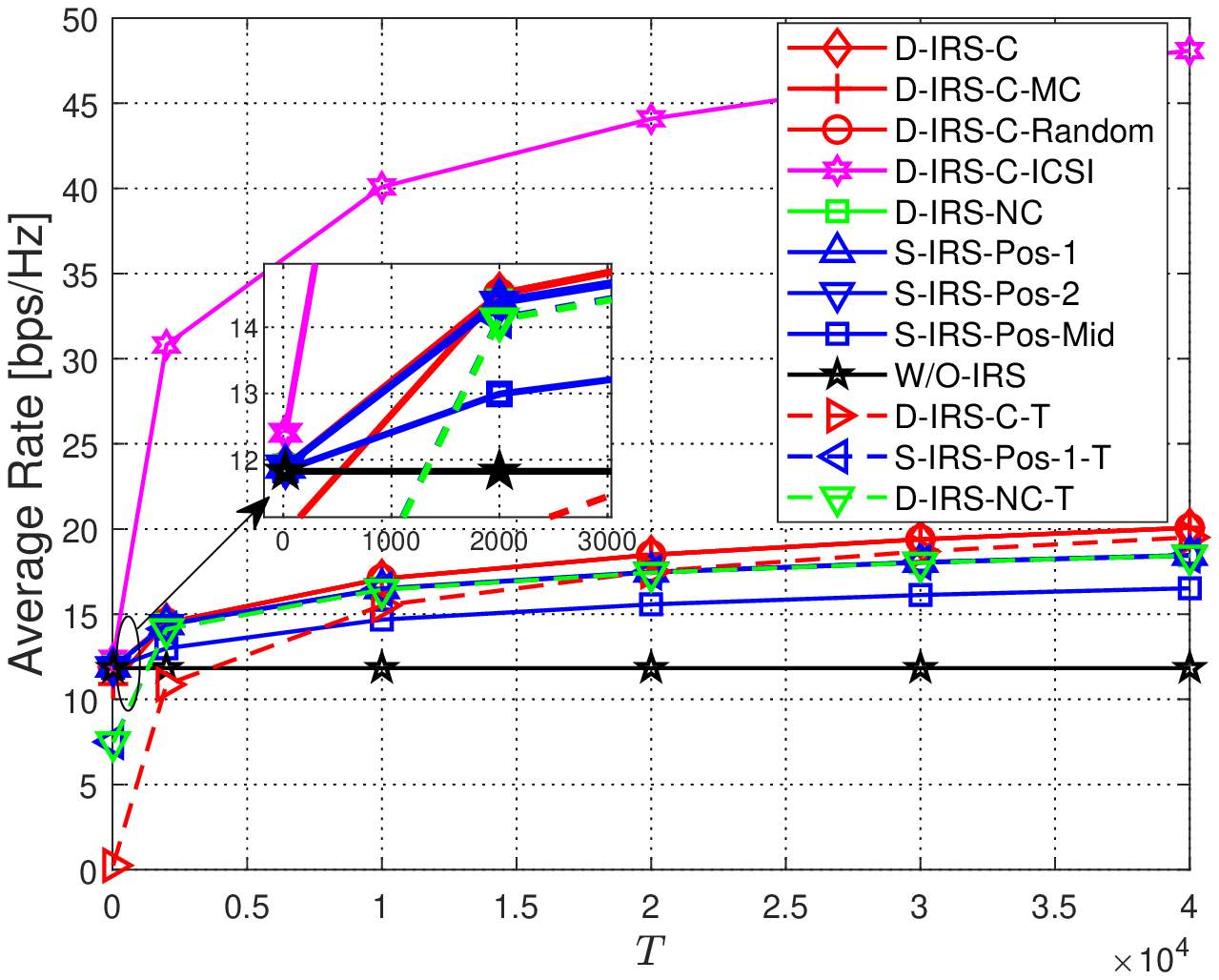}}}
		\end{center}
		\vspace{-7mm}
		\caption{\small{\mcthr{Average rate versus $T$ at $T_1=T_2=T/2$.}}}
		\label{fig:averageT}
		\vspace{-3mm}
		\begin{center}
			\subfigure[\small{\mcthr{Results for general  regime.}}]
			{\resizebox{5.4 cm}{!}{\includegraphics{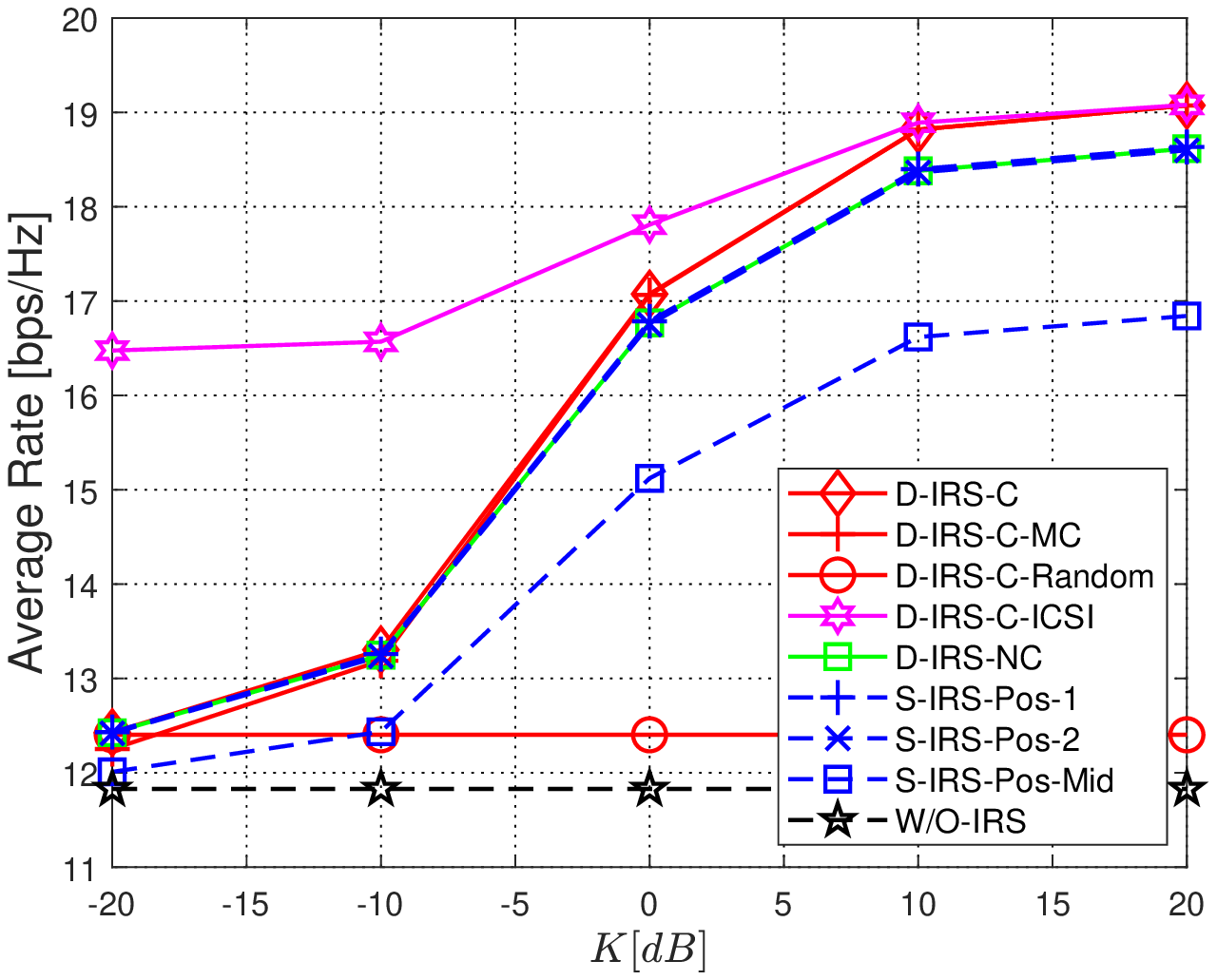}}}\quad
			\subfigure[\small{\mcthr{Results for special  regimes.}}]
			{\resizebox{5.4 cm}{!}{\includegraphics{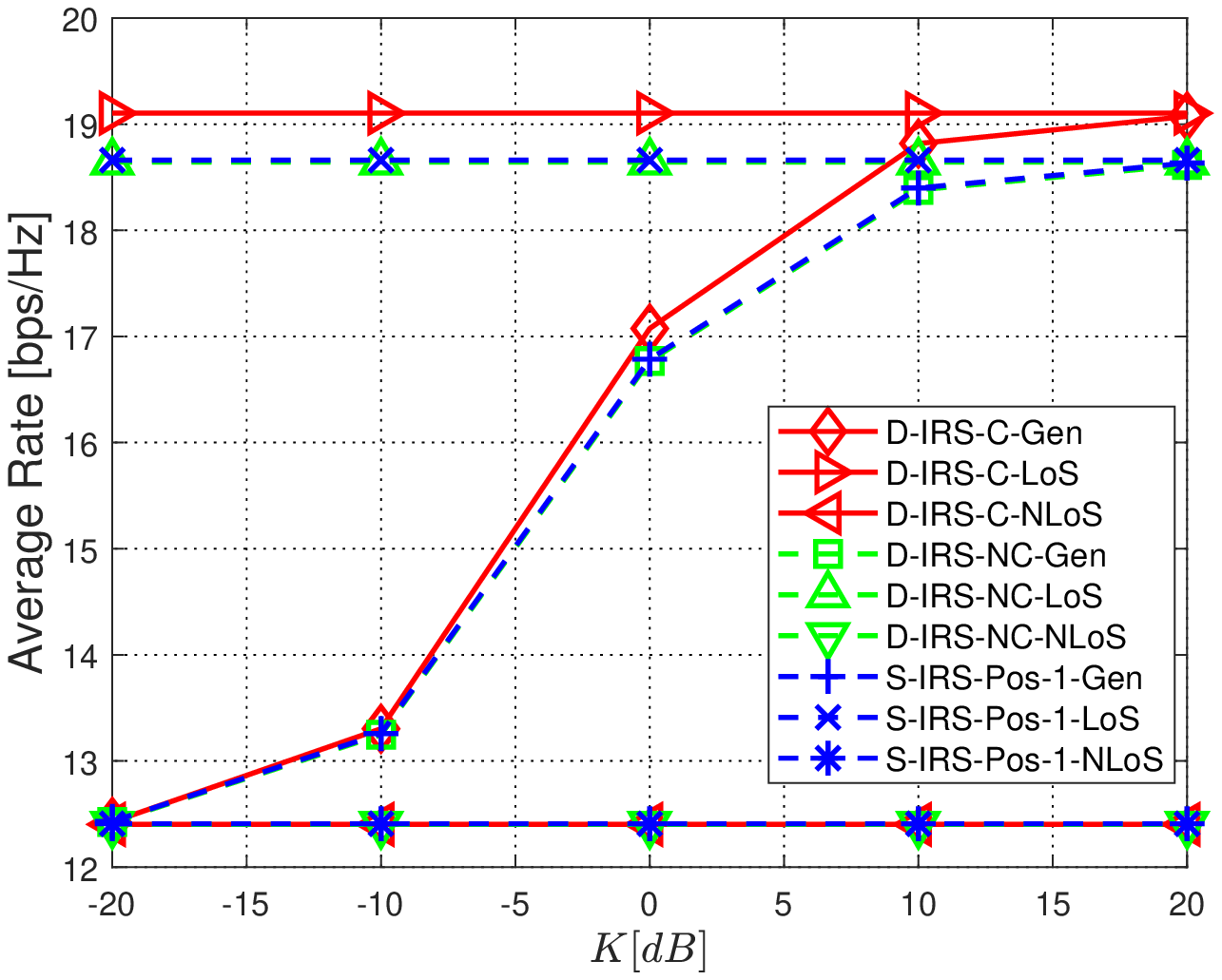}}} \quad
		\end{center}
		\vspace{-7mm}
		\caption{\small{\mcthr{Average rate versus $K$.}}}
		\label{fig:averageK}
	\end{figure}

	Fig.~\ref{fig:averagePS}, Fig.~\ref{fig:averageTS},  Fig.~\ref{fig:averageT1}, \mcthr{Fig.~\ref{fig:averagedxdy}}, Fig.~\ref{fig:averageT},  and Fig.~\ref{fig:averageK} illustrate the average rate versus the transmit power $P_S$,  number of antennas at the BS $T_S$,  number of IRS 1's reflecting elements $T_1$, \mcthr{IRS locations}, total number of reflecting elements in the system $T$, and Rician factor $K$, respectively. In these figures, {\em D-IRS-C-MC}  represents the numerical expectation of the rate of {\em D-IRS-C}, and each other curve represents the analytical upper bound of the average rate of a scheme based on Jensen's inequality. From Fig.~\ref{fig:averagePS}, Fig.~\ref{fig:averageTS}, Fig.~\ref{fig:averageT1}, Fig.~\ref{fig:averageT}, and Fig.~\ref{fig:averageK} (a), we can see that \mcthr{{\em D-IRS-C-MC} and {\em D-IRS-C}} are very close to each other, \mcthr{indicating that the upper bound, $\log_2\left( 1+ \frac{P_S}{\sigma^2} \gamma( \bm \phi_1, \bm \phi_2) \right)$, is a good approximation of $C(\bm \phi_1, \bm \phi_2)$.} In Fig.~\ref{fig:averageT}, \mcthr{{\em D-IRS-C-T}, {\em D-IRS-NC-T}, and {\em S-IRS-Pos-1-T}} represent the  asymptotic results for \mcthr{{\em D-IRS-C}, {\em D-IRS-NC}, and {\em S-IRS-Pos-1}} at large $T$, respectively. For each of them, the gap between the general result at any $T$ and asymptotic result at large $T$ decreases with $T$, which is in accordance with Theorem~\ref{thm:doubleAsyT}, Lemma~\ref{thm:doubleAsyTLargeK}, Lemma~\ref{thm:doubleAsyTSmallK}, \mcthr{Theorem~\ref{thm:doubleAsyTnon}}, and Lemma~\ref{thm:singleAsyT}. \mcthr{In Fig.~\ref{fig:averageK} (b), the gap between {\em D-IRS-C-Gen} and {\em D-IRS-C-LoS} ({\em D-IRS-C-NLoS}) decreases (increases) with $K$, in accordance with Corollary~\ref{col:analyLargeK} (Corollary~\ref{col:analySmallK}); the gap between {\em D-IRS-NC-Gen} and {\em D-IRS-NC-LoS} ({\em D-IRS-NC-NLoS}) decreases (increases) with $K$, in accordance with Corollary~\ref{col:analyLargeKnon} (Corollary~\ref{col:analySmallKnon}); and the gap between {\em S-IRS-Pos-1-Gen} and {\em S-IRS-Pos-1-LoS} ({\em S-IRS-Pos-1-NLoS}) decreases (increases) with $K$, in accordance with Lemma~\ref{thm:analyAvgPowSgl}.}
	
	From Fig.~\ref{fig:averagePS} and Fig.~\ref{fig:averageTS}, we see that the average rate of each scheme increases with $P_S$ and $T_S$, respectively.   
	From Fig.~\ref{fig:averageT1}, we see that the average rate of the proposed solution is maximized when the two IRSs have the same number of elements, mainly due to the symmetric channel setup.
	\mcthr{From Fig.~\ref{fig:averagedxdy}, we see that the average rates of {\em D-IRS-C}, {\em D-IRS-NC}, and {\em S-IRS-Pos-1} are maximized when IRS 1 and IRS 2 are placed closest to the BS and user, respectively.} 
	From Fig. \ref{fig:averageT}, we observe that the average rate of each scheme for IRS-assisted systems increases with $T$, mainly  due to the increment of reflecting signal power.
	Fig. \ref{fig:averageK} shows that the average rate of each quasi-static phase shift design increases with $K$, mainly  due to the increment of the channel power of each LoS component.
	
	From Fig.~\ref{fig:averagePS}, Fig.~\ref{fig:averageTS}, Fig.~\ref{fig:averageT1}, Fig.~\ref{fig:averageT},  and Fig.~\ref{fig:averageK}, we can see that \mcthr{{\em D-IRS-C} outperforms {\em D-IRS-C-Random}, {\em D-IRS-Pos-Mid}, and {\em D-IRS-NC} at any total number of reflecting elements and  outperforms {\em D-IRS-NC}, {\em S-IRS-Pos-1}, and {\em S-IRS-Pos-2} as long as the total number of reflecting elements is large enough.} Specifically, the gain of \mcthr{{\em D-IRS-C}} over {\em D-IRS-C-Random} comes from the optimization of the quasi-static phase shifts, and the  gain of \mcthr{{\em D-IRS-C}} over {\em D-IRS-NC} derives from the effective utilization of the inter-IRS channel \mcthr{ that can convey signals from the BS to the user.} \mcthr{The performance gains of { \em D-IRS-C} over {\em D-IRS-NC}, {\em S-IRS-Pos-1}, and {\em S-IRS-Pos-2} depend on the average channel power of the cascaded channel $(S1,12,2U)$, which increases with $T_1$, $T_2$ (as shown in Fig.~\ref{fig:averageT}) and $K_{S1}$, $K_{12}$, $K_{2U}$ (as shown in Fig.~\ref{fig:averageK}).} Furthermore, \mcthr{{\em D-IRS-C-ICSI}} outperforms  \mcthr{{\em D-IRS-C}} at the sacrifice of \mcthr{increased channel estimation and phase adjustment costs and overall computational complexity. Thus, quasi-static phase shift design has practical sense as long as the LoS components are sufficiently large.}

	\section{Conclusion} \label{sec:Conclusion}
	This paper investigated the analysis and optimization of the quasi-static phase shift design for \mcthr{D-IRS-C}. Furthermore, this paper compared the \mcthr{optimal quasi-static phase shift design of D-IRS-C with those  of}  \mcthr{D-IRS-NC} and \mcthr{S-IRS}. Both analytical and numerical results demonstrate notable gains of the proposed solutions over the existing solutions and reveal insights into designing practical IRS-assisted systems. \mcthr{This work can be extended to IRS-cooperatively-assisted systems with multiple IRSs and users.}

	%	\appendices
\section*{Appendix A: Proof of Lemma~\ref{lem:Influence}} \label{proof:Influence}
First, we prove Statement (i). As $\gamma\left( \bm \phi_1,\bm \phi_2\right)  = \mathbb{E} \left[\| \mathbf{h}_e(\bm\phi_1,\bm\phi_2) \|_2^2 \right]$, with $\mathbf{h}_e(\bm\phi_1,\bm\phi_2)$  given by \eqref{eq:he}, it is sufficient to show $\mathbf{h}^H_e(\bm\phi_1,\bm\phi_2)\overset{d}{\sim} \mathbf{h}_{SU}^H+ \mathbf{h}_{2U}^H \diag(\mathbf{v}_2^H) \mathbf{H}_{S2} + (\mathbf{h}_{1U}^H + \mathbf{h}_{2U}^H \diag(\mathbf{v}_2^H) \mathbf{H}_{12})\mathbf{H}_{S1}$, which is irrelevant to $\mathbf{v}_1$. By \eqref{eq:he}, it is equivalent to show $\mathbf{H}_{S1} \overset{d}{\sim}\diag(\mathbf{v}_1^H)\mathbf{H}_{S1}$ or $\mathbf{H}_{12}\overset{d}{\sim}\mathbf{H}_{12}\diag(\mathbf{v}_1^H)$ and $\mathbf{h}_{1U}^H\overset{d}{\sim}\mathbf{h}_{1U}^H\diag(\mathbf{v}_1^H)$. First, we show that $\mathbf{H}_{S1} \overset{d}{\sim}\diag(\mathbf{v}_1^H)\mathbf{H}_{S1}$ if $K_{S1} = 0$. If $K_{S1} = 0$, then $\mathbf{H}_{S1} = \sqrt{\alpha_{S1}}\tilde{\mathbf{H}}_{S1}$.
As all elements of $\tilde{\mathbf{H}}_{S1}$ are i.i.d.  $\mathcal{CN}(0,1)$, $\mathrm{vec}(\mathbf{H}_{S1}) \sim \mathcal{CN}(0, \alpha_{S1}\mathbf{I}_{T_ST_1})$. Besides, we have:
\begin{align}\label{eq:HS1equilavent}
	\mathrm{vec}(\diag(\mathbf{v}_1^H)\mathbf{H}_{S1}) \stackrel{(a)}{=} (\mathbf{I}_{T_S}\otimes\diag(\mathbf{v}_1^H)) \mathrm{vec}(\mathbf{H}_{S1}) \stackrel{(b)}{\sim} \mathcal{CN}(0, \alpha_{S1}\mathbf{I}_{T_ST_1}).
\end{align}
where (a) is due to \cite[Theorem 7.16]{Schott1997Matrix}, and (b) is due to \cite[(A.26)]{Tse2005WireComm} and the fact that $\mathbf{I}_{T_S}\otimes\diag(\mathbf{v}_1^H)$ is an unitary matrix (noting that $\diag(\mathbf{v}_1^H)$ is an unitary matrix). Thus, we have $\mathbf{H}_{S1} \overset{d}{\sim}\diag(\mathbf{v}_1^H)\mathbf{H}_{S1}$. Next, we show $\mathbf{H}_{12}\overset{d}{\sim}\mathbf{H}_{12}\diag(\mathbf{v}_1^H)$ and $\mathbf{h}_{1U}^H\overset{d}{\sim}\mathbf{h}_{1U}^H\diag(\mathbf{v}_1^H)$ if $K_{12}=K_{1U} =0$.  If $K_{12}=K_{1U} =0$, then $\mathbf{H}_{12} = \sqrt{\alpha_{12}}\tilde{\mathbf{H}}_{12}$ and  $\mathbf{h}_{1U}^H = \sqrt{\alpha_{1U}}\tilde{\mathbf{h}}_{1U}^H$. As all elements of $\tilde{\mathbf{H}}_{12}$ and $\tilde{\mathbf{h}}_{1U}^H$ are i.i.d.  $\mathcal{CN}(0,1)$, $\mathrm{vec}(\mathbf{H}_{12}) \sim \mathcal{CN}(0, \alpha_{12}\mathbf{I}_{T_1T_2})$ and $\mathbf{h}_{1U}^H \sim \mathcal{CN}(0, \alpha_{1U}\mathbf{I}_{T_1})$. Besides, we have:
\begin{align}
	& \mathrm{vec}(\mathbf{H}_{12}\diag(\mathbf{v}_1^H)) \stackrel{(c)}{=} ( \diag(\mathbf{v}_1^H)\otimes \mathbf{I}_{T_2}) \mathrm{vec}(\mathbf{H}_{12}) \stackrel{(d)}{\sim} \mathcal{CN}(0, \alpha_{12}\mathbf{I}_{T_1T_2}),\label{eq:h2Uequilavent}
	\\
	& \mathbf{h}_{1U}^H\diag(\mathbf{v}_1^H) \stackrel{(e)}{\sim} \mathcal{CN}(0, \alpha_{1U}\mathbf{I}_{T_1}).\label{eq:H12equilavent}
\end{align}
where (c) is due to \cite[Theorem 7.16]{Schott1997Matrix}, (d) is due to \cite[A.26]{Tse2005WireComm} and the fact that $\diag(\mathbf{v}_1^H)\otimes \mathbf{I}_{T_2}$ is unitary matrix, and (e) is due to the fact that $\diag(\mathbf{v}_1^H)$ is an unitary matrix. Thus, we have $\mathbf{H}_{12}\overset{d}{\sim}\mathbf{H}_{12}\diag(\mathbf{v}_1^H)$ and $\mathbf{h}_{1U}^H\overset{d}{\sim}\mathbf{h}_{1U}^H\diag(\mathbf{v}_1^H)$. Therefore, we complete the proof of Statement (i).

Next, we prove Statement (ii). Similarly, it is sufficient to show $\mathbf{h}_{2U}\overset{d}{\sim} \mathbf{H}_{2U}\diag(\mathbf{v}_2^H)$ or $\mathbf{H}_{12}\overset{d}{\sim}\diag(\mathbf{v}_2^H)\mathbf{H}_{12}$ and $\mathbf{H}_{S2}\overset{d}{\sim}\diag(\mathbf{v}_2^H)\mathbf{H}_{S2}$.
Following the proof for Statement (i), we can show that $\mathbf{h}_{2U}\overset{d}{\sim} \mathbf{H}_{2U}\diag(\mathbf{v}_2^H)$ if $K_{2U} = 0$, and $\mathbf{H}_{12}\overset{d}{\sim}\diag(\mathbf{v}_2^H)\mathbf{H}_{12}$ and $\mathbf{H}_{S2}\overset{d}{\sim}\diag(\mathbf{v}_2^H)\mathbf{H}_{S2}$ if $K_{12} = K_{S2} = 0$. Therefore, we can show Statement (ii).

\section*{Appendix B: Proof of Theorem~\ref{thm:analyGeneralK}} \label{proof:analyGeneralK}
In what follows, we derive the expression of $\gamma(\bm \phi_1, \bm \phi_2)$ for all $K_{S1},K_{S2},K_{12},K_{1U},K_{2U},K_{SU}\geq 0$, based on which we can readily obtain $\gamma^{(0)}$, $\gamma^{(1)}(\bm \phi_1)$, $\gamma^{(2)}(\bm \phi_2)$, and  $\gamma^{(3)}(\bm \phi_1, \bm \phi_2)$ given by \eqref{eq:gamma0}, \eqref{eq:gamma1}, \eqref{eq:gamma2}, and \eqref{eq:gamma3}, respectively. Note that for all $K_{S1},K_{S2},K_{12},K_{1U},K_{2U},K_{SU}\geq 0$, we have:
\begin{small}
	\begin{align}\label{eq:gammag2}
		\gamma(\bm \phi_1, \bm \phi_2) & = \mathbb{E} \left[ \|\mathbf{h}_e(\bm \phi_1, \bm \phi_2)\|_2^2\right] \nonumber
		\stackrel{(a)}{=}\mathbb{E} \left[\|
		\underbrace{\mathbf{h}_{SU}^H}_{\triangleq\mathbf{x}_0^H} + \sum_{l\in\mathcal{L}}\underbrace{\mathbf{h}_{lU}^H \diag(\mathbf{v}_l^H) \mathbf{H}_{Sl}}_{\triangleq \mathbf{x}_l^H} + \underbrace{\mathbf{h}_{2U}^H \diag(\mathbf{v}_2^H) \mathbf{H}_{12} \diag(\mathbf{v}_1^H) \mathbf{H}_{S1}}_{\triangleq\mathbf{x}_3^H}\|_2^2
		\right] \nonumber
		\\& =\sum_{i=0}^{3} \mathbb{E} \big[	\| \mathbf{x}_i^H \|_2^2 \big] +2\sum_{i = 0}^{2}\sum_{j = i+1}^{3}  \mathfrak{R} \left\{\mathbb{E}\big[ \mathbf{x}_j^H\mathbf{x}_i \big]\right\},
	\end{align}
\end{small}where $(a)$ is due to \eqref{eq:he}. Thus, it remains to calculate $\mathbb{E} \big[\| \mathbf{x}_i^H \|_2^2 \big], i = 0,1,2,3$ and $\mathbb{E}\big[\mathbf{x}_j^H\mathbf{x}_i \big],i=0,1,2, j= i+1,..,3$. As all elements of  $\tilde{\mathbf{h}}_{1U}$, $\tilde{\mathbf{h}}_{2U}$, $\tilde{\mathbf{h}}_{SU}$, $\tilde{\mathbf{H}}_{S1}$, $\tilde{\mathbf{H}}_{S2}$, and $\tilde{\mathbf{H}}_{12}$ are i.i.d. $\mathcal{CN}(0,1)$, we can easily show that multiplying any of them with an independent random matrix or a constant matrix producing a random matrix (or vector) with zero mean. This fact will be used in the following proof.

(i) We have:
\begin{align}
	%X_0^2
	\mathbb{E} \big[\| \mathbf{x}_0^H \|_2^2 \big] & \stackrel{(a)}{=} \mathbb{E} \left[\| \sqrt{L_{\overline{SU}}}\bar{\mathbf{h}}_{SU}^H+ \sqrt{L_{\widetilde{SU}}}\tilde{\mathbf{h}}_{SU}^H\|^2_2 \right]
	\stackrel{(b)}{=} L_{\overline{Sl}} \| \bar{\mathbf{h}}_{SU}^H\|^2_2 + L_{\widetilde{Sl}} \mathbb{E} \left[ \| \tilde{\mathbf{h}}_{SU}^H\|^2_2\right] \stackrel{(c)}{=}\alpha_{SU}T_S, \label{eq:x0x0}
\end{align}
where $(a)$ is due to \eqref{eq:hab}, $(b)$ is due to $\mathbb{E}[\bar{\mathbf{h}}_{SU}^H \tilde{\mathbf{h}}_{SU}]=0$, and $(c)$ is due to $\| \bar{\mathbf{h}}_{SU}^H\|^2_2 = T_S$ and $ \mathbb{E}\left[ \| \tilde{\mathbf{h}}_{SU}^H\|^2_2\right] = T_S$.

(ii) For $l \in \mathcal{L}$, we have:
\begin{align}
	\mathbb{E}\left[ \| \tilde{\mathbf{h}}_{lU}^H \diag(\mathbf{v}_l^H) \bar{\mathbf{H}}_{Sl} \|_2^2\right] & = \mathbb{E}\left[ \mathrm{Tr}\left(\tilde{\mathbf{h}}_{lU}^H \diag(\mathbf{v}_l^H) \bar{\mathbf{H}}_{Sl}\bar{\mathbf{H}}_{Sl}^H\diag(\mathbf{v}_l)\tilde{\mathbf{h}}_{lU} \right)\right] \nonumber
	\\
	& \stackrel{(a)}{=}\mathbb{E}\left[ \mathrm{Tr}\left( \bar{\mathbf{H}}_{Sl}\bar{\mathbf{H}}_{Sl}^H\tilde{\mathbf{h}}_{lU} \tilde{\mathbf{h}}_{lU}^H \right)\right] \stackrel{(b)}{=}\mathrm{Tr}\left( \bar{\mathbf{H}}_{Sl}\bar{\mathbf{H}}_{Sl}^H\mathbb{E}\left[\tilde{\mathbf{h}}_{lU} \tilde{\mathbf{h}}_{lU}^H \right]\right) \nonumber
	\\
	& \stackrel{(c)}{=}\mathrm{Tr}\left( \bar{\mathbf{H}}_{Sl}\bar{\mathbf{H}}_{Sl}^H\right)  = T_ST_l, \label{eq:ii1}
\end{align}
where $(a)$ is due to $\tilde{\mathbf{h}}_{lU}^H \diag(\mathbf{v}_l^H) \overset{d}{\sim}\tilde{\mathbf{h}}_{lU}^H$ (which is shown in Appendix A) and $\mathrm{Tr}(\mathbf{A}\mathbf{B}\mathbf{C}) = \mathrm{Tr}(\mathbf{B}\mathbf{C}\mathbf{A})$, and $(e)$ is due to $\mathbb{E} \left[\tilde{\mathbf{h}}_{lU} \tilde{\mathbf{h}}_{lU}^H\right] = \mathbf{I}_{T_l}$. Similarly, we can show:
\begin{align}
	\mathbb{E}\left[ \| \bar{\mathbf{h}}_{lU}^H \diag(\mathbf{v}_l^H) \tilde{\mathbf{H}}_{Sl} \|_2^2\right]  = T_ST_l, \quad
	\mathbb{E}\left[ \| \tilde{\mathbf{h}}_{lU}^H \diag(\mathbf{v}_l^H) \tilde{\mathbf{H}}_{Sl} \|_2^2\right] = T_ST_l. \label{eq:ii2}
\end{align}
Thus, we have:
\begin{align}
	%X_l^2
	\mathbb{E} \big[\| \mathbf{x}_l^H \|_2^2 \big] & \stackrel{(d)}{=} \mathbb{E} \big[\| ( \sqrt{L_{\overline{lU}}}\bar{\mathbf{h}}_{lU}^H +\sqrt{L_{\widetilde{lU}}}\tilde{\mathbf{h}}_{lU}^H)\diag(\mathbf{v}_l^H)(\sqrt{L_{\overline{Sl}}}\bar{\mathbf{H}}_{Sl}^H +\sqrt{L_{\widetilde{Sl}}}\tilde{\mathbf{H}}_{Sl}^H)  \|_2^2 \big] \nonumber
	\\
	& \stackrel{(e)}{=} L_{\overline{Sl},\overline{lU}} \| \bar{\mathbf{h}}_{lU}^H \diag(\mathbf{v}_l^H) \bar{\mathbf{H}}_{Sl} \|_2^2 + L_{\widetilde{Sl},\overline{lU}} \mathbb{E}\left[ \| \tilde{\mathbf{h}}_{lU}^H \diag(\mathbf{v}_l^H) \bar{\mathbf{H}}_{Sl} \|_2^2\right] \nonumber
	\\
	& \quad + L_{\overline{Sl},\widetilde{lU}}\mathbb{E}\left[ \| \bar{\mathbf{h}}_{lU}^H \diag(\mathbf{v}_l^H) \tilde{\mathbf{H}}_{Sl} \|_2^2\right] +L_{\widetilde{Sl},\widetilde{lU}}\mathbb{E}\left[ \| \tilde{\mathbf{h}}_{lU}^H \diag(\mathbf{v}_l^H) \tilde{\mathbf{H}}_{Sl} \|_2^2\right] \nonumber
	\\
	& \stackrel{(f)}{=}                                                  L_{\overline{Sl},\overline{lU}} \| \bar{\mathbf{h}}_{lU}^H \diag(\mathbf{v}_l^H) \bar{\mathbf{H}}_{Sl} \|_2^2 + (L_{\widetilde{Sl},\overline{lU}} + L_{\overline{Sl},\widetilde{lU}} +L_{\widetilde{Sl},\widetilde{lU}})T_ST_l \nonumber
	\\
	& \stackrel{(g)}{=}                                                L_{\overline{Sl},\overline{lU}} \mathbf{v}_l^H  \diag{(\bar{\mathbf{h}}_{lU}^H)} \bar{\mathbf{H}}_{Sl} \bar{\mathbf{H}}_{Sl}^H \diag{(\bar{\mathbf{h}}_{lU})}  \mathbf{v}_l + (L_{\widetilde{Sl},\overline{lU}} + L_{\overline{Sl},\widetilde{lU}} +L_{\widetilde{Sl},\widetilde{lU}})T_ST_l, \label{eq:xlxl}
\end{align}
where $(d)$ is due to \eqref{eq:Hab} and \eqref{eq:hab}, $(e)$ is due to $\mathbb{E}\left[\tilde{\mathbf{h}}_{lU}^H \diag(\mathbf{v}_l^H) \bar{\mathbf{H}}_{Sl} \right]= \mathbf{0}_{1 \times T_S}$,
$
\mathbb{E}\left[\tilde{\mathbf{h}}_{lU}^H \diag(\mathbf{v}_l^H) \tilde{\mathbf{H}}_{Sl} \right]= \mathbf{0}_{1 \times T_S}$, and
$\mathbb{E}\left[\tilde{\mathbf{h}}_{lU}^H \diag(\mathbf{v}_l^H) \tilde{\mathbf{H}}_{Sl} \right] = \mathbf{0}_{1 \times T_S}$, $(f)$ is due to \eqref{eq:ii1} and \eqref{eq:ii2},  and $(g)$ is due to $ \bar{\mathbf{h}}_{lU}^H \diag(\mathbf{v}_l^H ) = \mathbf{v}_l^H  \diag{(\bar{\mathbf{h}}_{lU}^H)}$.

(iii) Similarly to the derivation of \eqref{eq:ii1}, we have:
\begin{align}
	& \mathbb{E}\left[\| \tilde{\mathbf{h}}_{2U}^H \diag{(\mathbf{v}_2^H)} \bar{\mathbf{H}}_{12} \diag{(\mathbf{v}_1^H)} \bar{\mathbf{H}}_{S1}\|_2^2\right] =\|\bar{\mathbf{H}}_{12} \diag{(\mathbf{v}_1^H)} \bar{\mathbf{H}}_{S1}\|_F^2, \nonumber
	\\
	& \mathbb{E}\left[\| \bar{\mathbf{h}}_{2U}^H \diag{(\mathbf{v}_2^H)} \bar{\mathbf{H}}_{12} \diag{(\mathbf{v}_1^H)} \tilde{\mathbf{H}}_{S1}\|_2^2\right] = T_S\|\bar{\mathbf{h}}_{2U}^H \diag{(\mathbf{v}_2^H)} \bar{\mathbf{H}}_{12}\|_2^2,\nonumber
	\\
	& \mathbb{E}\left[ \|\tilde{\mathbf{h}}_{2U}^H \diag{(\mathbf{v}_2^H)} \bar{\mathbf{H}}_{12} \diag{(\mathbf{v}_1^H)} \tilde{\mathbf{H}}_{S1}\|_2^2\right]= \mathbb{E}\left[ \|\bar{\mathbf{h}}_{2U}^H \diag{(\mathbf{v}_2^H)} \tilde{\mathbf{H}}_{12} \diag{(\mathbf{v}_1^H)} \bar{\mathbf{H}}_{S1}\|_2^2 \right] \nonumber
	\\
	& =\mathbb{E}\left[ \|\tilde{\mathbf{h}}_{2U}^H \diag{(\mathbf{v}_2^H)} \tilde{\mathbf{H}}_{12} \diag{(\mathbf{v}_1^H)} \bar{\mathbf{H}}_{S1}\|_2^2 \right]=\mathbb{E}\left[ \|\tilde{\mathbf{h}}_{2U}^H \diag{(\mathbf{v}_2^H)} \tilde{\mathbf{H}}_{12} \diag{(\mathbf{v}_1^H)} \tilde{\mathbf{H}}_{S1} \|_2^2\right]  \nonumber
	\\
	& =\mathbb{E}\left[ \|\tilde{\mathbf{h}}_{2U}^H \diag{(\mathbf{v}_2^H)} \tilde{\mathbf{H}}_{12} \diag{(\mathbf{v}_1^H)} \tilde{\mathbf{H}}_{S1}\|_2^2 \right] = T_ST_1T_2.\label{eq:iii1}
\end{align}
Thus, we have:
\begin{small}
	\begin{align}
		%X_3^2
		\mathbb{E} \big[\| \mathbf{x}_3^H \|_2^2 \big] & \stackrel{(a)}{=} \mathbb{E} \left[ \|(\sqrt{L_{\overline{2U}}}\bar{\mathbf{h}}_{2U}^H +\sqrt{L_{\widetilde{2U}}}\tilde{\mathbf{h}}_{2U}^H )\diag{(\mathbf{v}_2^H)} (\sqrt{L_{\overline{12}}}\bar{\mathbf{H}}_{12} +\sqrt{L_{\widetilde{12}}}\tilde{\mathbf{H}}_{12})\diag{(\mathbf{v}_1^H)} (\sqrt{L_{\overline{S1}}}\bar{\mathbf{H}}_{S1} + \sqrt{L_{\widetilde{S1}}}\tilde{\mathbf{H}}_{S1})\|_2^2 \right]\nonumber
		\\
		& \stackrel{(b)}{=} L_{\overline{S1},\overline{12},\overline{2U}} \|\bar{\mathbf{h}}_{2U}^H \diag{(\mathbf{v}_2^H)} \bar{\mathbf{H}}_{12} \diag{(\mathbf{v}_1^H)} \bar{\mathbf{H}}_{S1} \|_2^2 + L_{\overline{S1},\overline{12},\widetilde{2U}} \mathbb{E}\left[ \|\tilde{\mathbf{h}}_{2U}^H \diag{(\mathbf{v}_2^H)} \bar{\mathbf{H}}_{12} \diag{(\mathbf{v}_1^H)} \bar{\mathbf{H}}_{S1} \|_2^2\right] \nonumber
		\\
		& \quad +L_{\widetilde{S1},\overline{12},\overline{2U}}\mathbb{E}\left[ \|\bar{\mathbf{h}}_{2U}^H \diag{(\mathbf{v}_2^H)} \bar{\mathbf{H}}_{12} \diag{(\mathbf{v}_1^H)} \tilde{\mathbf{H}}_{S1} \|_2^2\right] +L_{\widetilde{S1},\overline{12},\widetilde{2U}} \mathbb{E}\left[ \|\tilde{\mathbf{h}}_{2U}^H \diag{(\mathbf{v}_2^H)} \bar{\mathbf{H}}_{12} \diag{(\mathbf{v}_1^H)} \tilde{\mathbf{H}}_{S1} \|_2^2\right] \nonumber
		\\
		& \quad+L_{\overline{S1},\widetilde{12},\overline{2U}}\mathbb{E}\left[ \|\bar{\mathbf{h}}_{2U}^H \diag{(\mathbf{v}_2^H)} \tilde{\mathbf{H}}_{12} \diag{(\mathbf{v}_1^H)} \bar{\mathbf{H}}_{S1} \|_2^2\right] +L_{\overline{S1},\widetilde{12},\widetilde{2U}}\mathbb{E}\left[ \|\tilde{\mathbf{h}}_{2U}^H \diag{(\mathbf{v}_2^H)} \tilde{\mathbf{H}}_{12} \diag{(\mathbf{v}_1^H)} \bar{\mathbf{H}}_{S1} \|_2^2\right] \nonumber
		\\
		& \quad+L_{\widetilde{S1},\widetilde{12},\overline{2U}}\mathbb{E}\left[ \|\tilde{\mathbf{h}}_{2U}^H \diag{(\mathbf{v}_2^H)} \tilde{\mathbf{H}}_{12} \diag{(\mathbf{v}_1^H)} \tilde{\mathbf{H}}_{S1} \|_2^2\right] + L_{\widetilde{S1},\widetilde{12},\widetilde{2U}}\mathbb{E}\left[ \|\tilde{\mathbf{h}}_{2U}^H \diag{(\mathbf{v}_2^H)} \tilde{\mathbf{H}}_{12} \diag{(\mathbf{v}_1^H)} \tilde{\mathbf{H}}_{S1} \|_2^2\right] \nonumber
		\\
		& \stackrel{(c)}{=} L_{\overline{S1},\overline{12},\overline{2U}}
		\|\bar{\mathbf{h}}_{2U}^H \diag{(\mathbf{v}_2^H)} \bar{\mathbf{H}}_{12} \diag{(\mathbf{v}_1^H)} \bar{\mathbf{H}}_{S1} \|_2^2  +L_{\overline{S1},\overline{12},\widetilde{2U}} \| \bar{\mathbf{H}}_{12}\diag(\mathbf{v}_1^H) \bar{\mathbf{H}}_{S1} \|_F^2
		\nonumber
		\\
		& \quad +L_{\widetilde{S1},\overline{12},\overline{2U}}T_S \| \bar{\mathbf{h}}_{2U}\diag(\mathbf{v}_2^H) \bar{\mathbf{H}}_{12}\|_2^2		\nonumber
		\\
		& \quad +\big(L_{\widetilde{S1},\overline{12},\widetilde{2U}} L_{\overline{S1},\widetilde{12},\overline{2U}}
		+L_{\overline{S1},\widetilde{12},\widetilde{2U}}+L_{\widetilde{S1},\widetilde{12},\overline{2U}}+L_{\widetilde{S1},\widetilde{12},\widetilde{2U}}\big)T_ST_1T_2 \nonumber
		\\
		& \stackrel{(d)}{=}                                                 L_{\overline{S1},\overline{12},\overline{2U}}T_S \mathbf{v}_2^H  \left(\diag(\bar{\mathbf{h}}_{2U}^H) \bar{\mathbf{H}}_{12}\diag(\mathbf{a}_{A,S1})\right)\mathbf{v}_1^*\mathbf{v}_1^T \left(\diag(\bar{\mathbf{h}}_{2U}^H) \bar{\mathbf{H}}_{12}\diag(\mathbf{a}_{A,S1})\right)^H \mathbf{v}_2  \nonumber
		\\
		& \quad +L_{\overline{S1},\overline{12},\widetilde{2U}}T_2 \mathbf{v}_1^H  \diag(\mathbf{a}^H_{D,12})\bar{\mathbf{H}}_{S1}\bar{\mathbf{H}}_{S1}^H\diag(\mathbf{a}_{D,12})  \mathbf{v}_1 +L_{\widetilde{S1},\overline{12},\overline{2U}}T_S\mathbf{v}_2^H  \diag(\bar{\mathbf{h}}_{2U}^H)\bar{\mathbf{H}}_{12}\bar{\mathbf{H}}_{12}^H\diag(\bar{\mathbf{h}}_{2U}) \mathbf{v}_2 \nonumber
		\\
		& \quad +\big(L_{\widetilde{S1},\overline{12},\widetilde{2U}} +L_{\overline{S1},\widetilde{12},\overline{2U}}
		+L_{\overline{S1},\widetilde{12},\widetilde{2U}}+L_{\widetilde{S1},\widetilde{12},\overline{2U}}+L_{\widetilde{S1},\widetilde{12},\widetilde{2U}}\big)T_ST_1T_2, \label{eq:x3x3}
	\end{align}
\end{small}
where $(a)$ is due to \eqref{eq:Hab} and \eqref{eq:hab}, $(b)$ is due to
\begin{align*}
	& \mathbb{E}\left[ \tilde{\mathbf{h}}_{2U}^H \diag{(\mathbf{v}_2^H)} \bar{\mathbf{H}}_{12} \diag{(\mathbf{v}_1^H)} \bar{\mathbf{H}}_{S1}\right] =\mathbb{E}\left[ \bar{\mathbf{h}}_{2U}^H \diag{(\mathbf{v}_2^H)} \bar{\mathbf{H}}_{12} \diag{(\mathbf{v}_1^H)} \tilde{\mathbf{H}}_{S1}\right]
	\\
	& =\mathbb{E}\left[ \tilde{\mathbf{h}}_{2U}^H \diag{(\mathbf{v}_2^H)} \bar{\mathbf{H}}_{12} \diag{(\mathbf{v}_1^H)} \tilde{\mathbf{H}}_{S1}\right] =\mathbb{E}\left[ \bar{\mathbf{h}}_{2U}^H \diag{(\mathbf{v}_2^H)} \tilde{\mathbf{H}}_{12} \diag{(\mathbf{v}_1^H)} \bar{\mathbf{H}}_{S1} \right]
	\\
	& =\mathbb{E}\left[ \tilde{\mathbf{h}}_{2U}^H \diag{(\mathbf{v}_2^H)} \tilde{\mathbf{H}}_{12} \diag{(\mathbf{v}_1^H)} \bar{\mathbf{H}}_{S1} \right] =\mathbb{E}\left[ \tilde{\mathbf{h}}_{2U}^H \diag{(\mathbf{v}_2^H)} \tilde{\mathbf{H}}_{12} \diag{(\mathbf{v}_1^H)} \tilde{\mathbf{H}}_{S1} \right]
	\\
	& =\mathbb{E}\left[ \tilde{\mathbf{h}}_{2U}^H \diag{(\mathbf{v}_2^H)} \tilde{\mathbf{H}}_{12} \diag{(\mathbf{v}_1^H)} \tilde{\mathbf{H}}_{S1} \right] = \mathbf{0}_{1\times T_S},
\end{align*}
$(c)$ is due to \eqref{eq:iii1}, and $(d)$ is due to $\bar{\mathbf{h}}_{2U}^H \diag{(\mathbf{v}_2^H)} =  \mathbf{v}_2^H\diag{(\bar{\mathbf{h}}_{2U}^H)}$ and $\diag{(\mathbf{v}_1^H)} \mathbf{a}_{A,S1} = \diag{( \mathbf{a}_{A,S1})}\mathbf{v}_1^*$.

(iv) For $l \in \mathcal{L}$, we have:
\begin{align}
	%X_l^H * X_0
	\mathbb{E}\big[\mathbf{x}_l^H\mathbf{x}_0  \big] & \stackrel{(a)}{=} \mathbb{E}\big[( \sqrt{L_{\overline{lU}}}\bar{\mathbf{h}}_{lU}^H +\sqrt{L_{\widetilde{lU}}}\tilde{\mathbf{h}}_{lU}^H)\diag(\mathbf{v}_l^H)(\sqrt{L_{\overline{Sl}}}\bar{\mathbf{H}}_{Sl}^H +\sqrt{L_{\widetilde{Sl}}}\tilde{\mathbf{H}}_{Sl}^H) (\sqrt{L_{\overline{SU}}} \bar{\mathbf{h}}_{SU} + \sqrt{L_{\widetilde{SU}}} \tilde{\mathbf{h}}_{SU})  \big] \nonumber
	\\
	& \stackrel{(b)}{=}\sqrt { L_{\overline{SU}}  L_{\overline{Sl},\overline{lU}} }  \bar{\mathbf{h}}_{lU}^H \diag(\mathbf{v}_l^H) \bar{\mathbf{H}}_{Sl} \bar{\mathbf{h}}_{SU} \stackrel{(c)}{=}\sqrt { L_{\overline{SU}}  L_{\overline{Sl},\overline{lU}} } \mathbf{v}_l^H   \diag(\bar{\mathbf{h}}_{lU}^H   )\bar{\mathbf{H}}_{Sl}\bar{\mathbf{h}}_{SU}, \label{eq:xlx0}
\end{align}
where $(a)$ is due to \eqref{eq:Hab} and \eqref{eq:hab}, $(b)$ is due to $\mathbb{E}\left[\tilde{\mathbf{h}}_{lU}^H \diag(\mathbf{v}_l^H) \bar{\mathbf{H}}_{Sl}\right] =\mathbb{E}\left[\bar{\mathbf{h}}_{lU}^H \diag(\mathbf{v}_l^H) \tilde{\mathbf{H}}_{Sl}\right]=\mathbb{E}\left[\tilde{\mathbf{h}}_{lU}^H \diag(\mathbf{v}_l^H) \tilde{\mathbf{H}}_{Sl}\right]=\mathbf{0}_{1\times T_S}$ and $\mathbb{E} \left[ \tilde{\mathbf{h}}_{SU} \right] = \mathbf{0}_{T_S}$, $(c)$ is due to $\bar{\mathbf{h}}_{lU}^H \diag(\mathbf{v}_l^H) = \mathbf{v}_l^H   \diag(\bar{\mathbf{h}}_{lU}^H)$. Similarly, we can show:
\begin{align}
	% X_3^H * X_0
	\mathbb{E}\big[\mathbf{x}_3^H\mathbf{x}_0 \big]=  & \sqrt{L_{\overline{SU}}  L_{\overline{S1},\overline{12},\overline{2U}} } \bar{\mathbf{h}}_{2U}^H \diag(\mathbf{v}_2^H) \bar{\mathbf{H}}_{12}\diag(\mathbf{v}_1^H) \bar{\mathbf{H}}_{S1} \bar{\mathbf{h}}_{SU} \nonumber
	\\
	=                                                 & \sqrt{L_{\overline{SU}}  L_{\overline{S1},\overline{12},\overline{2U}} } \mathbf{v}_2^H  \diag(\bar{\mathbf{h}}_{2U}^H) \bar{\mathbf{H}}_{12} \diag(\bar{\mathbf{H}}_{S1} \bar{\mathbf{h}}_{SU})  \mathbf{v}_1^*, \label{eq:x3x0}
	\\
	% X_2^H * X_1
	\mathbb{E}\big[\mathbf{x}_2^H\mathbf{x}_1  \big]= & \sqrt{ L_{\overline{S1},\overline{1U}} L_{\overline{S2},\overline{2U}} } \bar{\mathbf{h}}_{2U}^H \diag(\mathbf{v}_2^H) \bar{\mathbf{H}}_{S2}  \bar{\mathbf{H}}_{S1}^H \diag(\mathbf{v}_1) \bar{\mathbf{h}}_{1U} \nonumber
	\\
	=                                                 & \sqrt{ L_{\overline{S1},\overline{1U}} L_{\overline{S2},\overline{2U}} }\mathbf{v}_2^H  \diag(\bar{\mathbf{h}}_{2U}^H)  \bar{\mathbf{H}}_{S2}  \bar{\mathbf{H}}_{S1}^H \diag(\bar{\mathbf{h}}_{1U}) \mathbf{v}_1, \label{eq:x2x1}
	\\
	%X_3^H * X_1
	\mathbb{E}\big[ \mathbf{x}_3^H\mathbf{x}_1 \big]
	=                                                 & \sqrt{L_{\overline{S1},\overline{1U}}L_{\overline{S1},\overline{12},\overline{2U}}}\bar{\mathbf{h}}_{2U}^H \diag(\mathbf{v}_2) \bar{\mathbf{H}}_{12}\diag(\mathbf{v}_1^H) \bar{\mathbf{H}}_{S1} \bar{\mathbf{H}}_{S1}^H\diag(\mathbf{v}_1) \bar{\mathbf{h}}_{1U}
	\\
	& +\sqrt{L_{\widetilde{S1},\overline{1U}} L_{\widetilde{S1},\overline{12},\overline{2U}}}T_S \bar{\mathbf{h}}_{2U}^H \diag(\mathbf{v}_2^H) \bar{\mathbf{H}}_{12} \bar{\mathbf{h}}_{1U} \nonumber
	\\
	=                                                 & \mathbf{v}_1^H \diag(\mathbf{v}_2^H \diag(\bar{\mathbf{h}}_{2U}^H) \bar{\mathbf{H}}_{12}) \left(\sqrt{L_{\overline{S1},\overline{1U}}L_{\overline{S1},\overline{12},\overline{2U}}}\bar{\mathbf{H}}_{S1} \bar{\mathbf{H}}_{S1}^H \diag( \bar{\mathbf{h}}_{1U})\mathbf{v}_1\right)   \nonumber
	\\
	& + \sqrt{L_{\widetilde{S1},\overline{1U}} L_{\widetilde{S1},\overline{12},\overline{2U}}}T_S\mathbf{v}_2^H \diag(\bar{\mathbf{h}}_{2U}^H) \bar{\mathbf{H}}_{12} \bar{\mathbf{h}}_{1U}, \label{eq:x3x1}
	\\
	%X_3^H*X_2
	\mathbb{E}\big[\mathbf{x}_3^H\mathbf{x}_2 \big]=  & \sqrt{L_{\overline{S2},\overline{2U}} L_{\overline{S1},\overline{12},\overline{2U}}}\bar{\mathbf{h}}_{2U}^H \diag(\mathbf{v}_2^H) \bar{\mathbf{H}}_{12} \diag(\mathbf{v}_1^H) \bar{\mathbf{H}}_{S1} \bar{\mathbf{H}}_{S2}^H \diag(\mathbf{v}_2) \bar{\mathbf{h}}_{2U} \nonumber
	\\
	& +\sqrt{L_{\overline{S2},\widetilde{2U}} L_{\overline{S1},\overline{12},\widetilde{2U}}} \mathrm{Tr}\left(\bar{\mathbf{H}}_{12}\diag(\mathbf{v}_1^H) \bar{\mathbf{H}}_{S1} \bar{\mathbf{H}}_{S2}^H \right) \nonumber
	\\
	=                                                 & \mathbf{v}_1^H \diag(\mathbf{v}_2^H \diag(\bar{\mathbf{h}}_{2U}^H) \bar{\mathbf{H}}_{12}) \left(\sqrt{L_{\overline{S2},\overline{2U}} L_{\overline{S1},\overline{12},\overline{2U}}} \bar{\mathbf{H}}_{S1} \bar{\mathbf{H}}_{S2}^H \diag( \bar{\mathbf{h}}_{2U})\mathbf{v}_2\right)
	\nonumber
	\\
	& + \sqrt{L_{\overline{S2},\widetilde{2U}} L_{\overline{S1},\overline{12},\widetilde{2U}}}\mathbf{v}_1^H  \mathrm{diag}\left(\mathbf{a}^H_{D,12}\right)
	\bar{\mathbf{H}}_{S1}
	\bar{\mathbf{H}}_{S2}^H\mathbf{a}_{A,12}, \label{eq:x3x2}
\end{align}
%	where $(r)$ is due to $\mathbb{E}\left[\tilde{\mathbf{h}}_{2U}^H \diag(\mathbf{v}_2^H) \bar{\mathbf{H}}_{12} \diag(\mathbf{v}_1^H) \bar{\mathbf{H}}_{S1} \bar{\mathbf{H}}_{S2}^H \diag(\mathbf{v}_2) \tilde{\mathbf{h}}_{2U}\right] = \mathrm{Tr}\left(\bar{\mathbf{H}}_{12} \diag(\mathbf{v}_1^H) \bar{\mathbf{H}}_{S1} \bar{\mathbf{H}}_{S2}^H\right)$. $(s)$ is due to $\mathrm{Tr}\left(\bar{\mathbf{H}}_{12} \diag(\mathbf{v}_1^H) \bar{\mathbf{H}}_{S1} \bar{\mathbf{H}}_{S2}^H\right) = \mathrm{Tr}\left(\mathbf{a}_{A,12}\mathbf{a}^H_{D,12} \diag(\mathbf{v}_1^H) \bar{\mathbf{H}}_{S1} \bar{\mathbf{H}}_{S2}^H\right) = \mathbf{v}_1^H \diag(\mathbf{a}^H_{D,12}) \bar{\mathbf{H}}_{S1} \bar{\mathbf{H}}_{S2}^H\mathbf{a}_{A,12}$.

Finally, substituting \eqref{eq:x0x0}-\eqref{eq:x3x2} into \eqref{eq:gammag2}, we can obtain the expression of $\gamma(\bm \phi_1, \bm \phi_2)$ for all values of the Rician factors. Based on the general expression of $\gamma(\bm \phi_1, \bm \phi_2)$, we can readily obtain $\gamma^{(0)}$, $\gamma^{(1)}(\bm \phi_1)$, $\gamma^{(2)}(\bm \phi_2)$, and  $\gamma^{(3)}(\bm \phi_1, \bm \phi_2)$. Therefore, we complete the proof of Theorem~\ref{thm:analyGeneralK}.

\section*{Appendix C: Proof of Theorem~\ref{thm:caselGeneralSpecialCase}} \label{proof:caselGeneralSpecialCase}
Before proving Theorem~\ref{thm:caselGeneralSpecialCase}, we first show the following lemma.
\begin{Lem}\label{lem:premilarly}
	Suppose $\mathbf{A} = \mathbf{x} \mathbf{y}^H \in \mathbb{C}^{M \times N}$ with $\mathbf{x} \in \mathbb{C}^{M}$ and $\mathbf{y} \in \mathbb{C}^{N}$, $\mathbf{v} \triangleq \left( e^{-j \phi_{n}}\right)_{n\in\mathcal{N}} \in \mathbb{C}^{N}$ with $\mathcal{N} \triangleq\{ n=1,...,N\} $, $\mathbf{b} \in \mathbb{C}^{M}$, and $c_1,c_2\in\mathbb{R}$. The unique optimal solution of the problem
	\begin{equation}\label{prob:premilarly}
		\begin{aligned}
			\max_{\bm \phi} & \qquad  	c_1 \mathbf{v}^H\mathbf{A}^H\mathbf{A}\mathbf{v} + c_2 \mathfrak{R} \left\{\mathbf{v}^H \mathbf{A}^H\mathbf{b}\right\}
			\\
			\mathrm{s.t.}   & \qquad 	\phi_{n} \in [0,2\pi), n \in \mathcal{N},
		\end{aligned}
	\end{equation}
	is $\bm \phi^\star = \Lambda \left( - \angle(\mathbf{y}) - \angle(\mathbf{x}^H\mathbf{b}) \mathbf{1}_{N}\right)$.
\end{Lem}
\begin{IEEEproof}
	We have:
	\begin{align}\label{eq:Rxleqx}
		\mathfrak{R} \left\{   \mathbf{v}^H\mathbf{y} \mathbf{x}^H\mathbf{b}\right\} \overset{(a)}{\leq } |\mathbf{v}^H\mathbf{y} \mathbf{x}^H\mathbf{b}|=|\mathbf{v}^H\mathbf{y}|| \mathbf{x}^H\mathbf{b}|,
	\end{align}
	where $(a)$ is due to $\mathfrak{R}\{x\} \leq |x|, x \in \mathbb{C}$ and holds with equality if and only if:
	\begin{align} \label{eq:Lemma102}
		\angle(\mathbf{v}^H \mathbf{y}) + \angle(\mathbf{x}^H\mathbf{b}) = 0.
	\end{align}
	
	Besides, we have:
	\begin{align}\label{eq:triangle}
		|\mathbf{v}^H \mathbf{y}| =\left| \sum_{n=1}^{N} e^{j \phi_{n}} |y_n| e^{j  \angle(y_n)}\right|= \left| \sum_{n=1}^{N} |y_n| e^{j ( \phi_{n} + \angle(y_n))}\right| \stackrel{(b)}{\leq}  \sum_{n=1}^{N} |y_n|,
	\end{align}
	where $(b)$ is due to the triangle inequality and  holds with equality if and only if
	\begin{align}\label{eq:Lemma101}
		\bm \phi = -\angle(\mathbf{y}) + \psi\mathbf{1}_N,\psi \in \mathbb{R}.
	\end{align}
	
	Thus, we have:
	\begin{align*}
		c_1\mathbf{v}^H\mathbf{A}^H\mathbf{A}\mathbf{v} + c_2 \mathfrak{R} \left\{\mathbf{v}^H \mathbf{A}^H\mathbf{b}\right\} & \stackrel{(c)}{=}
		c_1 |\mathbf{v}^H\mathbf{y}|^2 \| \mathbf{x} \|_2^2   + c_2 \mathfrak{R} \left\{   \mathbf{v}^H\mathbf{y} \mathbf{x}^H\mathbf{b}\right\}
		\\
		& \stackrel{(d)}{\leq}  c_1 |\mathbf{v}^H\mathbf{y}|^2 \| \mathbf{x} \|_2^2 + c_2 |\mathbf{v}^H\mathbf{y}|| \mathbf{x}^H\mathbf{b}|
		\\
		& \stackrel{(e)}{\leq}  c_1  \left(\sum_{n=1}^{N} |y_n|\right)^2 \| \mathbf{x} \|_2^2+ c_2  \left(\sum_{n=1}^{N} |y_n|\right)|\mathbf{x}^H\mathbf{b}| ,
	\end{align*}
	where $(c)$ is due to $\mathbf{A} = \mathbf{x} \mathbf{y}^H$, $(d)$ is due to  \eqref{eq:Rxleqx}, and $(e)$ is due to \eqref{eq:triangle}.  $(d)$ and $(e)$ both hold with equality if and only if $\bm \phi$ satisfied  \eqref{eq:Lemma101} and \eqref{eq:Lemma102}. By noting that $\phi_{n} \in [0,2\pi), n \in \mathcal{N}$, the optimal solution of the problem in \eqref{prob:premilarly} is $\bm \phi^\star = \Lambda \left( - \angle(\mathbf{y}) - \angle(\mathbf{x}^H\mathbf{b}) \mathbf{1}_{N}\right)$.\end{IEEEproof}

Now, we prove Theorem~\ref{thm:caselGeneralSpecialCase}. First, we prove Statement (i).
Consider $K_{12}=0$. By \eqref{eq:gamma1}, we have: $$\gamma^{(1)}(\bm \phi_1) =L_{\overline{S1},\overline{1U}} \mathbf{v}_1^H \mathbf{G}_{11}^{H}\mathbf{G}_{11}\mathbf{v}_1
+2\sqrt{L_{\overline{SU}}L_{\overline{S1},\overline{1U}}}\mathfrak{R}\{\mathbf{v}_1^H \mathbf{G}_{11}^{H}\mathbf{g}_{11}\} + \gamma^{(0)}, $$ where $\mathbf{G}_{11} \triangleq   \bar{\mathbf{H}}_{S1}^H \mathrm{diag}\left(\bar{\mathbf{h}}_{1U}\right)$ and $\mathbf{g}_{11} \triangleq \mathbf{h}_{SU}$. By \eqref{eq:Hbar}, \eqref{eq:hbar}, and Lemma~\ref{lem:premilarly}, we have $\bm \phi_1^{\star} =\Lambda ( -\angle( \diag(\mathbf{a}^H_{D,1U}) \mathbf{a}_{A,S1}) $ $- \angle(\mathbf{a}_{D,S1}^H \mathbf{a}_{D,SU})\mathbf{1}_{T_1}) \overset{(e)}{=}\Lambda \left( -\bm \Delta_{\overline{S1},\overline{1U}} - \angle(r_{\overline{S1},\overline{SU}}) \mathbf{1}_{T_1}\right)$, where $(e)$ is due to \eqref{eq:Delta} and \eqref{eq:ralab}.
Then, consider $K_{1U}=0$. By \eqref{eq:gamma1}, we have: $$\gamma^{(1)}(\bm \phi_1)=  L_{\overline{S1},\overline{12},\widetilde{2U}}T_2 \mathbf{v}_1^H \mathbf{G}_{12}^{H}\mathbf{G}_{12} \mathbf{v}_1 + 2\sqrt{L_{\overline{S2},\widetilde{2U}}L_{\overline{S1},\overline{12},\widetilde{2U}}} \mathfrak{R}\left\{\mathbf{v}_1^H \mathbf{G}_{12}^{H}\mathbf{g}_{12}\right\}+\gamma^{(0)},$$ where $\mathbf{G}_{12} \triangleq \bar{\mathbf{H}}_{S1}^H \diag(\mathbf{a}_{D,12})$ and $\mathbf{g}_{12} \triangleq \bar{\mathbf{H}}_{S2}^H\mathbf{a}_{A,12}$. By \eqref{eq:Hbar}, \eqref{eq:hbar}, and Lemma~\ref{lem:premilarly}, we have $\bm \phi_1^{\star} =\Lambda \big( -\angle(\diag(\mathbf{a}^H_{D,12})$ $ \mathbf{a}_{A,S1}) - \angle(\mathbf{a}_{D,S1}^H \mathbf{a}_{D,S2}\mathbf{a}^H_{A,S2}\mathbf{a}_{A,12}) \mathbf{1}_{T_1}\big) \overset{(f)}{=} \Lambda (-\bm \Delta_{\overline{S1},\overline{12}} - \angle(r_{\overline{S1},\overline{S2}}$ $r_{\overline{S2},\overline{12}})\mathbf{1}_{T_1})$, where $(f)$ is due to \eqref{eq:Delta}, \eqref{eq:ralab}, and \eqref{eq:rablb}.

Next, we prove Statement (ii). Consider $K_{12}=0$. By \eqref{eq:gamma2}, we have: $$\gamma^{(2)}(\bm \phi_2) =L_{\overline{S2},\overline{2U}} \mathbf{v}_2^H \mathbf{G}_{21}^{H}\mathbf{G}_{21}\mathbf{v}_2+2\sqrt{L_{\overline{SU}}L_{\overline{S2},\overline{2U}}}\mathfrak{R}\{\mathbf{v}_2^H \mathbf{G}_{21}^{H}\mathbf{g}_{21}\} + \gamma^{(0)},$$ where $\mathbf{G}_{21} \triangleq   \bar{\mathbf{H}}_{S2}^H \mathrm{diag}\left(\bar{\mathbf{h}}_{2U}\right)$ and $\mathbf{g}_{21} \triangleq \mathbf{h}_{SU}$. By \eqref{eq:Hbar}, \eqref{eq:hbar}, and Lemma~\ref{lem:premilarly}, we have $\bm \phi_2^{\star} =\Lambda ( -\angle( \diag(\mathbf{a}^H_{D,2U}) \mathbf{a}_{A,S2})$ $ - \angle(\mathbf{a}_{D,S2}^H \mathbf{a}_{D,SU})\mathbf{1}_{T_2}) \overset{(g)}{=}\Lambda \left( -\bm \Delta_{\overline{S2},\overline{2U}} - \angle(r_{\overline{S2},\overline{SU}})\mathbf{1}_{T_2}\right)$, where $(g)$ is due to \eqref{eq:Delta} and \eqref{eq:ralab}.
Then, consider $K_{S2}=0$. By \eqref{eq:gamma2}, we have: $$\gamma^{(2)}(\bm \phi_2)=  L_{\widetilde{S1},\overline{12},\overline{2U}}T_S \mathbf{v}_2^H \mathbf{G}_{22}^{H}\mathbf{G}_{22} \mathbf{v}_2 + 2\sqrt{L_{\widetilde{S1},\overline{1U}}L_{\widetilde{S1},\overline{12},\overline{2U}}}T_S \mathfrak{R}\left\{\mathbf{v}_2^H \mathbf{G}_{22}^{H}\mathbf{g}_{22}\right\}+\gamma^{(0)},$$ with $\mathbf{G}_{22} \triangleq \bar{\mathbf{H}}_{12}^H \diag(\mathbf{a}_{D,12})$ and $\mathbf{g}_{12} \triangleq \bar{\mathbf{h}}_{1U}$. By \eqref{eq:Hbar}, \eqref{eq:hbar}, and Lemma~\ref{lem:premilarly}, we have $\bm \phi_1^{\star} =\Lambda \big( -\angle(\diag(\mathbf{a}^H_{D,12}) \mathbf{a}_{A,12})$ $ - \angle(\mathbf{a}_{D,12}^H \mathbf{a}_{D,1U})\mathbf{1}_{T_2} \big) \overset{(h)}{=} \Lambda (-\bm \Delta_{\overline{12},\overline{2U}}$ $ - \angle(r_{\overline{12},\overline{1U}}) \mathbf{1}_{T_2})$, where $(h)$ is due to \eqref{eq:Delta} and \eqref{eq:ralab}.

Therefore, we complete the proof of Theorem~\ref{thm:caselGeneralSpecialCase}.

\section*{Appendix D: Proof of Lemma~\ref{lem:caselGeneralCDslu}} \label{proof:caselGeneralCDslu} By \eqref{eq:gamma1}, \eqref{eq:gamma2}, \eqref{eq:Al}, and \eqref{eq:bl}, we have:
\begin{small}
	\begin{align*}
		& \gamma^{(l)}(\phi_{l,t}, \bm \phi_{l,-t}) \overset{}{=} \left(\sum_{k \in \mathcal{T}_l, k\ne t \atop i \in \mathcal{T}_l, i\ne t}A_{l,k,i} e^{j(\phi_{l,k}-\phi_{l,i})} + A_{l,t,t}\right)+ \left(e^{j\phi_{l,t}}\sum_{k \in \mathcal{T}_l, k \ne t} A_{l,t,k} e^{-j\phi_{l,k}} + e^{-j\phi_{l,t}}\sum_{k \in \mathcal{T}_l, k \ne t} A_{l,k,t} e^{j\phi_{l,k}}\right)
		\\
		& \qquad \qquad \qquad \quad + 2\mathfrak{R}\left\{\left( e^{j\phi_{l,t}} b_{l,t} + \sum_{k \in \mathcal{T}_l, k\ne t} e^{j\phi_{l,k}} b_{l,k}\right) \right\}  + \gamma^{(0)}
		\\
		& \overset{(a)}{=} 2\mathfrak{R}\left\{ e^{j\phi_{l,t}}\left( \sum_{k \in \mathcal{T}_l, k \ne t} A_{l,t,k} e^{-j\phi_{l,k}}+ b_{l,t}\right) + \sum_{k \in \mathcal{T}_l, k\ne t} e^{j\phi_{l,k}} b_{l,k} \right\}
		+ \sum_{k \in \mathcal{T}_l, k\ne t \atop i \in \mathcal{T}_l, i\ne t}A_{l,k,i} e^{j(\phi_{l,k}-\phi_{l,i})} + A_{l,t,t} + \gamma^{(0)},
	\end{align*}
\end{small}
where $(a)$ is due to  $\mathbf{A}_l = \mathbf{A}_l^H$. As $\bm \phi_{l,-t}$ is given, to solve Problem~\ref{prob:caselGeneralCD} is equivalent to solve:
\begin{align*}
	\max_{\phi_{l,t}\in[0,2\pi)} \mathfrak{R}\left\{e^{j\phi_{l,t}}\left( \sum_{k \in \mathcal{T}_l, k \ne t} A_{l,t,k} e^{-j\phi_{l,k}}+ b_{l,t}\right)\right\}.
\end{align*}

Besides, we have:
\begin{small}
	\begin{align*}
		\mathfrak{R}\left\{e^{j\phi_{l,t}}\left( \sum_{k \in \mathcal{T}_l, k \ne t} A_{l,t,k} e^{-j\phi_{l,k}}+ b_{l,t}\right)\right\} \overset{(b)}{\leq} \left|e^{j\phi_{l,t}}\left( \sum_{k \in \mathcal{T}_l, k \ne t} A_{l,t,k} e^{-j\phi_{l,k}}+ b_{l,t}\right)\right| \overset{}{=} \left| \sum_{k \in \mathcal{T}_l, k \ne t} A_{l,t,k} e^{-j\phi_{l,k}}+ b_{l,t}\right|,
	\end{align*}
\end{small}where $(b)$ is due to $\mathfrak{R}\left\{ x\right\} \leq |x|$, $x \in \mathbb{C}$ and  holds  with equality if and only if $\phi_{l,t} = -\angle\left(\sum_{k \in \mathcal{T}_l, k \ne t} A_{l,t,k} e^{-j\phi_{l,k}}+ b_{l,t}\right)$. By noting that $\phi_{l,t}\in[0,2\pi)$, the unique optimal solution of the equivalent problem is $\phi_{l,t}^\dagger = \Lambda \left(-\angle\left( \sum_{k \in \mathcal{T}_l, k \ne t} A_{l,t,k} e^{-j\phi_{l,k}}+ b_{l,t}\right)\right)$. Therefore, we complete the proof of Lemma~\ref{lem:caselGeneralCDslu}.

\section*{Appendix E: Proof of Theorem~\ref{thm:case3GeneralSpecialCase}} \label{proof:case3GeneralSpecialCase}
First, we prove statement (i). Consider $K_{S2}=K_{1U}=0$. We first have:
\begin{align}
	\left|\mathbf{v}_1^H \diag(\mathbf{a}^H_{D,12}) \mathbf{a}_{A,S1}\right| & \overset{(a)}{=} \left|\sum_{t \in \mathcal{T}_1}e^{j\left(\phi_{1,t}+\Delta_{\overline{S1},\overline{12},t}\right)} \right| \overset{(b)}{\leq} T_1, \label{eq:Ea}
	\\
	\left|\mathbf{v}_2^H \diag(\mathbf{h}^H_{2U}) \mathbf{a}_{A,12}\right|   & \overset{(c)}{=} \left|\sum_{t \in \mathcal{T}_2}e^{j\left(\phi_{2,t}+\Delta_{\overline{12},\overline{2U},t}\right)} \right| \overset{(d)}{\leq} T_2,\label{eq:Eb}
\end{align}
where $(a)$ and $(c)$ are due to \eqref{eq:Delta}, and $(b)$ and $(d)$ are due to the triangle inequality. Note that $(b)$ and $(d)$ hold with equality if and only if:
\begin{align}
	\bm \phi_1 & = -\bm \Delta_{\overline{S1},\overline{12}} + \psi_1\mathbf{1}_{T_1}, \psi_1 \in \mathbb{R},\label{eq:statement11}
	\\
	\bm \phi_2 & = -\bm \Delta_{\overline{12},\overline{2U}} + \psi_2\mathbf{1}_{T_2}, \psi_2 \in \mathbb{R}.\label{eq:statement12}
\end{align}
Then, we have:
\begin{align}
	\gamma^{(3)}(\bm \phi_1,\bm \phi_2)
	& \overset{(e)}{=} L_{\overline{S1},\overline{12},\widetilde{2U}}T_ST_2\left|\mathbf{v}_1^H \diag(\mathbf{a}^H_{D,12}) \mathbf{a}_{A,S1}\right|^2 + L_{\widetilde{S1},\overline{12},\overline{2U}} T_ST_1\left| \mathbf{v}_2^H \diag(\mathbf{h}_{2U}^H)\mathbf{a}_{A,12}\right|^2 \nonumber
	\\
	& \quad +L_{\overline{S1},\overline{12},\overline{2U}}T_S \left| \mathbf{v}_2^H \diag(\mathbf{h}_{2U}^H)\mathbf{a}_{A,12}\mathbf{a}^H_{D,12} \diag(\mathbf{a}_{A,S1}) \mathbf{v}_1^*\right|^2 \nonumber
	\\
	& \quad +2\sqrt{L_{\overline{SU}}  L_{\overline{S1},\overline{12},\overline{2U}} } \mathfrak{R}\left\{\mathbf{v}_2^H \diag(\mathbf{h}_{2U}^H)\mathbf{a}_{A,12}\mathbf{a}^H_{D,12} \diag(\mathbf{a}_{A,S1}) \mathbf{v}_1^* r_{\overline{S1},\overline{SU}} \right\} +\gamma^{(0)} \nonumber
	\\
	& \stackrel{(f)}{\leq} L_{\overline{S1},\overline{12},\widetilde{2U}}T_ST_2\left|\mathbf{v}_1^H \diag(\mathbf{a}^H_{D,12}) \mathbf{a}_{A,S1}\right|^2 + L_{\widetilde{S1},\overline{12},\overline{2U}} T_ST_1\left| \mathbf{v}_2^H \diag(\mathbf{h}_{2U}^H)\mathbf{a}_{A,12}\right|^2 \nonumber
	\\
	& \quad +L_{\overline{S1},\overline{12},\overline{2U}}T_S \left| \mathbf{v}_2^H \diag(\mathbf{h}_{2U}^H)\mathbf{a}_{A,12} \right|^2 \left| \mathbf{v}_1^H \diag(\mathbf{a}^H_{D,12}) \mathbf{a}_{A,S1} \right|^2 \nonumber
	\\
	& \quad+2\sqrt{L_{\overline{SU}}  L_{\overline{S1},\overline{12},\overline{2U}} } \left| \mathbf{v}_2^H \diag(\mathbf{h}_{2U}^H)\mathbf{a}_{A,12} \right| \left| \mathbf{v}_1^H \diag(\mathbf{a}^H_{D,12}) \mathbf{a}_{A,S1} \right| \left|r_{\overline{S1},\overline{SU}} \right| +\gamma^{(0)} \nonumber
	\\
	& \stackrel{(g)}{\leq} L_{\overline{S1},\overline{12},\widetilde{2U}}T_ST_2T_1^2 + L_{\widetilde{S1},\overline{12},\overline{2U}} T_ST_1T_2^2
	+L_{\overline{S1},\overline{12},\overline{2U}}T_S T_1^2 T_2^2 \nonumber
	\\
	& \quad +2\sqrt{L_{\overline{SU}}  L_{\overline{S1},\overline{12},\overline{2U}} } T_1 T_2 \left|r_{\overline{S1},\overline{SU}} \right| +\gamma^{(0)}, \label{eq:Egamma3}
\end{align}
where $(e)$ is due to \eqref{eq:gamma3}, $(f)$ is due to $\mathbf{a}^H_{D,12} \diag(\mathbf{a}_{A,S1}) \mathbf{v}_1^* =\mathbf{v}_1^H \diag(\mathbf{a}^H_{D,12}) \mathbf{a}_{A,S1}$ and $\mathfrak{R}\{x\} \leq |x|,x \in \mathbb{C}$, and $(g)$ is due to \eqref{eq:Ea} and \eqref{eq:Eb}. Note that  $(f)$ holds with equality if and only if
\begin{align}
	\angle \left(\mathbf{v}_2^H \diag(\mathbf{h}_{2U}^H)\mathbf{a}_{A,12} \right)+\angle\left( \mathbf{v}_1^H \diag(\mathbf{a}^H_{D,12}) \mathbf{a}_{A,S1}\right) +  \angle\left(r_{\overline{S1},\overline{SU}}\right) = 0.\label{eq:statement13}
\end{align}
Besides, note that $(g)$  holds with equality if and only if \eqref{eq:Ea} and \eqref{eq:Eb} hold with equality, i.e., \eqref{eq:statement11} and \eqref{eq:statement12} hold.  Thus, the upper bound in \eqref{eq:Egamma3} is achieved if and only if \eqref{eq:statement11}, \eqref{eq:statement12}, and  \eqref{eq:statement13} hold simultaneously. By substituting \eqref{eq:statement11} and \eqref{eq:statement12} into \eqref{eq:statement13}, we have $\psi_1 + \psi_2 + \angle\left(r_{\overline{S1},\overline{SU}}\right)=0$.  By noting that $\bm \phi_{l,t} \in [0,2\pi), t \in \mathcal{T}_l,l\in \mathcal{L}$ and by choosing $\psi_1 =- \frac{1}{2}\angle(r_{\overline{S1},\overline{SU}}) + \psi$ and $\psi_2 =- \frac{1}{2}\angle(r_{\overline{S1},\overline{SU}}) - \psi$ for all $\psi \in \mathbb{R}$, we can show that any $(\bm \phi_1^\star,\bm \phi_2^\star)$ with $\bm \phi_1^\star = \Lambda\left(-\bm \Delta_{\overline{S1},\overline{12}} + \psi\mathbf{1}_{T_1} - \frac{1}{2}\angle\left(r_{\overline{S1},\overline{SU}}\right)\mathbf{1}_{T_1}\right)$ and $\bm \phi_2^\star = \Lambda\left(-\bm \Delta_{\overline{12},\overline{2U}} - \psi\mathbf{1}_{T_2} - \frac{1}{2}\angle\left(r_{\overline{S1},\overline{SU}}\right)\mathbf{1}_{T_2}\right)$, $\psi \in \mathbb{R}$ is an optimal solution of Problem \ref{prob:caselGeneral}.

Next, we prove statement (ii). Consider $K_{SU}=K_{12}=0$. We first have:
\begin{align}
	\left|\mathbf{v}_l^H \diag(\mathbf{h}^H_{lU}) \mathbf{a}_{A,Sl}\right| \overset{(h)}{=} \left|\sum_{t \in \mathcal{T}_l}e^{j\left(\phi_{l,t}+\Delta_{\overline{Sl},\overline{lU},t}\right)} \right| \overset{(i)}{\leq} T_l, l \in \mathcal{L},\label{eq:El}
\end{align}
where $(h)$ is due to \eqref{eq:Delta}, $(i)$ is due to the triangle inequality and holds with equality if and \mbox{only if:}
\begin{align}\label{eq:statementl}
	\bm \phi_l = \Lambda\left(-\bm \Delta_{\overline{Sl},\overline{lU}} + \psi_l\right),  \psi_l \in \mathbb{R}.
\end{align}
Then, we have:
\begin{align}
	\gamma^{(3)}(\bm \phi_1,\bm \phi_2)  \overset{(j)}{=} & \sum_{l=1}^{2} L_{\overline{Sl},\overline{lU}}T_S\left|\mathbf{v}_l^H \diag(\bar{\mathbf{h}}_{lU}^H)\mathbf{a}_{A,Sl}\right|^2\nonumber
	\\&+2\sqrt{ L_{\overline{S1},\overline{1U}} L_{\overline{S2},\overline{2U}} }\mathfrak{R}\left\{\mathbf{v}_2^H  \mathrm{diag} \left(\bar{\mathbf{h}}_{2U}^H\right) \bar{\mathbf{H}}_{S2}  \bar{\mathbf{H}}_{S1}^H  \mathrm{diag} \left(\bar{\mathbf{h}}_{1U}\right)\mathbf{v}_1 \right\} + \gamma^{(0)} \nonumber
	\\
	\overset{(k)}{\leq}                                   & \sum_{l=1}^{2} L_{\overline{Sl},\overline{lU}}T_S\left|\mathbf{v}_l^H \diag(\bar{\mathbf{h}}_{lU}^H)\mathbf{a}_{A,Sl}\right|^2 + \gamma^{(0)}\nonumber
	\\
	& + 2\sqrt{ L_{\overline{S1},\overline{1U}} L_{\overline{S2},\overline{2U}} } \left|\mathbf{v}_2^H  \mathrm{diag} \left(\bar{\mathbf{h}}_{2U}^H\right) \mathbf{a}_{A,S2} \right|\left| \mathbf{v}_1^H   \mathrm{diag} \left(\bar{\mathbf{h}}_{1U}^H\right)\mathbf{a}_{A,S1}\right| \left|r_{\overline{S1},\overline{S2}}\right|\nonumber
	\\
	\overset{(l)}{\leq}                                   & \sum_{l=1}^{2} L_{\overline{Sl},\overline{lU}}T_ST_l^2
	+ 2\sqrt{ L_{\overline{S1},\overline{1U}} L_{\overline{S2},\overline{2U}} } T_1T_2 \left|r_{\overline{S1},\overline{S2}}\right|+ \gamma^{(0)},\label{eq:Egamma32}
\end{align}
where $(j)$ is due to \eqref{eq:gamma3}, $(k)$ is due to $\mathfrak{R}\{x\} \leq |x|,x \in \mathbb{C}$, and $(l)$ is due to \eqref{eq:El}. Note that $(k)$ holds with equality if and only if: \begin{align}\label{eq:statement21}
	\angle\left(\mathbf{v}_1^H \diag(\bar{\mathbf{h}}_{1U}^H)\mathbf{a}_{A,S1}\right)-\angle\left(\mathbf{v}_2^H \diag(\bar{\mathbf{h}}_{2U}^H)\mathbf{a}_{A,S2}\right)+\angle\left(r_{\overline{S1},\overline{S2}}\right)=0.
\end{align}
Besides, note that $(l)$ holds with equality if and only if \eqref{eq:El} holds with equality, i.e., \eqref{eq:statementl} holds. Thus, the upper bound in \eqref{eq:Egamma32} is achieved if and only if \eqref{eq:statementl} and \eqref{eq:statement21} hold simultaneously. By substituting \eqref{eq:statementl} into \eqref{eq:statement21}, we have $\psi_1 - \psi_2 +\angle\left(r_{\overline{S1},\overline{S2}}\right)=0$. By noting that $\bm \phi_{l,t} \in [0,2\pi), t \in \mathcal{T}_l,l\in \mathcal{L}$ and by choosing $\psi_1= \psi$ and  $\psi_2= \psi + \angle(r_{\overline{S1},\overline{S2}})$ for all $\psi \in \mathbb{R}$, we can show that any $(\bm \phi_1^\star,\bm \phi_2^\star)$ with $\phi^{\star}_1 = \Lambda(-\bm \Delta_{\overline{S1},\overline{1U}}+\psi\mathbf{1}_{T_1})$, $\phi^{\star}_2=\Lambda(-\bm \Delta_{\overline{S2},\overline{2U}}+\psi\mathbf{1}_{T_2} + \angle(r_{\overline{S1},\overline{S2}})\mathbf{1}_{T_2})$, $\psi \in \mathbb{R}$ is an optimal solution of Problem~\ref{prob:caselGeneral}.

Therefore, we complete the proof of Theorem~\ref{thm:case3GeneralSpecialCase}.
%	where the equality holds when $\angle\left( \mathbf{v}_2^H  \mathrm{diag} \left(\bar{\mathbf{h}}_{2U}^H\right)\mathbf{a}_{A,S2}\right) -\angle\left(  \mathbf{v}_1^H   \mathrm{diag} \left(\bar{\mathbf{h}}_{1U}^H\right)\mathbf{a}_{A,S1}\right)  + \angle\left(r_{\overline{S1},\overline{S2}}\right) = 0$. For all $\psi \in \mathbb{R}$, Problem \ref{prob:case3General} for Case 3 with $K_{SU}=K_{12}=0$ can be decomposed as the following two problems.
%	\begin{align*}
%		\max_{\bm \phi_1} & \quad \left|\mathbf{v}_1^H   \mathrm{diag} \left(\bar{\mathbf{h}}_{1U}^H\right)\mathbf{a}_{A,S1}\right|
%		\\
%		\mathrm{s.t.}     & \quad \phi_{1,t},t\in\mathcal{T}_1
%		\\
%		& \quad \angle\left(  \mathbf{v}_1^H   \mathrm{diag} \left(\bar{\mathbf{h}}_{1U}^H\right)\mathbf{a}_{A,S1}\right) -\psi =0
%		\\
%		\max_{\bm \phi_2} & \quad \left|\mathbf{v}_2^H  \mathrm{diag} \left(\bar{\mathbf{h}}_{2U}^H\right)\mathbf{a}_{A,S2}\right|
%		\\
%		\mathrm{s.t.}     & \quad \phi_{2,t},t\in\mathcal{T}_2
%		\\
%		& \quad \angle\left( \mathbf{v}_2^H  \mathrm{diag} \left(\bar{\mathbf{h}}_{2U}^H\right)\mathbf{a}_{A,S2}\right) + \angle\left(r_{\overline{S1},\overline{S2}}\right) - \psi = 0
%	\end{align*}
%	
%	By the triangle inequality, it is not difficult to show that the optimal solution to the above two problems are
%	$
%	\phi^{\star}_1 = \Lambda(-\bm \Delta_{\overline{S1},\overline{1U}}+\psi\mathbf{1}_{T_1})$ and $\phi^{\star}_2=\Lambda(-\bm \Delta_{\overline{S2},\overline{2U}}+\psi\mathbf{1}_{T_2} + \angle(r_{\overline{S1},\overline{S2}})\mathbf{1}_{T_2}).
%	$

\section*{Appendix F: Proof of Lemma~\ref{lem:case3GeneralCDslu}} \label{proof:case3GeneralCDslu}
By \eqref{eq:gamma3}, \eqref{eq:Cl}, and \eqref{eq:dl}, we have:
\begin{small}
	\begin{align*}
		& \gamma^{(3)}(\phi_{l,t}, \bm \phi_{l,-t},\phi_{-l}) \overset{}{=} \left(\sum_{k \in \mathcal{T}_l, k\ne t \atop i \in \mathcal{T}_l, i\ne t}C_{l,k,i} e^{j(\phi_{l,k}-\phi_{l,i})} + C_{l,t,t}\right)+ \left(e^{j\phi_{l,t}}\sum_{k \in \mathcal{T}_l, k \ne t} C_{l,t,k} e^{-j\phi_{l,k}} + e^{-j\phi_{l,t}}\sum_{k \in \mathcal{T}_l, k \ne t} C_{l,k,t} e^{j\phi_{l,k}}\right)
		\\
		& \qquad \qquad \qquad\qquad \quad\quad + 2\mathfrak{R}\left\{\left( e^{j\phi_{l,t}} d_{l,t} + \sum_{k \in \mathcal{T}_l, k\ne t} e^{j\phi_{l,k}} d_{l,k}\right) \right\}  + \gamma^{(0)}
		\\
		& \overset{(a)}{=} 2\mathfrak{R}\left\{ e^{j\phi_{l,t}}\left( \sum_{k \in \mathcal{T}_l, k \ne t} C_{l,t,k} e^{-j\phi_{l,k}}+ d_{l,t}\right) + \sum_{k \in \mathcal{T}_l, k\ne t} e^{j\phi_{l,k}} d_{l,k} \right\}
		+ \sum_{k \in \mathcal{T}_l, k\ne t \atop i \in \mathcal{T}_l, i\ne t}C_{l,k,i} e^{j(\phi_{l,k}-\phi_{l,i})} + C_{l,t,t} + \gamma^{(0)},
	\end{align*}
\end{small}where $(a)$ is due to $\mathbf{C}_l=\mathbf{C}_l^H$. As $\left(\bm \phi_{l,-t},\phi_{-l}\right)$ is given, to solve Problem~\ref{prob:case3GeneralCD} is equivalent to solve:
\begin{align*}
	\max_{\phi_{l,t}\in[0,2\pi)} \mathfrak{R}\left\{e^{j\phi_{l,t}}\left( \sum_{k \in \mathcal{T}_l, k \ne t} C_{l,t,k} e^{-j\phi_{l,k}}+ d_{l,t}\right)\right\}.
\end{align*}
Besides, we have:
\begin{align*}
	\mathfrak{R}\left\{e^{j\phi_{l,t}}\left( \sum_{k \in \mathcal{T}_l, k \ne t} C_{l,t,k} e^{-j\phi_{l,k}}+ d_{l,t} \right)\right\} \overset{(b)}{\leq} \left| \sum_{k \in \mathcal{T}_l, k \ne t} C_{l,t,k} e^{-j\phi_{l,k}}+ d_{l,t}\right|,
\end{align*}
where $(b)$ is due to $\mathfrak{R} \{ x\} \leq |x|$, $x \in \mathbb{C}$ and holds with equality if and only if $\phi_{l,t} = -\angle\left(\sum_{k \in \mathcal{T}_l, k \ne t} C_{l,t,k} e^{-j\phi_{l,k}}+ d_{l,t}\right)$. By noting that $\phi_{l,t}\in[0,2\pi)$, the unique optimal solution of the equivalent problem is $\phi_{l,t}^\dagger = \Lambda \left(-\angle\left( \sum_{k \in \mathcal{T}_l, k \ne t} C_{l,t,k} e^{-j\phi_{l,k}}+ d_{l,t}\right)\right)$. Therefore, we complete the proof of Lemma~\ref{lem:case3GeneralCDslu}.

\section*{Appendix G: Proof of Theorem~\ref{thm:doubleAsyT}} \label{proof:doubleAsyT}
Before proving Theorem~\ref{thm:doubleAsyT}, we first show:
\begin{align}
	\| \mathbf{v}_l^H \diag(\bar{\mathbf{h}}_{lU}^H ) \bar{\mathbf{H}}_{Sl} \|_2                                 & = \| \mathbf{v}_l^H \diag(\bar{\mathbf{h}}_{lU}^H ) \mathbf{a}_{A,Sl} \mathbf{a}^H_{D,Sl} \|_2 \nonumber
	\\
	& \overset{(a)}{\leq}  |\mathbf{v}_l^H \diag(\mathbf{a}_{D,lU}^H ) \mathbf{a}_{A, Sl} | \|\mathbf{a}^H_{D, Sl}\|_2 \overset{(b)}{\leq} \sqrt{T_S}T_l, \label{eq:vlHsl}
	\\
	|\mathbf{v}_2^H  \diag(\bar{\mathbf{h}}_{2U}^H) \bar{\mathbf{H}}_{12}\diag(\mathbf{a}_{A,S1})\mathbf{v}_1^*| & = \|\mathbf{v}_2^H  \diag(\mathbf{a}_{D,2U}^H)
	\mathbf{a}_{A,12} \mathbf{a}^H_{D,12}\diag(\mathbf{a}_{A,S1})\mathbf{v}_1^*\|_2 \nonumber
	\\
	& \overset{(c)}{\leq} |\mathbf{v}_2^H   \diag(\mathbf{a}_{D,2U}^H)
	\mathbf{a}_{A,12}| | \mathbf{a}^H_{D,12}\diag(\mathbf{a}_{A,S1})\mathbf{v}_1^*|\overset{(d)}{\leq} T_1T_2, \label{eq:v1H12v2}
	\\
	\| \mathbf{v}_1^H \diag(\mathbf{a}^H_{D,12}) \bar{\mathbf{H}}_{S1} \|_2                                      & =\| \mathbf{v}_1^H \diag(\mathbf{a}^H_{D,12}) \mathbf{a}_{A,S1} \mathbf{a}^H_{D,S1}\|_2 \nonumber
	\\
	& \overset{(e)}{\leq} | \mathbf{v}_1^H \diag(\mathbf{a}^H_{D,12}) \mathbf{a}_{A,S1}| \| \mathbf{a}^H_{D,S1}\|_2 \overset{(f)}{\leq} \sqrt{T_S}T_1,\label{eq:v1HS12}
	\\
	\| \mathbf{v}_2^H \diag(\bar{\mathbf{h}}_{2U}) \bar{\mathbf{H}}_{12}\|_2                                     & = \| \mathbf{v}_2^H \diag(\bar{\mathbf{h}}_{2U}) \mathbf{a}_{A,12} \mathbf{a}^H_{D,12}\|_2 \nonumber
	\\
	& \overset{(g)}{\leq} | \mathbf{v}_2^H \diag(\bar{\mathbf{h}}_{2U}) \mathbf{a}_{A,12}| \| \mathbf{a}^H_{D,12}\|_2\overset{(h)}{\leq} \sqrt{T_1}T_2,\label{eq:v2H12}
\end{align}
where $(a)$, $(c)$, $(e)$, and $(g)$ are due to Cauchy-Schwartz inequality, and $(b)$, $(d)$, $(f)$, and $(h)$ are due to the triangle inequality. Based on \eqref{eq:vlHsl},  \eqref{eq:v1H12v2}, \eqref{eq:v1HS12}, and \eqref{eq:v2H12}, we prove Theorem~\ref{thm:doubleAsyT} below.

First, consider Case 0. As $\lim_{T_1,T_2\rightarrow \infty} \frac{\gamma^{(0)}  }{(L_{\widetilde{S1},\overline{12},\widetilde{2U}} +L_{\overline{S1},\widetilde{12},\overline{2U}}+L_{\overline{S1},\widetilde{12},\widetilde{2U}}+L_{\widetilde{S1},\widetilde{12},\overline{2U}}+L_{\widetilde{S1},\widetilde{12},\widetilde{2U}})T_ST_1T_2}=1,$ we can show $\gamma^{(0)}       \stackrel{T_1,T_2\rightarrow\infty}{\sim} (L_{\widetilde{S1},\overline{12},\widetilde{2U}} +L_{\overline{S1},\widetilde{12},\overline{2U}}+L_{\overline{S1},\widetilde{12},\widetilde{2U}}+L_{\widetilde{S1},\widetilde{12},\overline{2U}}+L_{\widetilde{S1},\widetilde{12},\widetilde{2U}})T_ST_1T_2$.

Next, consider Case 1. For all $\bm\phi_1$, we have:
\begin{small}
	\begin{align}
		& \gamma^{(1)}(\bm\phi_1) \overset{(i)}{=} L_{\overline{S1},\overline{1U}} \| \mathbf{v}_1^H \diag(\bar{\mathbf{h}}_{1U}^H ) \bar{\mathbf{H}}_{S1} \|_2^2  + L_{\overline{S1},\overline{12},\widetilde{2U}}T_2 \left\| \mathbf{v}_1^H \diag(\mathbf{a}^H_{D,12}) \bar{\mathbf{H}}_{S1} \right\|_2^2+ \gamma^{(0)}
		\nonumber
		\\
		& \qquad\qquad\quad + 2\mathfrak{R} \left\{ \mathbf{v}_1^H \left(\sqrt{L_{\overline{SU}}L_{\overline{S1},\overline{1U}}}  \diag(\bar{\mathbf{h}}_{1U}^H ) \bar{\mathbf{H}}_{S1} \bar{\mathbf{h}}_{SU} +\sqrt{L_{\overline{S2},\widetilde{2U}}L_{\overline{S1},\overline{12},\widetilde{2U}}} \mathrm{diag}\left(\mathbf{a}_{D,12}^H\right) \bar{\mathbf{H}}_{S1} \bar{\mathbf{H}}_{S2}^H\mathbf{a}_{A,12}\right)\right\}
		\nonumber
		\\
		& \stackrel{(j)}{\leq}   L_{\overline{S1},\overline{1U}} \| \mathbf{v}_1^H \diag(\bar{\mathbf{h}}_{1U}^H ) \bar{\mathbf{H}}_{S1} \|_2^2  + L_{\overline{S1},\overline{12},\widetilde{2U}}T_2 \left\| \mathbf{v}_1^H \diag(\mathbf{a}^H_{D,12}) \bar{\mathbf{H}}_{S1} \right\|_2^2+ \gamma^{(0)}
		\nonumber
		\\
		& \quad + 2\sqrt{L_{\overline{SU}}L_{\overline{S1},\overline{1U}}} \| \mathbf{v}_1^H\diag(\bar{\mathbf{h}}_{1U}^H ) \bar{\mathbf{H}}_{S1} \|_2 \|\mathbf{a}_{D,SU}\|_2
		+ 2 \sqrt{L_{\overline{S2},\widetilde{2U}} L_{\overline{S1},\overline{12},\widetilde{2U}}} \|\mathbf{v}_1^H \mathrm{diag}\left(\mathbf{a}_{D,12}^H\right)\bar{\mathbf{H}}_{S1}\|_2 \|\mathbf{a}_{D,S2}\|_2 |r_{S2,12}|
		\nonumber
		\\
		& \stackrel{(k)}{\leq} L_{\overline{S1},\overline{1U}} T_ST_1^2  + L_{\overline{S1},\overline{12},\widetilde{2U}}T_ST_1^2T_2 + 2\sqrt{L_{\overline{SU}}L_{\overline{S1},\overline{1U}}} T_ST_1+ 2 \sqrt{L_{\overline{S2},\widetilde{2U}} L_{\overline{S1},\overline{12},\widetilde{2U}}} T_ST_1T_2 + \gamma^{(0)} \triangleq \gamma^{(1)\star}_{ub}, \label{eq:gamma1starub}
	\end{align}
\end{small}where $(i)$ is due to \eqref{eq:gamma1}, $(j)$ is due to $\mathfrak{R}\{x\}\leq |x|, x \in \mathbb{C}$ and Cauchy-Swartz inequality, and $(k)$ is due to \eqref{eq:vlHsl} and \eqref{eq:v1HS12}. On the other hand, we have:
\begin{small}
	\begin{align}
		& \gamma^{(1)}\left(\bm \hat{\phi}_1\right) = L_{\overline{S1},\overline{1U}} \| \hat{\mathbf{v}}_1^H \diag(\bar{\mathbf{h}}_{1U}^H ) \bar{\mathbf{H}}_{S1} \|_2^2  + L_{\overline{S1},\overline{12},\widetilde{2U}}T_2 \left\| \hat{\mathbf{v}}_1^H \diag(\mathbf{a}^H_{D,12}) \bar{\mathbf{H}}_{S1} \right\|_2^2+ \gamma^{(0)}
		\nonumber
		\\
		& \qquad \qquad \quad + 2\mathfrak{R} \left\{ \hat{\mathbf{v}}_1^H \left(\sqrt{L_{\overline{SU}}L_{\overline{S1},\overline{1U}}}  \diag(\bar{\mathbf{h}}_{1U}^H ) \bar{\mathbf{H}}_{S1} \bar{\mathbf{h}}_{SU} +\sqrt{L_{\overline{S2},\widetilde{2U}}} \mathrm{diag}\left(\mathbf{a}_{D,12}^H\right) \bar{\mathbf{H}}_{S1} \bar{\mathbf{H}}_{S2}^H\mathbf{a}_{A,12}\right)\right\}
		\nonumber
		\\
		& \stackrel{(l)}{\geq} L_{\overline{S1},\overline{12},\widetilde{2U}}T_2 \left\| \hat{\mathbf{v}}_1^H\diag(\mathbf{a}_{D,12}^H)\bar{\mathbf{H}}_{S1}\right\|_2^2 +\gamma^{(0)}
		- 2\sqrt{L_{\overline{SU}}L_{\overline{S1},\overline{1U}}} \| \hat{\mathbf{v}}_1^H\diag(\bar{\mathbf{h}}_{1U}^H ) \bar{\mathbf{H}}_{S1} \|_2 \|\mathbf{a}_{D,SU}\|_2
		\nonumber
		\\
		& \quad  - 2 \sqrt{L_{\overline{S2},\widetilde{2U}} L_{\overline{S1},\overline{12},\widetilde{2U}}} \|\hat{\mathbf{v}}_1^H \mathrm{diag}\left(\mathbf{a}_{D,12}^H\right)\bar{\mathbf{H}}_{S1}\|_2 \|\mathbf{a}_{D,S2}\|_2 |r_{S2,12}|
		\nonumber
		\\
		& \stackrel{(m)}{\geq} L_{\overline{S1},\overline{12},\widetilde{2U}}T_ST_1^2T_2-  2\sqrt { L_{\overline{SU}}  L_{\overline{S1},\overline{1U}} } T_1T_S - 2\sqrt{L_{\overline{S2},\widetilde{2U}} L_{\overline{S1},\overline{12},\widetilde{2U}}} T_1T_ST_2 + \gamma^{(0)}\triangleq \gamma^{(1)\star}_{lb},\label{eq:gamma1starlb}
	\end{align}
\end{small}where $\bm \hat{\phi}_1 \triangleq \Lambda \left(-\bm \Delta_{\overline{S1},\overline{12}} - \angle(r_{\overline{S1},\overline{S2}}r_{\overline{S2},\overline{12}})\mathbf{1}_{T_1}\right)$, $\hat{\mathbf{v}}_1 \triangleq  \left( e^{-j\hat{\phi}_{1,t}}\right)_{t \in \mathcal{T}_1}$, $(l)$ is due to $\mathfrak{R}\{ x\} \geq -|x|, x \in \mathbb{C}$ and Cauchy-Swartz inequality, and $(m)$ is due to \eqref{eq:vlHsl}, \eqref{eq:v1HS12}, and
$
\left\| \hat{\mathbf{v}}_1^H\diag(\mathbf{a}_{D,12}^H)\bar{\mathbf{H}}_{S1}\right\|_2^2
= T_S \left|
\sum_{t=1}^{T_1}e^{j(- \Delta_{\overline{S1},\overline{12},t} - \angle(r_{\overline{S1},\overline{S2}}r_{\overline{S2},\overline{12}}) +\Delta_{\overline{S1},\overline{12},t})} \right|_2^2=T_ST_1^2.
$
Clearly, we have $\gamma^{(1)\star}_{lb} \leq \gamma^{(1)\star}\leq \gamma^{(1)\star}_{ub}$. As
$
\lim_{T_1,T_2\rightarrow\infty} \frac{\gamma^{(1)\star}_{ub}}{L_{\overline{S1},\overline{12},\widetilde{2U}}T_ST_1^2T_2} = 1$ and $\lim_{T_1,T_2\rightarrow\infty} \frac{\gamma^{(1)\star}_{lb}}{L_{\overline{S1},\overline{12},\widetilde{2U}}T_ST_1^2T_2} = 1
$,
by the squeeze theorem, we have
$
\lim_{T_1,T_2\rightarrow\infty} \frac{\gamma^{(1)\star}}{L_{\overline{S1},\overline{12},\widetilde{2U}}T_ST_1^2T_2}=1.
$
Thus, we can show $\gamma^{(1)\star} \stackrel{T_1,T_2\rightarrow\infty}{\sim} L_{\overline{S1},\overline{12},\widetilde{2U}} T_ST_1^2T_2$.

Then, consider Case 2. For all $\bm \phi_2$, we have:
\begin{small}
	\begin{align}
		& \gamma^{(2)}(\bm\phi_2) \stackrel{(o)}{=} L_{\overline{S2},\overline{2U}} \| \mathbf{v}_2^H \diag(\bar{\mathbf{h}}_{2U}^H ) \bar{\mathbf{H}}_{S2} \|_2^2  + L_{\widetilde{S1},\overline{12},\overline{2U}}T_S \left\| \mathbf{v}_2^H \diag( \bar{\mathbf{h}}^H_{2U}) \bar{\mathbf{H}}_{12} \right\|_2^2+ \gamma^{(0)}
		\nonumber
		\\
		& \quad + 2\mathfrak{R}\left\{ \mathbf{v}_2^H \left( \sqrt{ L_{\overline{SU}}  L_{\overline{S2},\overline{2U}}} \mathrm{diag} \left( \bar{\mathbf{h}}_{2U}^H \right) \bar{\mathbf{H}}_{S2} \bar{\mathbf{h}}_{SU}+ \sqrt{L_{\widetilde{S1},\overline{1U}} L_{\widetilde{S1},\overline{12},\overline{2U}}}T_S \mathrm{diag} \left(\bar{\mathbf{h}}_{2U}^H\right) \bar{\mathbf{H}}_{12} \bar{\mathbf{h}}_{1U} \right)\right\}
		\nonumber
		\\
		& \stackrel{(p)}{\leq}   L_{\overline{S2},\overline{2U}} \| \mathbf{v}_2^H \diag(\bar{\mathbf{h}}_{2U}^H ) \bar{\mathbf{H}}_{S2} \|_2^2  + L_{\widetilde{S1},\overline{12},\overline{2U}}T_S \left\| \mathbf{v}_2^H \diag( \bar{\mathbf{h}}^H_{2U}) \bar{\mathbf{H}}_{12} \right\|_2^2+ \gamma^{(0)}
		\nonumber
		\\
		& \quad + 2\sqrt{L_{\overline{SU}}L_{\overline{S2},\overline{2U}}} \| \mathbf{v}_2^H\diag(\bar{\mathbf{h}}_{2U}^H ) \bar{\mathbf{H}}_{S2} \|_2 \|\mathbf{a}_{D,SU}\|_2
		+ 2 \sqrt{L_{\widetilde{S1},\overline{1U}} L_{\widetilde{S1},\overline{12},\overline{2U}}}T_S \|\mathbf{v}_2^H \mathrm{diag}\left( \bar{\mathbf{h}}_{2U}^H\right)\bar{\mathbf{H}}_{12}\|_2 \|\mathbf{a}_{D,1U}\|_2
		\nonumber
		\\
		& \stackrel{(q)}{\leq}  L_{\overline{S2},\overline{2U}} T_ST_2^2  + L_{\widetilde{S1},\overline{12},\overline{2U}}T_ST_1T_2^2 + 2\sqrt{L_{\overline{SU}}L_{\overline{S2},\overline{2U}}} T_ST_2+ 2 \sqrt{L_{\widetilde{S1},\overline{1U}} L_{\widetilde{S1},\overline{12},\overline{2U}}} T_ST_1T_2 + \gamma^{(0)}\triangleq\gamma^{(2)\star}_{ub}, \label{eq:gamma2starub}
	\end{align}
\end{small}where $(o)$ is due to \eqref{eq:gamma2}, $(p)$ is due to $\mathfrak{R}\{x\}\leq |x|, x \in \mathbb{C}$ and Cauchy-Schwartz inequality, and $(q)$ is due to \eqref{eq:vlHsl} and \eqref{eq:v2H12}.
On the other hand, we have:
\begin{small}
	\begin{align}
		& \gamma^{(2)}(\hat{\bm\phi}_2) = L_{\overline{S2},\overline{2U}} \| \hat{\mathbf{v}}_2^H \diag(\bar{\mathbf{h}}_{2U}^H ) \bar{\mathbf{H}}_{S2} \|_2^2  + L_{\widetilde{S1},\overline{12},\overline{2U}}T_S \left\| \hat{\mathbf{v}}_2^H \diag( \bar{\mathbf{h}}^H_{2U}) \bar{\mathbf{H}}_{12} \right\|_2^2+ \gamma^{(0)}
		\nonumber
		\\
		& \quad + 2\mathfrak{R}\left\{ \hat{\mathbf{v}}_2^H \left( \sqrt{ L_{\overline{SU}}  L_{\overline{S2},\overline{2U}}} \mathrm{diag} \left( \bar{\mathbf{h}}_{2U}^H \right) \bar{\mathbf{H}}_{S2} \bar{\mathbf{h}}_{SU}+ \sqrt{L_{\widetilde{S1},\overline{1U}} L_{\widetilde{S1},\overline{12},\overline{2U}}}T_S \mathrm{diag} \left(\bar{\mathbf{h}}_{2U}^H\right) \bar{\mathbf{H}}_{12} \bar{\mathbf{h}}_{1U} \right)\right\}
		\nonumber
		\\
		& \stackrel{(r)}{\geq}   L_{\widetilde{S1},\overline{12},\overline{2U}}T_S \left\| \hat{\mathbf{v}}_2^H \diag( \bar{\mathbf{h}}^H_{2U}) \bar{\mathbf{H}}_{12} \right\|_2^2+ \gamma^{(0)}
		\nonumber
		\\
		& \quad - 2\sqrt{L_{\overline{SU}}L_{\overline{S2},\overline{2U}}} \| \hat{\mathbf{v}}_2^H\diag(\bar{\mathbf{h}}_{2U}^H ) \bar{\mathbf{H}}_{S2} \|_2 \|\mathbf{a}_{D,SU}\|_2
		- 2 \sqrt{L_{\widetilde{S1},\overline{1U}} L_{\widetilde{S1},\overline{12},\overline{2U}}}T_S \|\hat{\mathbf{v}}_2^H \mathrm{diag}\left( \bar{\mathbf{h}}_{2U}^H\right)\bar{\mathbf{H}}_{12}\|_2 \|\mathbf{a}_{D,1U}\|_2
		\nonumber
		\\
		& \stackrel{(s)}{\geq} L_{\overline{S1},\overline{12},\widetilde{2U}}T_ST_1T_2^2 - 2\sqrt{L_{\overline{SU}}L_{\overline{S2},\overline{2U}}} T_ST_2- 2 \sqrt{L_{\widetilde{S1},\overline{1U}} L_{\widetilde{S1},\overline{12},\overline{2U}}} T_ST_1T_2 + \gamma^{(0)}\triangleq\gamma^{(2)\star}_{lb}, \label{eq:gamma2starlb}
	\end{align}
\end{small}where $\hat{\bm \phi_2} \triangleq \Lambda(-\bm \Delta_{\overline{12},\overline{2U}} - \angle(r_{\overline{12},\overline{1U}})\mathbf{1}_{T_2})$, $\hat{\mathbf{v}}_2 \triangleq  \left( e^{-j\hat{\phi}_{2,t}}\right)_{t \in \mathcal{T}_2}$, $(r)$ is due to $\mathfrak{R}\{ x\} \geq -|x|,x\in\mathbb{C}$ and Cauchy-Schwartz inequality, $(s)$ is due to \eqref{eq:vlHsl}, \eqref{eq:v2H12}, and
$\left\| \hat{\mathbf{v}}_2^H\diag(\bar{\mathbf{h}}_{2U}^H)\bar{\mathbf{H}}_{12}\right\|_2^2= T_1 |\sum_{t=1}^{T_2} $ $e^{j(- \Delta_{\overline{12},\overline{2U},t} - \angle(r_{\overline{12},\overline{1U}}) +\Delta_{\overline{12},\overline{2U},t})} |_2^2=T_1T_2^2.
$
Clearly, we have $\gamma^{(2)\star}_{lb} \leq \gamma^{(2)\star}\leq \gamma^{(2)\star}_{ub}$. As
$
\lim_{T_1,T_2\rightarrow\infty} $ $\frac{\gamma^{(2)\star}_{ub}}{L_{\overline{S1},\overline{12},\widetilde{2U}}T_ST_1T_2^2} = 1$ and $\lim_{T_1,T_2\rightarrow\infty} \frac{\gamma^{(2)\star}_{lb}}{L_{\overline{S1},\overline{12},\widetilde{2U}}T_ST_1T_2^2} = 1$, by the squeeze theorem, we have
$
\lim_{T_1,T_2\rightarrow\infty}$ $ \frac{\gamma^{(2)\star}}{L_{\overline{S1},\overline{12},\widetilde{2U}}T_ST_1T_2^2}=1,
$
Thus, we can show $\gamma^{(2)\star} \stackrel{T_1,T_2\rightarrow\infty}{\sim} L_{\overline{S1},\overline{12},\widetilde{2U}}T_ST_1T_2^2$.

Finally, consider Case 3. For all $\bm \phi_1$ and $\bm \phi_2$, we have:
\begin{align*}
	& \gamma^{(3)}(\bm \phi_1,\bm \phi_2) \stackrel{(t)}{\leq} \gamma^{(1)\star}_{ub} +\gamma^{(2)\star}_{ub} - \gamma^{(0)} + L_{\overline{S1},\overline{12},\overline{2U}}|\mathbf{v}_2^H  \diag(\bar{\mathbf{h}}_{2U}^H) \bar{\mathbf{H}}_{12}\diag(\mathbf{a}_{A,S1})\mathbf{v}_1^*|^2\| \mathbf{a}^H_{D,S1}\|_2^2
	\\
	& \quad +2\sqrt{L_{\overline{S1},\overline{1U}}L_{\overline{S1},\overline{12},\overline{2U}}} |\mathbf{v}_2^H  \diag(\bar{\mathbf{h}}_{2U}^H) \bar{\mathbf{H}}_{12}\diag(\mathbf{a}_{A,S1})\mathbf{v}_1^*|  \| \mathbf{a}^H_{D,S1}\|_2 \|\mathbf{v}_1^H\diag(\bar{\mathbf{h}}^H_{1U}) \bar{\mathbf{H}}_{S1}\|_2
	\\
	& \quad +2\sqrt{L_{\overline{S2},\overline{2U}} L_{\overline{S1},\overline{12},\overline{2U}}} |\mathbf{v}_2^H  \diag(\bar{\mathbf{h}}_{2U}^H) \bar{\mathbf{H}}_{12}\diag(\mathbf{a}_{A,S1})\mathbf{v}_1^*|  \| \mathbf{a}^H_{D,S1}\|_2 \|\mathbf{v}_2^H\diag(\bar{\mathbf{h}}^H_{2U}) \bar{\mathbf{H}}_{S2}\|_2
	\\
	& \quad +2\sqrt{L_{\overline{SU}}  L_{\overline{S1},\overline{12},\overline{2U}} } |\mathbf{v}_2^H  \diag(\bar{\mathbf{h}}_{2U}^H) \bar{\mathbf{H}}_{12}\diag(\mathbf{a}_{A,S1})\mathbf{v}_1^*| |r_{S1,SU}|
	\\
	& \quad +2 \sqrt{ L_{\overline{S1},\overline{1U}} L_{\overline{S2},\overline{2U}} } \|\mathbf{v}_2^H\diag(\bar{\mathbf{h}}^H_{2U}) \bar{\mathbf{H}}_{S2}\|_2 \|\mathbf{v}_1^H\diag(\bar{\mathbf{h}}^H_{1U}) \bar{\mathbf{H}}_{S1}\|_2
	\\
	& \stackrel{(u)}{\leq} \gamma^{(1)\star}_{ub} +\gamma^{(2)\star}_{ub} - \gamma^{(0)} + L_{\overline{S1},\overline{12},\overline{2U}}T_ST_1^2T_2^2 + 2\sqrt{L_{\overline{S1},\overline{1U}}L_{\overline{S1},\overline{12},\overline{2U}}T_S} T_1^2T_2
	\\
	& + 2\sqrt{L_{\overline{S2},\overline{2U}} L_{\overline{S1},\overline{12},\overline{2U}}T_S} T_1T_2^2 + 2(\sqrt{L_{\overline{SU}}  L_{\overline{S1},\overline{12},\overline{2U}} }+\sqrt{ L_{\overline{S1},\overline{1U}} L_{\overline{S2},\overline{2U}} })T_ST_1T_2 \triangleq \gamma^{(3)\star}_{ub},
\end{align*}
where $(t)$ is due to \eqref{eq:gamma1starub}, \eqref{eq:gamma2starub}, $\mathfrak{R}\{x\}\leq |x|, x \in \mathbb{C}$, and Cauchy-Schwartz inequality, and $(u)$ is due to \eqref{eq:vlHsl}-\eqref{eq:v2H12}.
On the other hand,  we have:
\begin{align*}
	& \gamma^{(3)}(\hat{\bm \phi}_1,\hat{\bm \phi}_2) \stackrel{(v)}{\geq} \gamma^{(1)\star}_{lb} +\gamma^{(2)\star}_{lb} - \gamma^{(0)} + L_{\overline{S1},\overline{12},\overline{2U}}|\hat{\mathbf{v}}_2^H  \diag(\bar{\mathbf{h}}_{2U}^H) \bar{\mathbf{H}}_{12}\diag(\mathbf{a}_{A,S1})\hat{\mathbf{v}}_1^*|^2\| \mathbf{a}^H_{D,S1}\|_2^2
	\\
	& \quad -2\sqrt{L_{\overline{S1},\overline{1U}}L_{\overline{S1},\overline{12},\overline{2U}}} |\hat{\mathbf{v}}_2^H  \diag(\bar{\mathbf{h}}_{2U}^H) \bar{\mathbf{H}}_{12}\diag(\mathbf{a}_{A,S1})\hat{\mathbf{v}}_1^*|  \| \mathbf{a}^H_{D,S1}\|_2 \|\hat{\mathbf{v}}_1^H\diag(\bar{\mathbf{h}}^H_{1U}) \bar{\mathbf{H}}_{S1}\|_2
	\\
	& \quad -2\sqrt{L_{\overline{S2},\overline{2U}} L_{\overline{S1},\overline{12},\overline{2U}}} |\hat{\mathbf{v}}_2^H  \diag(\bar{\mathbf{h}}_{2U}^H) \bar{\mathbf{H}}_{12}\diag(\mathbf{a}_{A,S1})\hat{\mathbf{v}}^*|  \| \mathbf{a}^H_{D,S1}\|_2 \|\hat{\mathbf{v}}_2^H\diag(\bar{\mathbf{h}}^H_{2U}) \bar{\mathbf{H}}_{S2}\|_2
	\\
	& \quad -2\sqrt{L_{\overline{SU}}  L_{\overline{S1},\overline{12},\overline{2U}} } |\hat{\mathbf{v}}_2^H  \diag(\bar{\mathbf{h}}_{2U}^H) \bar{\mathbf{H}}_{12}\diag(\mathbf{a}_{A,S1})\hat{\mathbf{v}}_1^*| |r_{S1,SU}|
	\\
	& \quad -2 \sqrt{ L_{\overline{S1},\overline{1U}} L_{\overline{S2},\overline{2U}} } \|\hat{\mathbf{v}}_2^H\diag(\bar{\mathbf{h}}^H_{2U}) \bar{\mathbf{H}}_{S2}\|_2 \hat{\mathbf{v}}_1^H\diag(\bar{\mathbf{h}}^H_{1U}) \bar{\mathbf{H}}_{S1}\|_2
	\\
	& \stackrel{(w)}{\geq} \gamma^{(1)\star}_{lb} +\gamma^{(2)\star}_{lb} - \gamma^{(0)} + L_{\overline{S1},\overline{12},\overline{2U}}T_ST_1^2T_2^2 - 2\sqrt{L_{\overline{S1},\overline{1U}}L_{\overline{S1},\overline{12},\overline{2U}} T_S}T_1^2T_2
	\\
	& - 2\sqrt{L_{\overline{S2},\overline{2U}} L_{\overline{S1},\overline{12},\overline{2U}}T_S} T_1T_2^2 - 2(\sqrt{L_{\overline{SU}}  L_{\overline{S1},\overline{12},\overline{2U}} }+\sqrt{ L_{\overline{S1},\overline{1U}} L_{\overline{S2},\overline{2U}} })T_ST_1T_2 \triangleq \gamma^{(3)\star}_{lb},
\end{align*}
where $(v)$ is due to \eqref{eq:gamma1starlb}, \eqref{eq:gamma2starlb}, $\mathfrak{R}\{ x\} \geq -|x|,x\in\mathbb{C}$, and Cauchy-Schwartz inequality, and $(w)$ is due to \eqref{eq:vlHsl}- \eqref{eq:v2H12}, and
$
\left|\hat{\mathbf{v}}_2^H  \diag(\bar{\mathbf{h}}_{2U}^H) \bar{\mathbf{H}}_{12}\diag(\mathbf{a}_{A,S1})\hat{\mathbf{v}}_1^*\right|
= \left| \sum_{t=1}^{T_1}e^{j(- \Delta_{\overline{S1},\overline{12},t} - \angle(r_{\overline{S1},\overline{S2}}r_{\overline{S2},\overline{12}}) +\Delta_{\overline{S1},\overline{12},t})}\right|^2 $ $\left|\sum_{t=1}^{T_2}e^{j(- \Delta_{\overline{12},\overline{2U},t} - \angle(r_{\overline{12},\overline{1U}}) +\Delta_{\overline{12},\overline{2U},t})}\right|^2 = T_1^2T_2^2.
$
Clearly, we have $\gamma^{(3)\star}_{lb} \leq \gamma^{(3)\star}\leq \gamma^{(3)\star}_{ub}$. As
$
\lim_{T_1,T_2\rightarrow\infty}$ $\frac{\gamma^{(3)\star}_{ub}}{L_{\overline{S1},\overline{12},\overline{2U}}T_ST_1^2T_2^2} =1$ and $\lim_{T_1,T_2\rightarrow\infty} \frac{\gamma^{(3)\star}_{lb}}{L_{\overline{S1},\overline{12},\overline{2U}}T_ST_1^2T_2^2} = 1$,
by the squeeze theorem, we have
$
\lim_{T_1,T_2\rightarrow\infty} \frac{\gamma^{(3)\star}}{L_{\overline{S1},\overline{12},\overline{2U}}T_ST_1^2T_2^2}=1,
$
Thus, we can show $\gamma^{(3)\star} \stackrel{T_1,T_2\rightarrow\infty}{\sim} L_{\overline{S1},\overline{12},\overline{2U}}T_ST_1^2T_2^2$.

Therefore, we complete the proof of Statement (i).
Furthermore, by substituting $T_1 = cT$ and $T_2=(1-c)T$ into Statement (i), we can easily show Statement (ii).

\section*{Appendix H: Proof of Lemma~\ref{lem:case3LargeK}} \label{proof:case3LargeK}
First, we prove Statement (i). By \eqref{eq:gamma3Infty}, we have $\bar{\gamma}^{(3)}(\bm \phi_1,\bm \phi_2) = \mathbf{v}_1^H\bar{\mathbf{G}}_1^H\bar{\mathbf{G}}_1\mathbf{v}_1 + 2\mathfrak{R}\{\bar{\mathbf{G}}_1^H \bar{\mathbf{g}}_1\} + \bar{\mathbf{g}}_1^H\bar{\mathbf{g}}_1$ where $\bar{\mathbf{G}}_1 \triangleq \bar{\mathbf{H}}_{S1}^H\diag(\sqrt{\alpha_{S1}\alpha_{1U}}\bar{\mathbf{h}}_{S1} + \sqrt{\alpha_{S1}\alpha_{12}\alpha_{2U} } \mathbf{B}_1^H \mathbf{v}_2) $ and $\bar{\mathbf{g}}_1^H \triangleq \sqrt{\alpha_{SU}} \bar{\mathbf{h}}_{SU}^H +\sqrt{\alpha_{S2}\alpha_{2U}}  \bar{\mathbf{h}}_{2U}^H$ $\diag( \mathbf{v}_2^H)\bar{\mathbf{H}}_{S2}$. By Lemma~\ref{lem:premilarly}, we can show that the unique optimal solution of Problem~\ref{prob:suboptphi1} is
\begin{align*}
	\bar{\bm \phi}_1^{\dagger} & = \Lambda \Big(-\angle\left( \diag\left( \sqrt{\alpha_{S1}\alpha_{1U}} \bar{\mathbf{h}}_{1U}^H + \sqrt{\alpha_{S1}\alpha_{1U}\alpha_{S1}\alpha_{12}\alpha_{2U}} \mathbf{v}_2^H \mathbf{B}_1\right) \mathbf{a}_{A,S1}\right) \nonumber
	\\
	& \quad\quad\ \  - \angle\big(\mathbf{a}^H_{D,S1} ( \sqrt{\alpha_{SU}} \bar{\mathbf{h}}_{SU} + \sqrt{\alpha_{S2}\alpha_{2U}}\bar{\mathbf{H}}_{S2}^H  \mathrm{diag}(\mathbf{v}_2)\bar{\mathbf{h}}_{2U})\big)\mathbf{1}_{T_1} \Big).
\end{align*}
Next, we prove Statement (ii). By \eqref{eq:gamma3}, we have:
\begin{align*}
	\bar{\gamma}^{(3)}(\phi_{2,t},\bm \phi_{2,-t},\bm \phi_1) \overset{(a)}{=} & 2\mathfrak{R}\left\{e^{j\phi_{2,t}}\left( \sum_{k \in \mathcal{T}_2, k \ne t} \bar{C}_{2,t,k} e^{-j\phi_{2,k}}+ \bar{d}_{2,t}\right)\right\} + \sum_{k \in \mathcal{T}_2, k\ne t}\sum_{ i \in \mathcal{T}_2, i\ne t}\bar{C}_{2,k,i} e^{j(\phi_{2,k}-\phi_{2,i})} \\&+\bar{C}_{2,t,t} + 2\mathfrak{R}\left\{ \sum_{k \in \mathcal{T}_2, k\ne t} e^{j\phi_{2,k}} \bar{d}_{2,k}\right\} + \gamma^{(0)},
\end{align*}
where $(a)$ is due to $\bar{\mathbf{C}}_l=\bar{\mathbf{C}}_l^H$. As $\left(\bm \phi_{2,-t},\bm \phi_1\right)$ is given, to solve Problem~\ref{prob:suboptphi2t} is equivalent to solve:
\begin{align*}
	\max_{\phi_{2,t}\in[0,2\pi)} \mathfrak{R}\left\{e^{j\phi_{l,t}}\left( \sum_{k \in \mathcal{T}_2, k \ne t} \bar{C}_{2,t,k} e^{-j\phi_{2,k}}+ \bar{d}_{2,t}\right)\right\}.
\end{align*}
Besides, we have:
$$\mathfrak{R}\left\{e^{j\phi_{2,t}}\left(\sum_{k \in \mathcal{T}_2, k \ne t} \bar{C}_{2,t,k} e^{-j\phi_{2,k}}+ \bar{d}_{2,t}\right)\right\} \overset{(b)}{\leq} \left| \sum_{k \in \mathcal{T}_2, k \ne t} \bar{C}_{2,t,k} e^{-j\phi_{2,k}}+ \bar{d}_{2,t}\right|,$$ where $(b)$ is due to $\mathfrak{R}\{x\}\leq|x|, x\in \mathbb{C}$ and holds with equality if and only if   $\phi_{2,t} = -\angle\big(\sum_{k \in \mathcal{T}_2, k \ne t} $ $\bar{C}_{2,t,k} e^{-j\phi_{2,k}}+ \bar{d}_{2,t}\big)$. By noting that $\phi_{2,t}\in[0,2\pi)$, the unique optimal solution of the equivalent problem is $\bar{\phi}_{2,t}^\dagger = \Lambda \left(-\angle\left( \sum_{k \in \mathcal{T}_2, k \ne t} \bar{C}_{2,t,k} e^{-j\phi_{2,k}}+ \bar{d}_{2,t}\right)\right)$. Therefore, we complete the proof of Lemma~\ref{lem:case3LargeK}.

	\bibliographystyle{IEEEtran}
	\bibliography{Double_IRS_ArXiv}
\end{document}